\documentclass[reprint,floatfix]{revtex4-2}

\usepackage{url}
\usepackage{amsmath}
\usepackage{graphicx}
\usepackage{dcolumn}
\usepackage{physics}
\usepackage{bm}
\usepackage{hyperref}
\usepackage{epsfig,colordvi}
\usepackage{hyperref}
\usepackage{graphicx}
\usepackage{dcolumn}
\usepackage{xspace}
\usepackage{times}
\usepackage{bm}
\usepackage{soul}
\usepackage{ulem}
\usepackage{xcolor}
\usepackage{color}
\usepackage{natbib}
\usepackage{float}
\usepackage{booktabs} 
\usepackage{lineno}

 \def\be{\begin{equation}}
 \def\ee{\end{equation}}
 \def\bea{\begin{eqnarray}}
 \def\eea{\end{eqnarray}}
 \def\bean{\begin{eqnarray*}}
 \def\eean{\end{eqnarray*}}

\newcommand{\EB}[1]{{\color{black}#1}}
\newcommand{\Yj}[1]{{\color{black}#1}}
\begin{document}

\title{Probing of EoS with clusters and hypernuclei}
\date{\today}

\author{
Yingjie Zhou$^{1}$, Susanne Gl{\"a}{\ss}el$^{3}$, Yue-Hang Leung$^{2}$, Viktar Kireyeu$^{4}$, Jiaxing Zhao$^{7,8}$, Hui Liu$^{11}$, Christoph Blume$^{3,1,7}$, Iouri Vassiliev$^{1}$, Vadim Voronyuk$^{4}$, Michael Winn$^{5}$, Norbert Herrmann$^{2,1}$, Yaping Wang$^{9}$, Nu Xu$^{9,10}$, J\"org Aichelin$^{5,6}$, Elena Bratkovskaya$^{1,7,8}$
}
\affiliation{$^{1}$ GSI Helmholtzzentrum f\"ur Schwerionenforschung GmbH, Planckstr. 1, 64291 Darmstadt, Germany}   
\affiliation{$^{2}$ Institute of Experimental Physics, Heidelberg University, Heidelberg, Germany}
\affiliation{$^{3}$ Institut f\"ur Kernphysik, Max-von-Laue-Str. 1, 60438 Frankfurt, Germany}
\affiliation{$^{4}$ Joint Institute for Nuclear Research, Joliot-Curie 6, 141980 Dubna, Moscow, region, Russia}
\affiliation{$^{5}$ SUBATECH, Nantes University, IMT Atlantique, IN2P3/CNRS, 4 rue Alfred Kastler, 44307 Nantes cedex 3, France}
\affiliation{$^{6}$ Frankfurt Institute for Advanced Studies, Ruth Moufang Str. 1, 60438 Frankfurt, Germany} 
\affiliation{$^{7}$ Helmholtz Research Academy Hessen for FAIR (HFHF),GSI Helmholtz Center for Heavy Ion Physics. Campus Frankfurt, 60438 Frankfurt, Germany}  
\affiliation{$^{8}$ Institut f\"ur Theoretische Physik, Johann Wolfgang Goethe University, Max-von-Laue-Str. 1, 60438 Frankfurt, Germany}
\affiliation{$^{9}$ Key Laboratory of Quark and Lepton Physics (MOE) and Institute of Particle Physics, Central China Normal University, Wuhan 430079, China}
\affiliation{$^{10}$ Institute of Modern Physics, Chinese Academy of Sciences, Lanzhou 730000, China}
\affiliation{$^{11}$ School of Science, Huzhou University, Huzhou, Zhejiang 313000, China}

\begin{abstract} 
The study of the nuclear equation-of-state (EoS) is a one of the primary goals of experimental and theoretical heavy-ion physics. The comparison of recent high statistics data from the STAR Collaboration with transport models provides a unique possibility to address this topic in a yet unexplored energy domain. Employing the microscopic \Yj{N-body} Parton-Hadron-Quantum-Molecular Dynamics (PHQMD) transport approach, which allows to describe the propagation and interactions of hadronic and partonic degrees of freedom including cluster and hyper-nucleus formation and dynamics, we investigate the influence of different EoS on bulk observables, the multiplicity, $p_T$ and rapidity distributions of protons, $\Lambda$s and clusters up to A=4 as well as their influence on the collective flow.  We explore three different EoS: two static EoS, dubbed 'soft' and 'hard',  which differ in the compressibility modulus, as well as a soft momentum dependent EoS.  We find that a soft momentum dependent EoS reproduces most baryon and cluster observables, including the flow observables, quantitatively, \EB{however, hard EOS show a similar trend.}

\end{abstract} 

\maketitle

\section{Introduction}
\label{sec:intro}

The quest for the equation of state (EoS) of strongly interacting matter is one of the major objectives of present day nuclear physics. Its knowledge is a cornerstone for the understanding of heavy-ion reactions, for the mass/radius relation of neutron stars and for the understanding of neutron star collisions. Therefore, in different fields efforts have been made to provide data, which may allow for determining the EoS. They include heavy-ion collisions \cite{HADES:2020lob,HADES:2022osk}, the search for massive neutron stars \cite{Miller:2019cac,Bogdanov:2019ixe,Bogdanov:2019qjb,Raaijmakers:2019qny} and since very recently also the study of gravitational waves emitted during neutron star mergers \cite{LIGOScientific:2018cki,LIGOScientific:2020aai}. 

On the theoretical side the EoS of nuclear matter can be calculated by the Brueckner-Dirac approach \cite{Gaitanos:2001hv} or by chiral effective theories \cite{Holt:2016pjb}. Both \Yj{rely}, however, on expansion schemes, which limit their predictive power to densities below or close to normal nuclear matter density.
Lattice gauge calculations should be the tool to extend these calculations towards higher densities or chemical potentials, $\mu$,  but the sign problem makes calculations difficult for $\mu \neq 0$ and at densities, which are reached in heavy ion collisions, the present efforts to determine the EoS in terms of a $\mu/T$ expansion around $\mu/T=0$ are not very reliable.

In this situation, in which the EoS cannot be inferred from underlying theories, the only systematic approach for its exploration is the comparison of experimental heavy-ion data with results of transport approaches, in which the EoS can be varied. By varying beam energy, system size and centrality of heavy ion reactions the EoS can be explored at densities up to about three times normal nuclear matter density, $\rho_0$, and up to temperatures of T=120 MeV. By this they cover the temperature-density $(T,\rho)$- region, which is also relevant for neutron star mergers. 

If one wants to follow this way it is obvious to investigate the most complete experiment in this kinematical sector. 
The STAR collaboration recently published data from Au+Au collisions at $\sqrt{s_{\rm{NN}}}=3$ GeV, where dense baryonic matter is expected to be formed. This data include the differential yields and azimuthal anisotropy of various particles, including hadrons (proton, $\Lambda$, $\Xi^{-}$ hyperons, etc.)~\cite{STAR:2023uxk, STAR:2021yiu, STAR:2024znc, STAR:2021yiu}, light clusters (deuteron, triton, ${}^{3}\rm{He}$, ${}^{4}\rm{He}$)~\cite{STAR:2023uxk, STAR:2021ozh}, and hypernuclei (${}^{3}_{\Lambda}\rm{H}$, ${}^{4}_{\Lambda}\rm{H}$)~\cite{STAR:2021orx,STAR:2022fnj}. The data is taken in 2018 as part of the Beam Energy Scan Phase II Program at the Relativistic Heavy Ion Collider (RHIC) located at the Brookhaven National Laboratory. 
This data is presently the most complete with respect to the phase space coverage, the statistics and the detector design.

In the past in heavy-ion collisions especially three observables have been identified, which are sensitive to the equation of state: subthreshold $K^+$ production \cite{Hartnack:2005tr,Hartnack:2011cn}, the in-plane $v_1$\cite{Molitoris:1986pp, Aichelin:1987ti} and the elliptic flow $v_2$ . The latter two can be obtained by expanding the azimuthal distribution of hadrons in a Fourier series
\begin{equation} 
\label{eq:azim}
\frac{dN}{d\phi}\propto 1 + 2 v_1 cos(\phi-\Psi_R) +2v_2 cos(2(\phi-\Psi_R)) + ...
\end{equation}
$\phi$ is the azimuthal angle of the particle measured with respect to the event plane (or a "reaction plane") $\Psi_R$. The flow coefficients $v_n, \ n=1,2,...$  are defined with respect to $\Psi_R$ as average over all particles in all events for a given centrality range \cite{Ollitrault:1997di,Poskanzer:1998yz}:
\begin{equation} 
\label{eq:vn}
\Yj{v_n = \langle \cos(n(\phi-\Psi_R))\rangle}.
\end{equation}  

The strategy to determine the nuclear EoS is then the following: In the transport approaches parametrized nucleon-nucleon potentials or nuclear mean field potentials are employed, which correspond in nuclear matter to a different EoS at zero temperature. Usually one limits the number of parameters to a minimum (3 out of which 2 are determined by the requirement that at $\rho_0$ the binding energy per nucleon is -16 MeV). The third parameter is traditionally expressed by the compressibility modulus K, 
\begin{equation}
K  = 9\rho\frac{d P}{d\rho}\vert_{\rho=\rho_0}=\left. 9 \rho^2
\frac{{\rm \partial}^2(E/A(\rho))}{({\rm \partial}\rho)^2} \right\vert_{\rho=\rho_0} ,\qquad
\end{equation}
which determines the curvature of E/A($\rho$) at $\rho_0$.  In studies of monopole vibrations \cite{Mekjian:2011wut}, which are sensitive to densities around normal nuclear matter density, a value of $K\approx 200$ MeV has been found, whereas the early Plastic Ball data \cite{Gustafsson:1984ka}, which are sensitive to much higher densities, could be explained by a larger compressibility modulus, $K \approx 380$ MeV. \cite{Molitoris:1986pp}.  An EoS with K = 200 MeV yields a weak repulsion against the compression of nuclear matter and thus describes "soft" matter (denoted by "S"). $K$ = 380 MeV causes a strong repulsion of nuclear matter under compression and is called a hard EoS, denoted by "H". More complex parametrizations have been employed as well, however, without exploring explicitly the parameter space \cite{Sahu:1998vz,Sahu:2002ku}.  

In most of the past studies static potentials have been employed due to the problem to introduce momentum dependent potentials in mean field calculations \cite{Welke:1988zz, Kireyeu:2024hjo}. Exceptions have been  QMD type calculations as well as the above mentioned HSD mean field calculation. The momentum dependence of the nucleon-nucleon interaction is important in heavy-ion collisions because, when projectile and target start to collide, the relative momentum is large and therefore the potential is much more repulsive than in neutron stars where the Fermi momentum is the momentum scale. Calculations for heavy ion collisions have shown that the momentum dependence of the potential increases strongly $v_1$ and $v_2$ in comparison with a static EoS with the same value of $K$ \cite{Kireyeu:2024hjo}.  

The goal of our study is to investigate the nuclear EoS and to obtain the constraints on the nucleon-nucleon momentum dependent potential. For that we  confront the entirety of the measured bulk observables (yields, rapidity and transverse momentum spectra) and flow harmonics of hadrons, clusters and hypernuclei produced in Au+Au collisions at invariant energies $\sqrt{s_{\rm{NN}}}=3$ GeV with the results of a transport approach for different EoS. As a tool for this investigation of the EoS we employ Parton-Hadron-Quantum-Molecular Dynamics (PHQMD)    \cite{Aichelin:2019tnk,Glassel:2021rod,Kireyeu:2022qmv,Coci:2023daq,Kireyeu:2024woo,Kireyeu:2024hjo}, a microscopic \Yj{N-body} approach for the description of hadron and cluster dynamics.
Our work is an extension of the recent PHQMD study on the influence of the EoS and momentum dependent potentials on baryon and cluster observables  measured by the HADES and FOPI collaborations at lower beam energies (1.2-1.5 AGeV).  The exploration of hadronic observables at an invariant energy of $\sqrt{s_{\rm{NN}}}=3$ GeV,
for which no data have been available before, allows  as well to study the influence of the momentum dependent potential.

Our paper is organized as follows: We start with the recall of the model description in Section II.  In Section III we define the centrality selection procedure within the PHQMD in line with the STAR procedure. In Section IV we present the PHQMD results for the bulk observables of hadrons: cluster yields and spectra. In Section V we present the results for the collective flow coefficients. We summarize our findings in Section VI. 

\section{Model description: PHQMD}

The Parton-Hadron-Quantum-Molecular Dynamics (PHQMD)   \cite{Aichelin:2019tnk,Glassel:2021rod,Kireyeu:2022qmv,Coci:2023daq,Kireyeu:2024woo} is  a microscopic \Yj{N-body} transport approach, which combines the  baryon propagation from the Quantum Molecular Dynamics (QMD)  model \cite{Aichelin:1991xy,Aichelin:1987ti,Aichelin:1988me,Hartnack:1997ez} and the dynamical properties and interactions in- and out-of-equilibrium of hadronic and partonic degrees-of-freedom of the Parton-Hadron-String-Dynamics (PHSD) approach~\cite{Cassing:2008sv,Cassing:2008nn,Cassing:2009vt,Bratkovskaya:2011wp,Linnyk:2015rco,Moreau:2019vhw}. Here we recall the basic concepts for the implementation of the potential in the PHQMD. 
\subsection{QMD Propagation}

The QMD equation-of-motions (EoM) for a N-body system (i.e. system of N interacting nucleons) are derived using the  Dirac-Frenkel-McLachlan  approach \cite{raab:2000,BROECKHOVE1988547}, which has  been developed in chemical physics and later applied to nuclear physics for QMD like models  \cite{Feldmeier:1989st,Aichelin:1991xy,Ono:1992uy,Hartnack:1997ez}. 
It is based on the variational principle for the Schr\"odinger equation, where 
the time evolution of the N-body wave function $\psi$  is obtained from the variation
\begin{equation}
\Yj{\delta \int_{t_1}^{t_2} dt \langle \psi(t)|i\frac{d}{dt}-H|\psi(t)\rangle = 0},
\label{varS}
\end{equation}
where $H$ is the N-body Hamiltonian \Yj{and $\langle \ldots \rangle$ means the expectation value with respect to the many-body wavefunction (same as the next equation)}. Eq. (\ref{varS}) can be formally solved by approximating the N-body wave function by the direct product of single particle "trial" wave functions (neglecting antisymmetrization) 
$\psi = \prod_i^N  \psi_i$. 
With the assumption that the wave functions have a Gaussian form and that the width of the wave function is time independent, one obtains two equations-of-motion for the time evolution of the centroids of the Gaussian single particle Wigner density, which resemble the EoM of a classical particle with the phase space coordinates ${\bf r_{i0},p_{i0}}$ \cite{Aichelin:1991xy}:
\begin{equation}
\dot{r}_{i0}=\frac{\partial\langle H \rangle}{\partial p_{i0}} \qquad
\dot{p}_{i0}=-\frac{\partial \langle H \rangle}{\partial r_{i0}} \quad .
\label{prop}
\end{equation}
We stress that the difference to the classical EoM is that here the expectation value of the quantal Hamiltonian is used and not a classical Hamiltonian.

In PHQMD the single-particle Wigner density of the Gaussian wave function of a nucleon,  $\psi_i$, is given by  
\begin{eqnarray} 
 &&f_i ({\bf r_i, p_i,r_{i0},p_{i0}},t) = \label{Wignerdens} \\
 &&=   \int d^3ye^{-i{\bf p}_i\cdot {\bf y}/\hbar} \psi_i({\bf r}_i-{{\bf y}\over2, },{\bf p}_{i0},{\bf r}_{i0})\psi_i^*({\bf r}_i+{{\bf y}\over2 },{\bf p}_{i0},{\bf r}_{i0}) \nonumber
 \\
 &&=\frac{1}{(2\pi \hbar)^3}
 {\rm e}^{-\frac{2}{L} ({\bf r_i} - {\bf r_{i0}} (t) )^2   }
 {\rm e}^{-\frac{L}{2\hbar^2} ({\bf p_i - p_{i0}} (t) )^2},\nonumber 
\end{eqnarray}
where the Gaussian width $L$ is taken as  $L=2.16$ fm$^2$. 
The corresponding single particle density at ${\bf r}$ is obtained by integrating the single-particle Wigner density over  momentum and summing up the contributions of all nucleons:
\begin{eqnarray}
\rho_{sp}({\bf r},t)= \sum_i
\int d{\bf p_i}   f ({\bf r, p_i,r_{i0},p_{i0}},t) \nonumber\\
= \sum_i\Big(\frac{2}{\pi L}  \Big)^{3/2}{\rm e}^{-\frac{2}{L} ({\bf r} - {\bf r_{i0}} (t) )^2}.
\label{rhosp}
\end{eqnarray}

Let us consider a system of N-nucleons.  $H$ is the sum of the Hamiltonians of the nucleons, composed of kinetic and two-body potential energy, which has a strong interaction and a Coulomb part
\begin{equation}
H = \sum_i H_i  = \sum_i  (T_i + \sum_{j\neq i}  V_{ij}).
\label{HTV}
\end{equation}

The total potential energy of nucleons in PHQMD has three parts, a local static Skyrme type interaction, a local momentum-dependent interaction and a Coulomb interaction
\begin{eqnarray}
V_{ij}&=& V({\bf r}_i, {\bf r}_j,{\bf r}_{i0},{\bf r}_{j0},{\bf p}_{i0},{\bf p}_{j0},t) \\
  &=& V_{\rm Skyrme\ loc}+ V_{\rm mom}+ V_{\rm Coul} \nonumber \\  
 &=&\frac{1}{2} t_1 \delta ({\bf r}_i - {\bf r}_j)  +  \frac{1}{\gamma+1}t_2 \delta ({\bf r}_i - {\bf r}_j)  \,     
    \rho_{int}^{\gamma-1}({\bf r}_{i0},{\bf r}_{j0},t) \nonumber \\  
&& + \frac{1}{2}V({\bf r}_{i},{\bf r}_{j},{\bf p}_{i0},{\bf p}_{j0})+ \frac{1}{2}  \frac{Z_i Z_j e^2}{|{\bf r}_i-{\bf r}_j|} . \nonumber
\label{eq:ep} 
\end{eqnarray}

The expectation value of the potential energy $V_{ij}$,
between the nucleons i and j is given by
\begin{eqnarray}
&&\langle V_{ij}({\bf r}_{i0},{\bf p}_{i0},{\bf r}_{j0},{\bf p}_{j0},t)\rangle  = \nonumber \\ 
&& =\int d^3r_id^3r_j d^3p_id^3p_j\ 
 V_{ij}({\bf r}_i, {\bf r}_j,{\bf p}_{i0},{\bf p}_{j0}) \nonumber \\ 
&&\times f ({\bf r}_i,{\bf p}_i,{\bf r}_{i0},{\bf p}_{i0},t) f({\bf r}_j, {\bf p}_j,{\bf r}_{j0},{\bf p}_{j0},t).
\label{Vpot}
\end{eqnarray}
and the interaction density is given by
\begin{eqnarray}
&& \rho_{\rm int}({\bf r}_{i0},t)=\sum_{j \neq i}\int d^3r_id^3r_jd^3p_id^3p_j \delta({\bf r}_i-{\bf r}_j )
\nonumber \\ 
&&\times f ({\bf r}_i, {\bf p}_i,{\bf r}_{i0},{\bf p}_{i0},t) f ({\bf r_j, p_j, r_{j0},p_{j0}},t).
\label{rhoint}
\end{eqnarray}

In order to extend  PHQMD to relativistic energies we take into account the Lorentz contraction of the initial nuclei. This is done in an approximate way, as explained in Ref. \cite{Aichelin:2019tnk}, by introducing a modified single-particle Wigner density for each nucleon $i$:
\bea
&& \tilde f (\mathbf{r}_i, \mathbf{p}_i,\mathbf{r}_{i0},\mathbf{p}_{i0},t) =  \label{fGam} \\
&& =\frac{1}{\pi^3} {\rm e}^{-\frac{2}{L} [ \mathbf{r}_{i}^T(t) - \mathbf{r}_{i0}^T (t) ]^2} 
  {\rm e}^{-\frac{2\gamma_{cm}^2}{L} [ \mathbf{r}_{i}^L(t) - \mathbf{r}_{i0}^L (t) ]^2}  \nonumber \\
&& \times {\rm e}^{-\frac{L}{2} [ \mathbf{p}_{i}^T(t) - \mathbf{p}_{i0}^T (t) ]^2} 
  {\rm e}^{-\frac{L}{2\gamma_{cm}^2} [ \mathbf{p}_{i}^L(t) - \mathbf{p}_{i0}^L (t) ]^2},  \nonumber 
\eea
which accounts for the Lorentz contraction of the nucleus in the beam $z$-direction
in coordinate and momentum space by  including $\gamma_{cm} =1/\sqrt{1-v_{cm}^2}$, where $v_{cm}$ is the velocity of projectile and target in the computational frame, which is the center-of-mass system of the heavy-ion collision.  
Accordingly, the interaction density modifies as 
\bea
 \tilde \rho_{int} (\mathbf{r}_{i0},t) 
 &\to & C  \sum_j \Big(\frac{1}{\pi L}\Big)^{3/2} \gamma_{cm} 
 {\rm e}^{-\frac{1}{L} [ \mathbf{r}_{i0}^T(t) - \mathbf{r}_{j0}^T (t) ]^2} \nonumber \\ 
 &&\times {\rm e}^{-\frac{\gamma_{cm}^2}{L} [ \mathbf{r}_{i0}^L(t) - \mathbf{r}_{j0}^L (t) ]^2}.  
 \label{densGam}
\eea
For the energies considered here the relativistic corrections are not important.


\subsection{Modeling the EoS within PHQMD }

\label{sec:PHQMD}
The potential, which a nucleons suffers while traveling through nuclear matter, is momentum-dependent, as can be inferred from its beam energy dependence observed in elastic pA scattering data \cite{Clark:2006rj,Cooper:1993nx}. Typically, these data are analyzed by comparing experimental results with solutions of the Dirac equation that include scalar ($U_s$) and vector potentials, the latter represented by its zero component ($U_0$). These analyses are detailed in Refs. \cite{Clark:2006rj,Cooper:1993nx}.

To derive a nucleon-nucleon potential, suitable for non-relativistic QMD-type calculations, we first calculate the Schrödinger equivalent potential - also called optical potential - $ U_{opt}$, as described in Ref. \cite{Jaminon:1980vk}.  
\begin{equation}
    U_{opt}(r,\epsilon) =U_s(r)+U_0(r)+\frac{1}{2m_N}(U_s^2(r)-U_0^2(r))+\frac{U_0(r)}{m}\epsilon , 
\end{equation}
where $\epsilon$   is the kinetic energy of the incoming proton in the target rest frame.
The optical potential has also an imaginary part which we neglect here because we treat collisions explicitly. 

\begin{figure}[h!]
    \centering
    \resizebox{0.45\textwidth}{!}{
\includegraphics{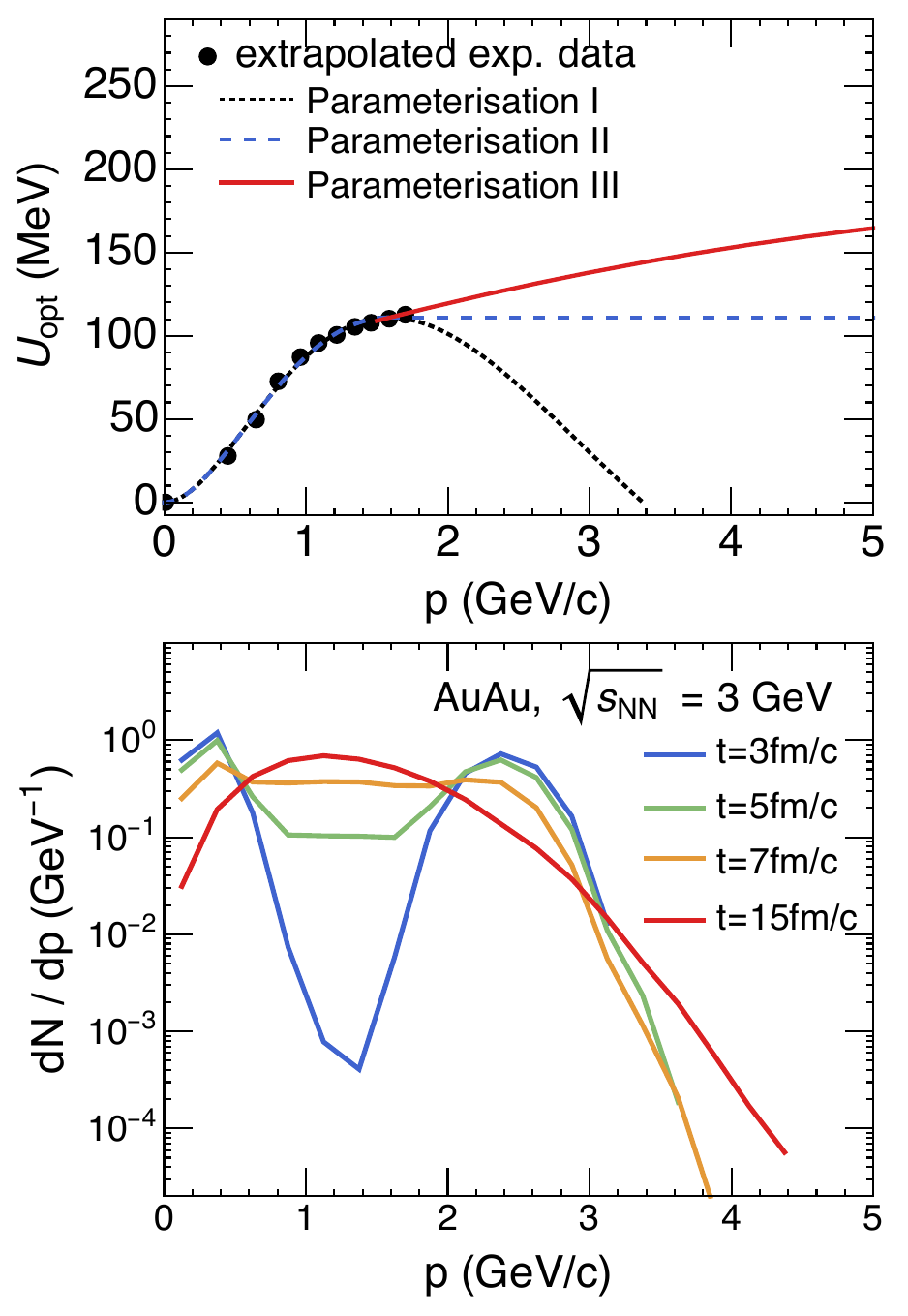}  }      
    \caption{Top: Schrödinger equivalent optical potential $U_{opt}$ versus the \Yj{absolute value of relative momentum $p$} of the proton extracted from pA collisions \cite{Hama:1990vr, Clark:2006rj,Cooper:1993nx}. Bottom; The distribution of the relative momenta
    among the nucleons for different times and for Au+Au collisions at $\sqrt{s_{\rm NN}}=3\rm GeV$.}
    \label{fig:uopt}
\end{figure}

With this potential we can rewrite the upper components of the Dirac equation 
\begin{equation}
    \Big(-i\boldsymbol \alpha\boldsymbol \nabla+\beta m +\beta U_s(r)+U_0(r)\Big)\phi=(\epsilon+m) \phi
\end{equation}
in form of a Schrödinger equation
\begin{equation}
 \big(\frac{-\nabla^2}{2m} + U_{opt}(r,\epsilon) \Big) \psi =\frac{2m\epsilon+\epsilon^2}{2m}\psi.  
\end{equation}
The analysis of the pA data at different beam energies shows that $U_s$ and $U_0$ depend on $\epsilon$, which yields also  a more complex $\epsilon$ dependence of $U_{opt}$. Our fit of $U_{opt}$ to these experimental data , normalized to $U_{opt}(p=0)=0$, is shown in Fig. \ref{fig:uopt} by the dotted line, alongside with $U_{opt}$ extracted from the pA scattering data analyzed via a Dirac equation\cite{Hama:1990vr}, presented as black dots. Beyond the proton kinetic energy of 1.04 GeV, no data are available. This introduces uncertainties for the extrapolation of the $U_{opt}(p)$ to large momenta $p$. This extrapolation is necessary in the PHQMD transport approach to describe heavy-ion collisions with $\sqrt{s} > 2.32$ GeV. In Section III we will demonstrate the consequences of these uncertainties of $U_{opt}(p)$ at large $p$ on the observables.

To judge the importance of the extrapolation of $U_{opt}$ to higher energies we show in the lower part of Fig. \ref{fig:uopt}  the distribution of the relative momentum between the nucleons at different times, from t=1 fm to t=15 fm. Initially
there is a peak for the momentum corresponding to the beam energy which lasts for a couple of fm/c. Starting from 7 fm/c the system approaches equilibrium
and the maximum of the distribution peaks at a value for which measured
values of $U_{opt}$ are available.

After having obtained the momentum dependence of the nuclear mean field we have to construct, in a second step, the momentum dependence of the two body interaction between two nucleons. The Schr\"odinger equivalent potential  is obtained by averaging the two body potential $V({\bf p},{\bf p}_1) $  over the Fermi distribution of the cold target nucleons,
\begin{equation}
    U_{opt}(p) =\frac{\int^{p_F} V({\bf p},{\bf p}_1) dp_1^3}{\frac{4}{3}\pi p_F^3}.
\end{equation}
 The experimental data are in the low momentum region with $p\lesssim 1.7\rm GeV$ and no more constraints in the high momentum region.  To study whether different parametrizations of $U_{opt}$ for high $p_T$  have an influence on the observables we perform also calculations with two other parameterizations of the momentum-dependent potential, which give the same results in the low momentum region, but differ for high momenta, shown in the top panel of Fig.~\ref{fig:uopt}. The  parameterization I (black dotted line) has the form of
\begin{equation}
\label{Umom12}
  V({\bf p, p}_1) =(a (\Delta p)^2 + b(\Delta p^4))\exp [-c \Delta p ].
\end{equation}
with $\Delta p =\sqrt{({\bf p}-{\bf p}_1)^2}$. The parameterization II (blue dashed line) is the same as the parameterization I for momenta $\le 2$ GeV$/c$, but with a constant for momentum $> 2$ GeV$/c$.
The parameterization III (red solid line) for $\Delta p<1.7\rm GeV$ is the same as the old one, while for $\Delta p>\rm 1.7$ GeV$/c$, we take, 
\begin{equation}
  V(\Delta p)=  d + e \Delta p + f\Delta p^2,
\end{equation}
which gives a slight increase of the potential in the region of interest in this study. The parameters of the parameterization are shown in Table~\ref{table_eos}.

In the Dirac analysis the vector and scalar mean field potentials depend roughly linearly on the baryon density in the nucleus \cite{Clark:2006rj, Cooper:1993nx}. This linear dependence can be reproduced if we assume in Eq. (\ref{eq:ep}),
 \begin{equation}
    V({\bf r}_1,{\bf r}_2,{\bf p}_{10},{\bf p}_{20})  = V({\bf p}_{10},{\bf p}_{20}) \delta({\bf r}_1-{\bf r}_2 ).
\end{equation}
because the $\delta$ function creates a linear density dependence when averaged over the wave functions. The  energy of the system is
\begin{eqnarray}
   E &=& \langle \psi(t) |(T+V) |\psi(t) \rangle\nonumber \\
&=&\sum_i[\langle i | \frac{p^2}{2m} |i\rangle + \sum_{i\neq j} \langle ij| V_{ij} |ij \rangle] \nonumber \\
&=& \int H(r) d^3r.
\label{eq:energy}
\end{eqnarray}
$|\psi(t)\rangle =\prod_i |\psi_i\rangle$ is the N-body wave function, which is taken, as said, as the direct product of the single particle wave functions of the nucleons. The momentum-dependent potential has been introduced in QMD type transport approaches in Ref. \cite{Aichelin:1987ti} and explored later in Refs. \cite{Aichelin:1991xy,LeFevre:2016vpp,Hillmann:2019wlt,Nara:2020ztb}
and for BUU type approaches in Ref. \cite{Gale:1987zz} and been widely applied in different forms \cite{Ko:1987gp,Weber:1993et,Sahu:1998vz,Danielewicz:1998vz,Sahu:1999mq,Sahu:2002ku,Cassing:2000bj,Buss:2011mx,Mohs:2020awg}.

\subsection{Relation of the potential  to the EoS of nuclear matter }

In infinite nuclear matter, momentum and position are not correlated and one can calculate  the equation-of-state of cold nuclear matter from the potential.  
In infinite matter the static part of the QMD potential is given, as in \cite{Aichelin:2019tnk}, by 
\begin{equation}
    V_{Skyrme\ stat} =\alpha\frac{\rho}{\rho_0}+\beta \Big(\frac{\rho}{\rho_0}\Big)^\gamma,
\end{equation}
to this the momentum-dependent part for cold nuclear matter is added, which  can be obtained by
\begin{equation}
    V_{mom}(p_F)=\frac{\int^{p_F}\int^{p_F} dp_1^3dp_2^3 V({\bf p_2}-{\bf p}_1) }{(\frac{4}{3}\pi p_F^3)^2}\frac{\rho}{\rho_0}.
\end{equation}

\begin{figure}[t]
    \centering
    \resizebox{0.45\textwidth}{!}{
        \includegraphics{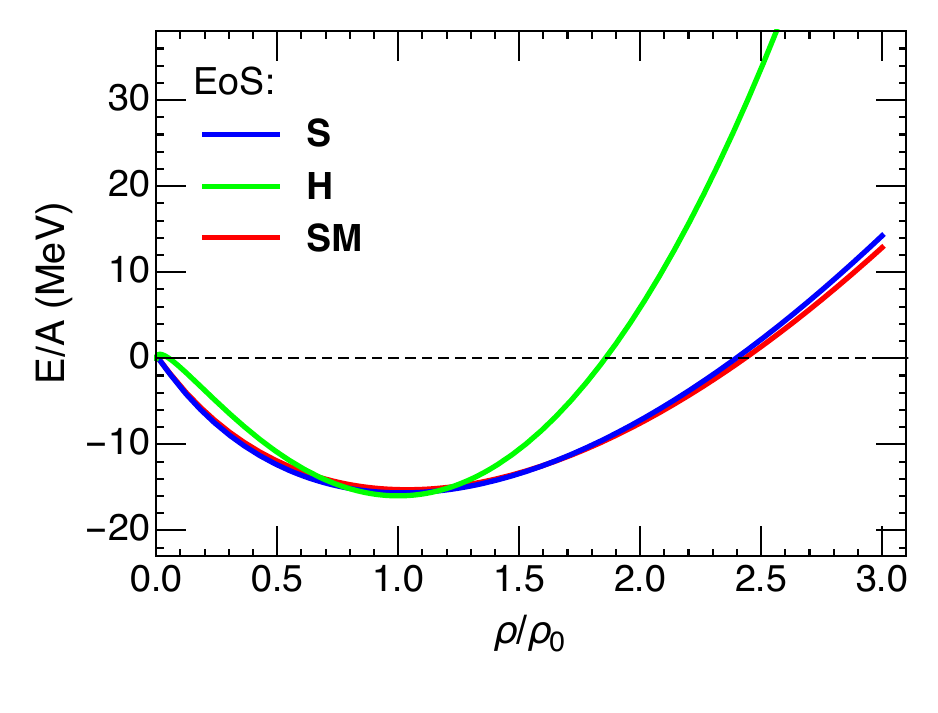}     }
    \caption{Equation-of-state for $T=0$ for the hard (green line), soft (blue line) and the soft momentum-dependent potential (red line). } 
    \label{fig:eos}
\end{figure}

The Fermi momentum is a function of the density and therefore one obtains the total strong interaction potential
\begin{equation}
    V_{Skyrme}(\rho) = V_{Skyrme \ stat}(\rho)+V_{mom}(\rho).
\end{equation}
To calculate the energy per nucleon, we introduce  $U=\int V(\rho) d\rho $. This allows to write
\begin{equation}
    \frac{E}{A}(\rho)=  \frac{3}{5}  E_{Fermi}(\rho) + \frac{U}{\rho}.
\end{equation}
As discussed, our equation contains the minimal number of 3 parameters $\alpha, \beta,\gamma$, which have to be determined.  They are obtained by the requirement that at normal nuclear density  $E/A = -16$ MeV. The compressibility modulus K is a free parameter. 
The hard (K = 380 MeV), the soft and the soft momentum-dependent (K=200 MeV) equations-of-state are illustrated in Fig. \ref{fig:eos} and the parameters for the three equations-of-state are presented in Table \ref{table_eos}.
For cold nuclear matter the soft and soft momentum-dependent EoS have by construction the same $\frac{E}{A}(\rho)$.

\begin{table}[h]
    \centering
      \begin{tabular}{c c c c c}
      \toprule
      EoS & $\alpha$ [GeV] & $\beta$ [GeV] & $\gamma$ & K [MeV]\\
      \midrule
      S  &  -0.3835 &  0.3295 & 1.15 & 200 \\
      H  &  -0.1253 &  0.071 &  2.0 & 380 \\
      SM & -0.478 & 0.4137 &  1.1 & 200 \\
      \cmidrule{2-4}
       ~ & $a$[GeV$^{-1}$]  & $b$ [GeV$^{-3}$] & $c$  [GeV$^{-1}$]   & ~\\
       ~ & 236.326 & -20.730 &  0.901 & ~ \\
        \cmidrule{2-4}
       ~ & $d$[GeV]  & $e$ & $f$  [GeV$^{-1}$]   & ~\\
       ~ & 72.237 & 27.085 &  -1.722 & ~ \\
      \bottomrule
      \end{tabular}
    \caption{Parameters of the potential used in PHQMD assuming that the momenta
     are given in GeV.} 
    \label{table_eos}
    \end{table}

\subsection{Cluster production in PHQMD}

Clusters can be identified in PHQMD by three different algorithms:

1) {\bf Coalescence mechanism:} 
 A proton and a neutron can form a deuteron if their distance at the time, when the last one of the two freezes out, is less than $|r_1-r_2| \le 3.575$ fm in coordinate space and less than $|p_1-p_2|\le 285$ MeV$/c$ in momentum space. For details of this cluster formation algorithm in PHQMD we refer to Ref. \cite{Kireyeu:2022qmv}. This method is not applied in this article.

2)  {\bf Kinetic mechanism:}  Deuterons can be created in catalytic hadronic reactions as $\pi NN \leftrightarrow \pi d$ and $NNN \leftrightarrow N d$  in different isospin channels. The quantum nature of the deuteron is considered through an excluded volume which forbids its production if another hadron is localized in this volume.  We project, furthermore, the relative momentum of the incoming nucleons onto the deuteron wave function in momentum space.  These quantum corrections  lead to a significant reduction of deuteron production by the kinetic mechanism, particularly at target/projectile  rapidities. The details of this algorithm are described in Ref. \cite{Coci:2023daq}.

3) {\bf Potential mechanism:} The attractive potential between baryons with a small relative momentum keeps them close together and can lead to a group of bound nucleons. At a given time $t$ a snapshot of the positions and momenta of all nucleons is recorded and the MST (Minimum Spanning Tree) clusterization algorithm is applied: two nucleons $i$ and $j$  with the corresponding isospin are considered as ``bound" to a deuteron or to a larger cluster $A>2$ if they fulfill the  condition \cite{Aichelin:1991xy} 
\begin{equation}\label{eq:MSTcond}
| \mathbf{r}_i^* - \mathbf{r}_j^* | < r_{clus} \, ,
\end{equation}
where on the left hand side the positions are boosted in the center-of-mass of the $ij$ pair. The maximal distance between cluster nucleons, $r_{clus}=4$ fm, corresponds roughly to the range of the attractive $NN$ potential. Additionally, in the aMST cluster algorithm \cite{Glassel:2021rod} the clusters have to be bound ($E_B>0$). It is important to highlight that MST serves as a tool for cluster recognition, not a mechanism for ‘building’ clusters, since the QMD transport model propagates baryons and not pre-formed clusters.

In the calculations presented here we employ, as detailed in Ref. \cite{Coci:2023daq}, a combination of 2) and 3) to identify clusters. 


\section{Model study: the influence of uncertainnesses in 
$U_{opt}(p)$ on observables}

\begin{figure}[h!]             
    \centering 
    \includegraphics[width=0.48\linewidth]{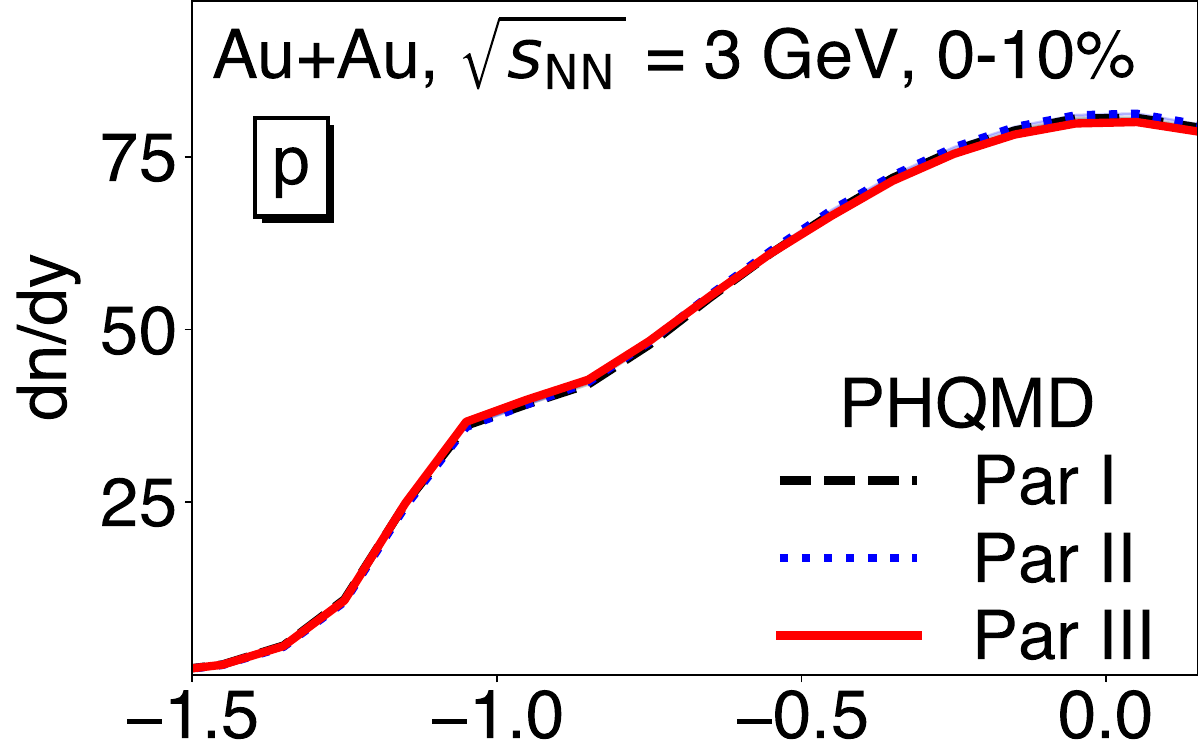}
    \includegraphics[width=0.48\linewidth]{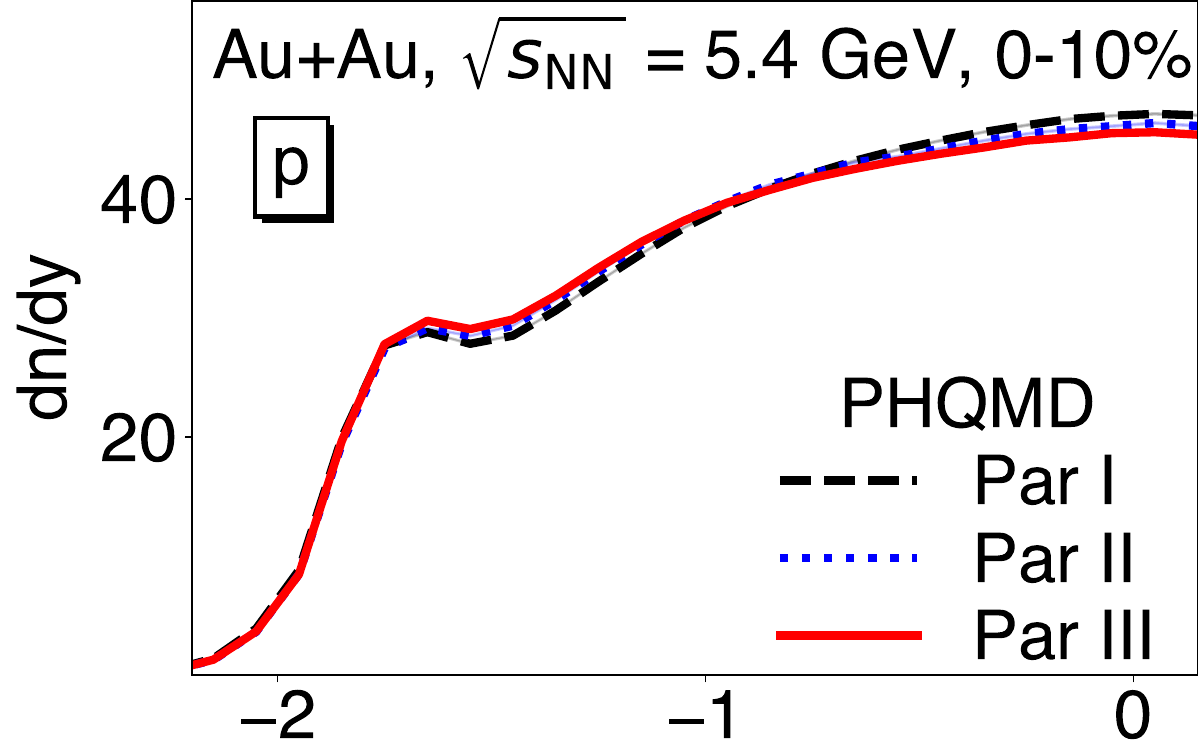}
    \includegraphics[width=0.48\linewidth]{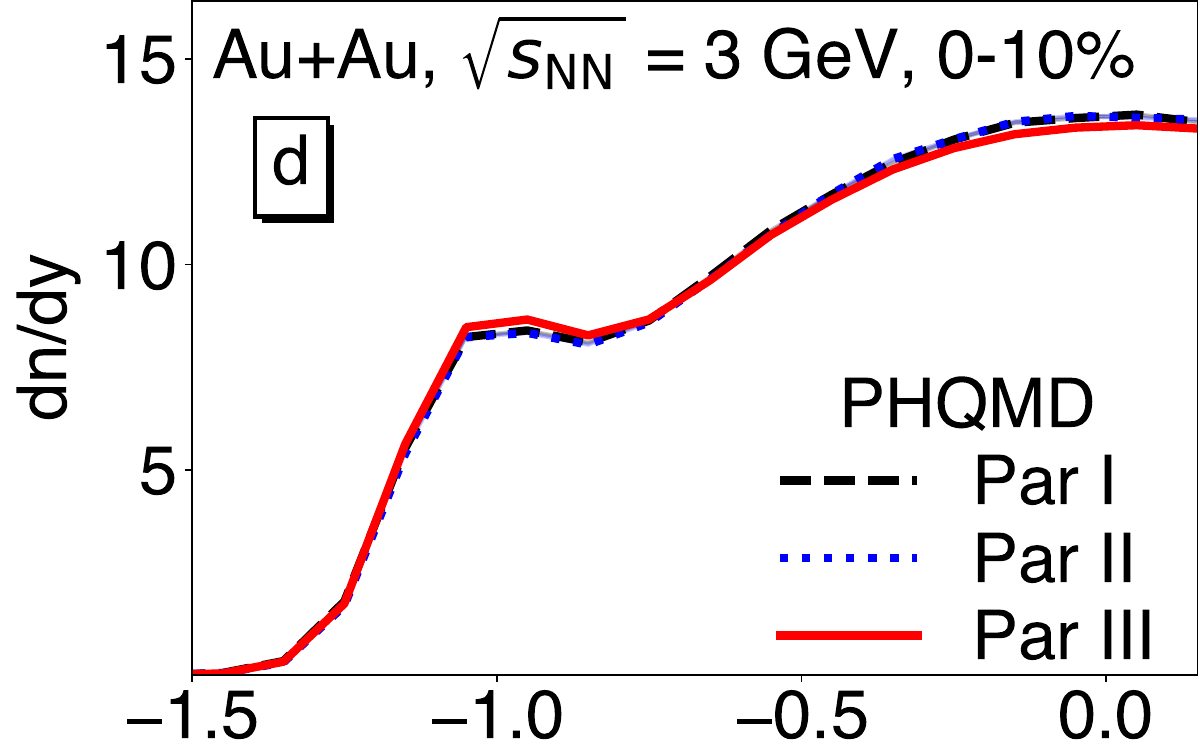}
    \includegraphics[width=0.48\linewidth]{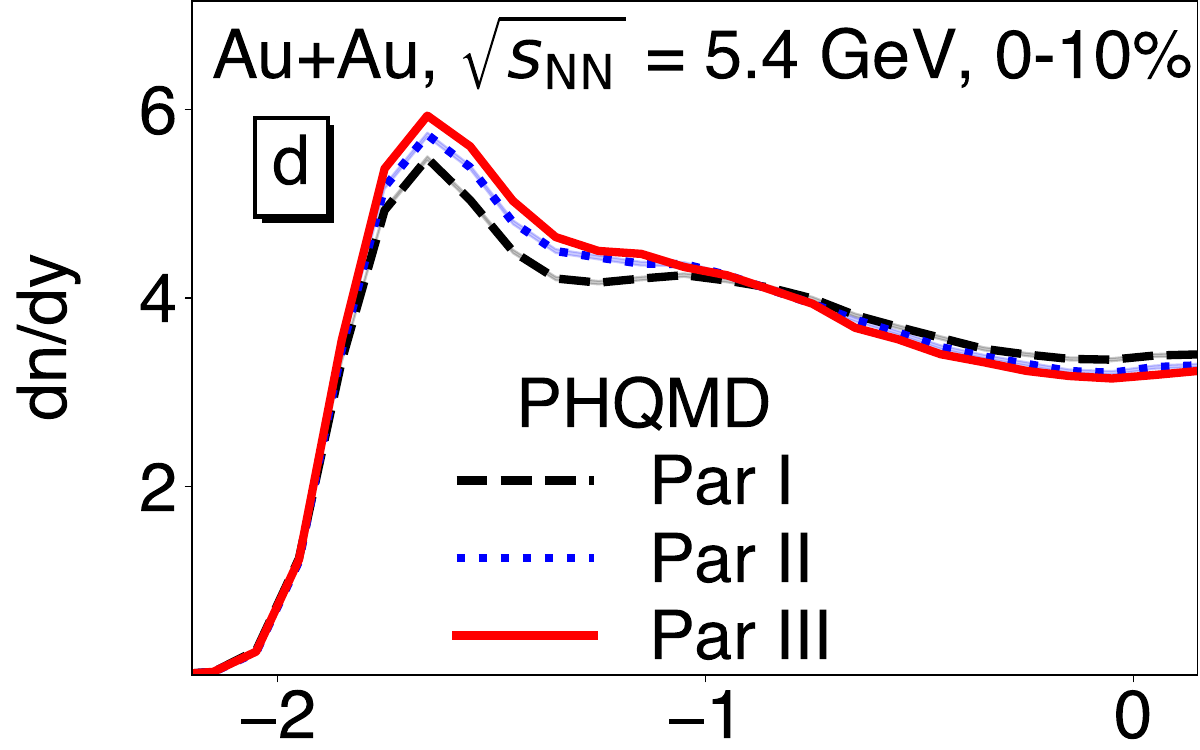}
    \includegraphics[width=0.48\linewidth]{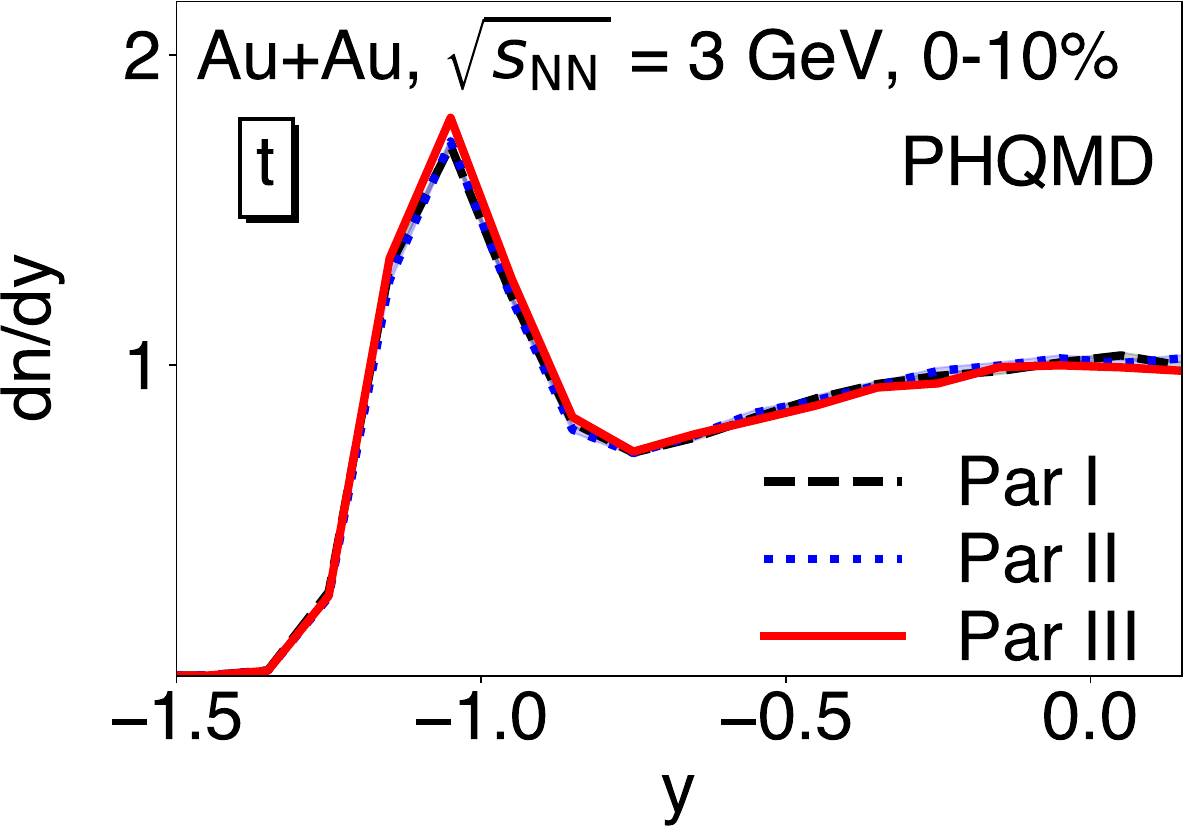}
    \includegraphics[width=0.48\linewidth]{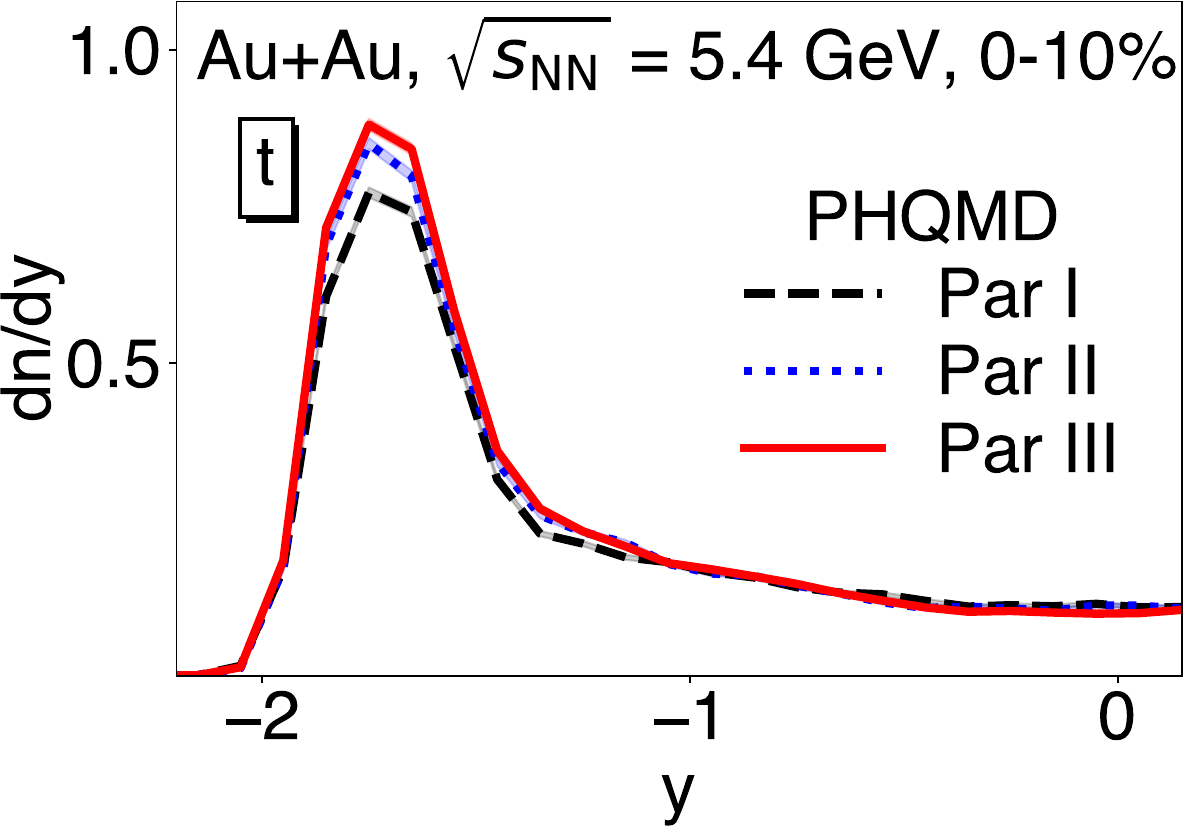}
    \caption{Rapidity distribution (integrated over all $p_T$) of protons (upper row), deuterons (middle row), and tritons (lower row)  from PHQMD simulations of Au+Au collisions at $\sqrt{s_{\rm{NN}}}=3$ GeV (left column) and 5.4 GeV (right column)  in the 0–10\% centrality class. Results are shown for the three different parameterizations of the nuclear optical potential (Parameterization I, II, and III) of Fig. \ref{fig:uopt}.}    
    \label{fig:dndy_optPotential}
\end{figure}

\begin{figure}[h!]             
    \centering 
    \includegraphics[width=0.48\linewidth]{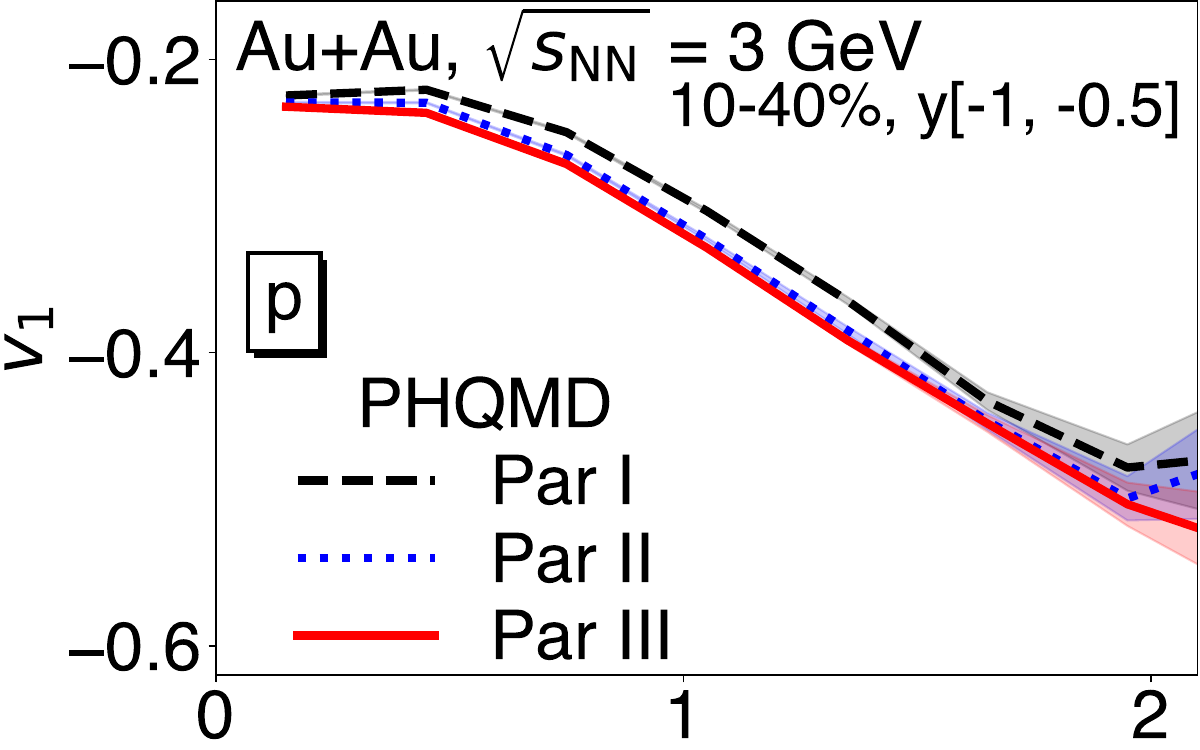}
    \includegraphics[width=0.48\linewidth]{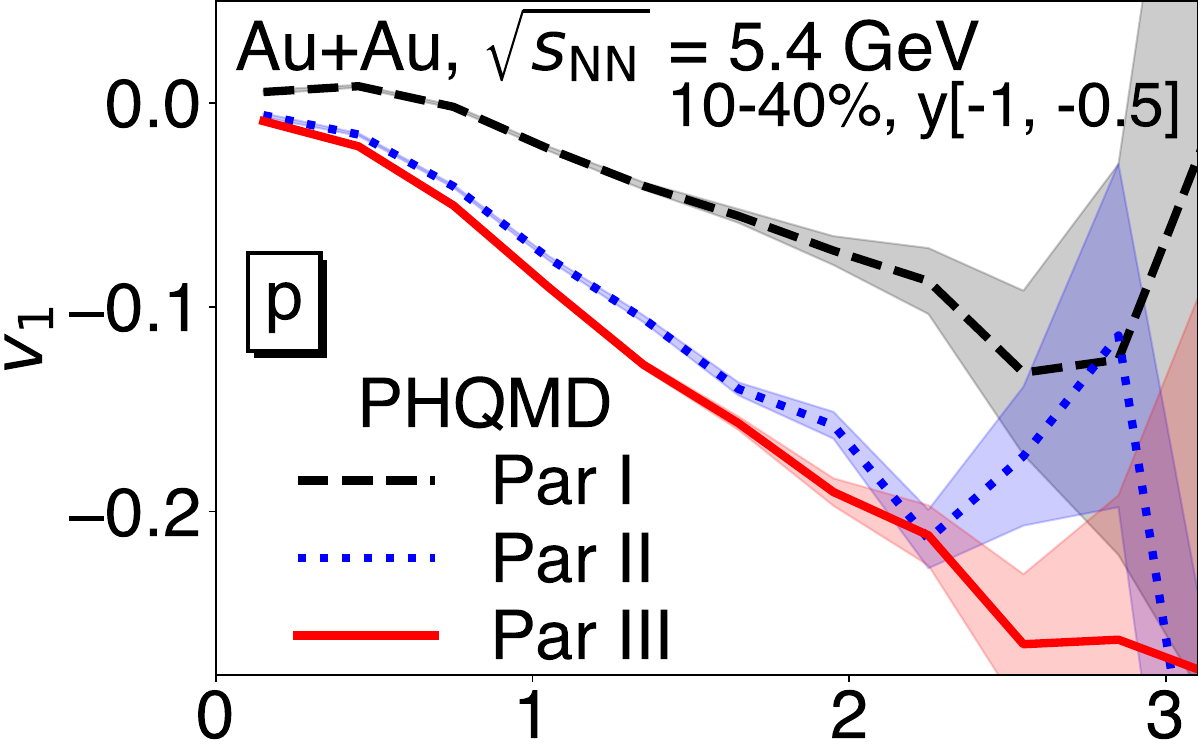}
    \includegraphics[width=0.48\linewidth]{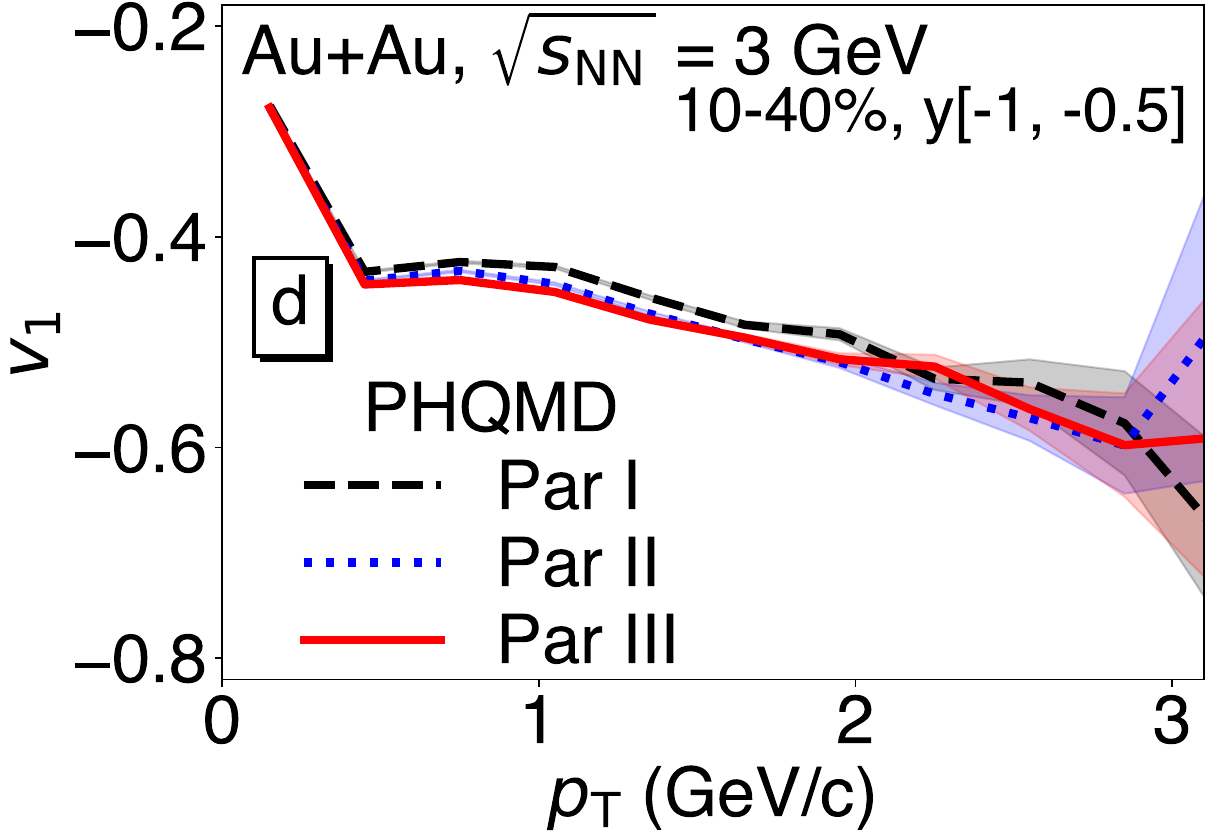}
    \includegraphics[width=0.48\linewidth]{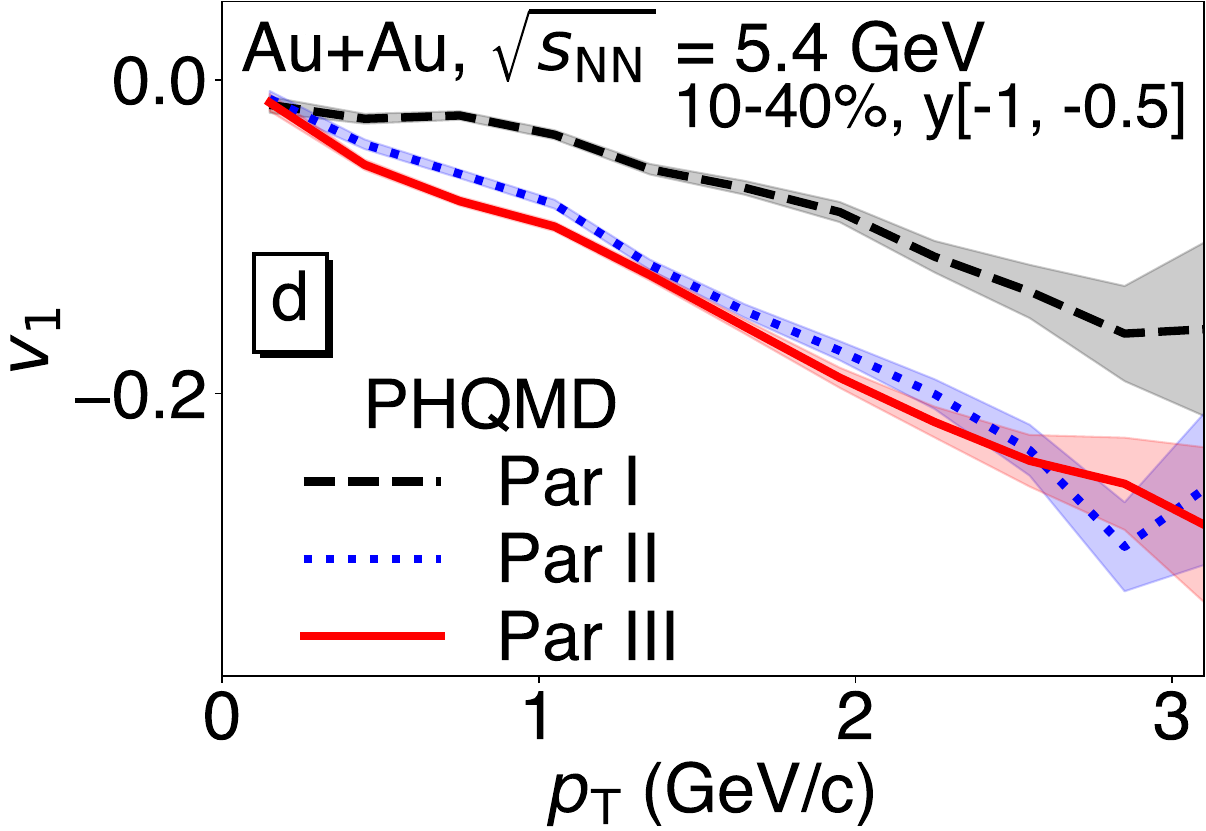}
    \caption{The in-plane flow $v_1(p_T)$  near target rapidity ($-1<y<-0.5$)  of protons and deuterons from PHQMD simulations of Au+Au collisions at $\sqrt{s_{\rm{NN}}}=3$ GeV (left column) and 5.4 GeV (right column)  in the 10–40\% centrality class. Results are shown for three different parameterizations of the nuclear optical potential (Parameterization I, II, and III).}
    \label{fig:v1pt_optPotential}
\end{figure}

\begin{figure}[h!]             
    \centering 
    \includegraphics[width=0.48\linewidth]{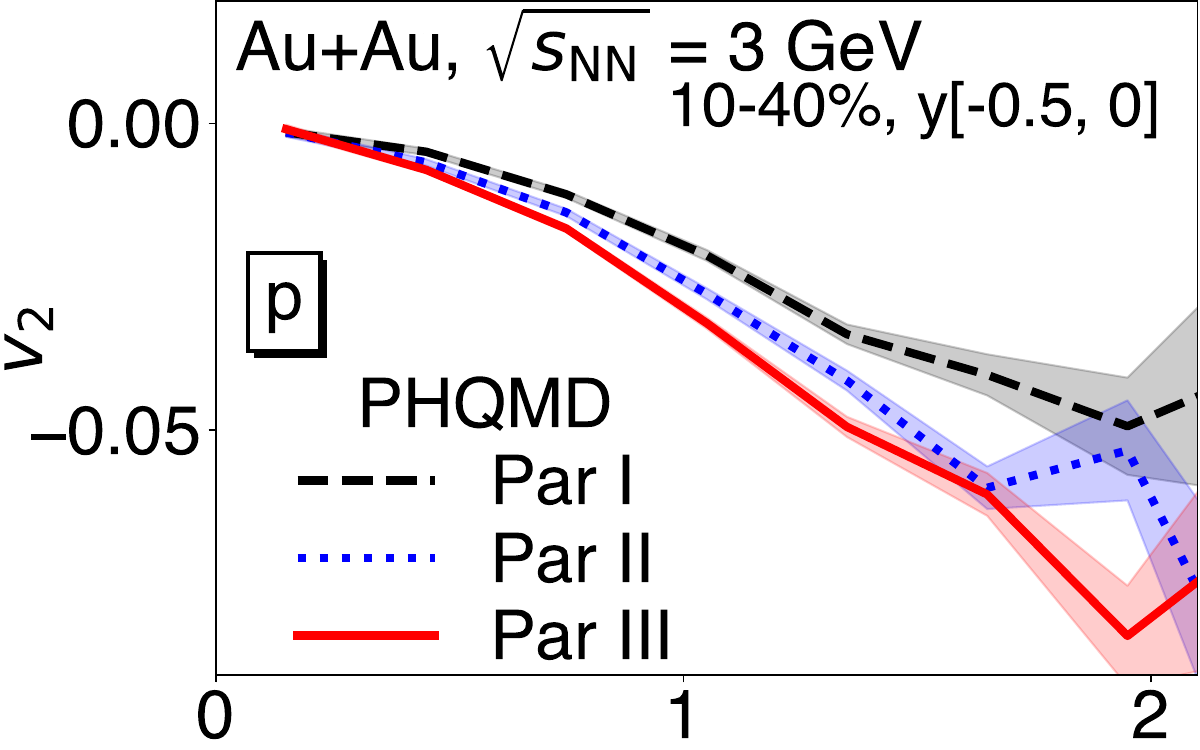}
    \includegraphics[width=0.48\linewidth]{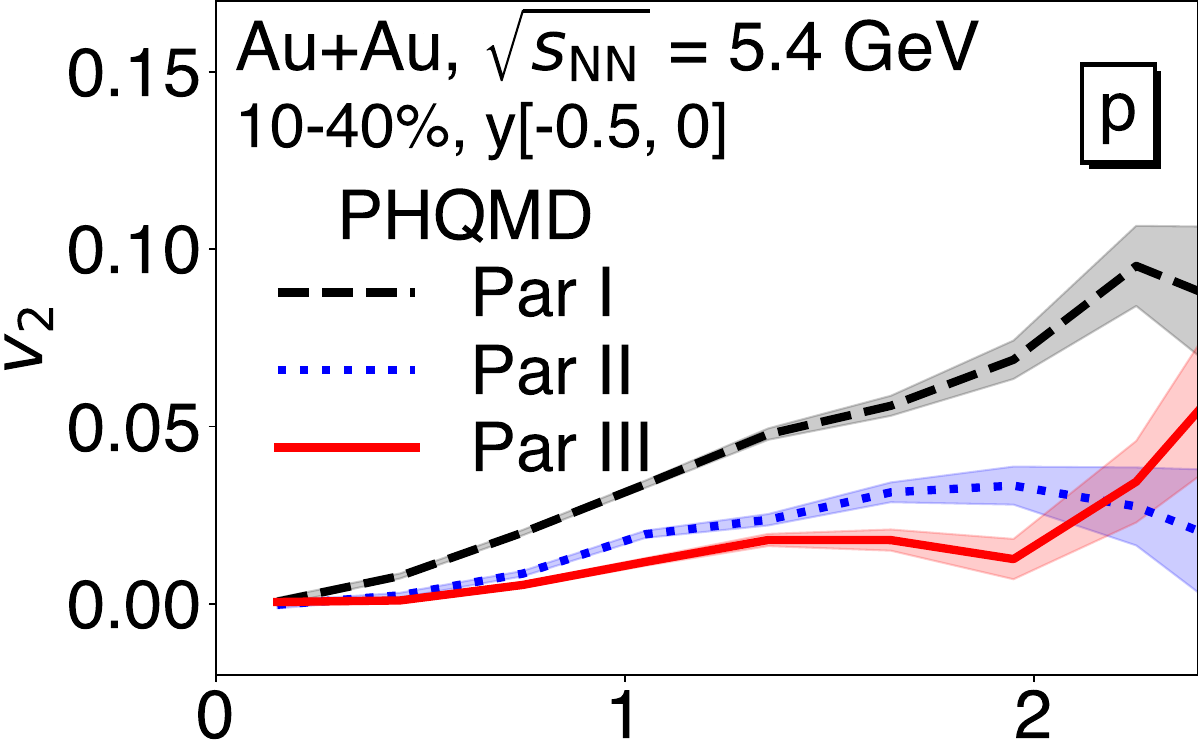}
    \includegraphics[width=0.48\linewidth]{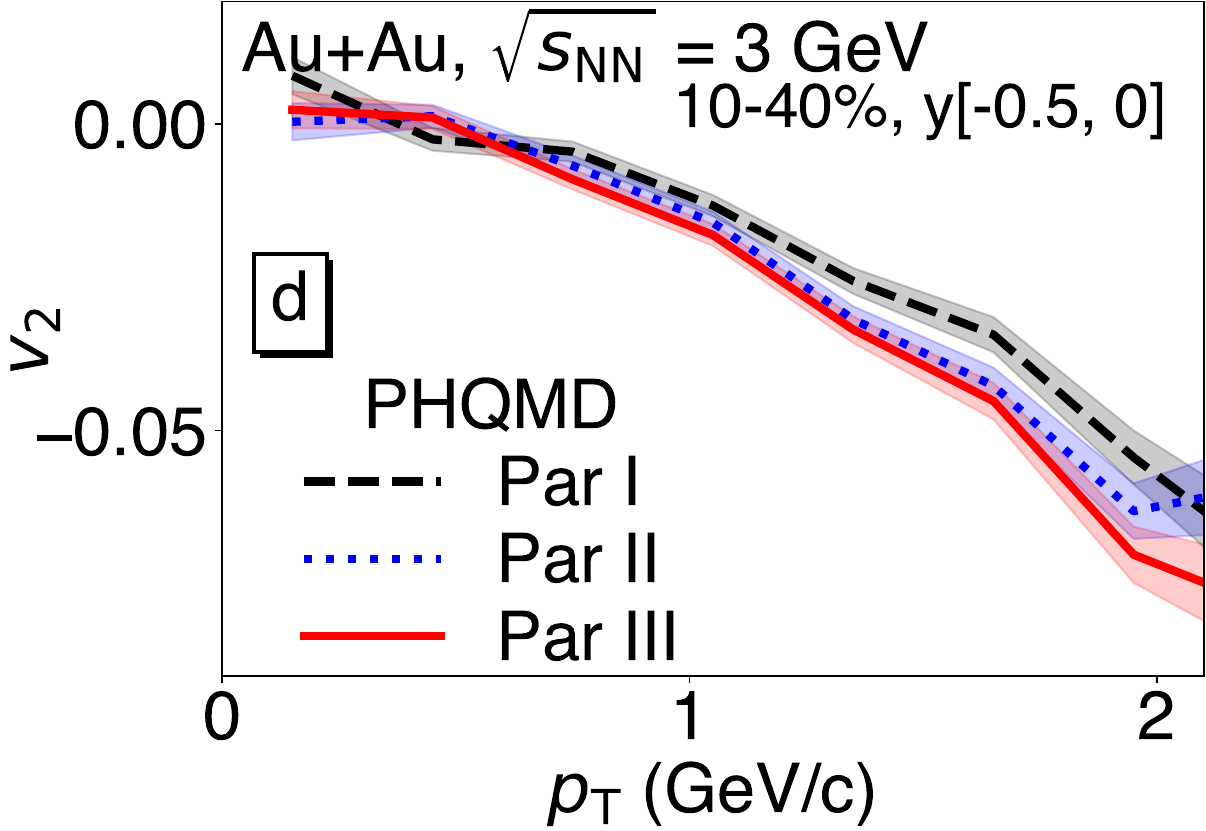}
    \includegraphics[width=0.48\linewidth]{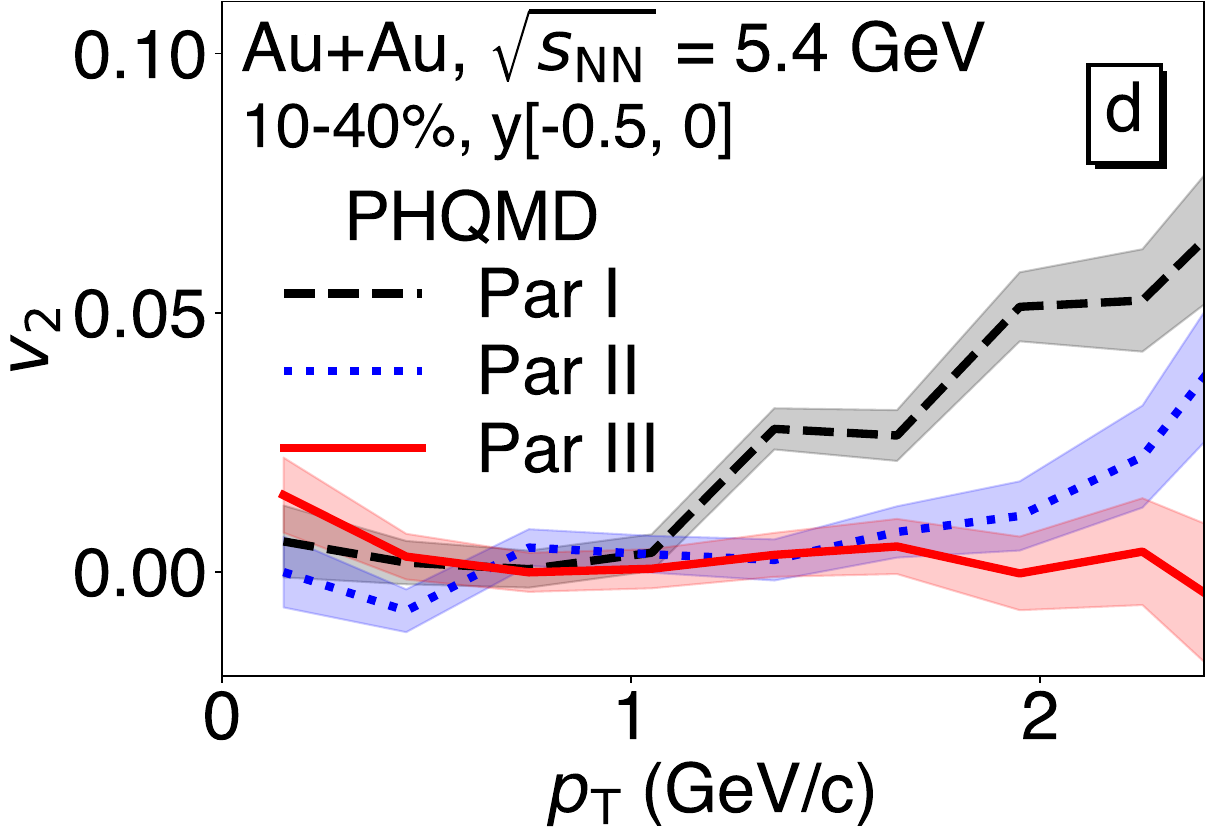}
    \caption{The elliptic flow $v_2(p_T)$ at mid-rapidity ($-0.5<y<0$) of protons and deuterons from PHQMD simulations of Au+Au collisions at $\sqrt{s_{\rm{NN}}}=3$ GeV (left column) and 5.4 GeV (right column)  in the 10–40\% centrality class. Results are shown for three different parameterizations of the nuclear optical potential (Parameterization I, II, and III).}
    \label{fig:v2pt_optPotential}
\end{figure}

Before coming to the comparison of the PHQMD results with the experimental data,  we report the results of a model study on the influence of uncertainties of the   $U_{opt}(p)$ on observables such as the rapidity distributions and the flow harmonics $v_1, v_2$ of protons and light clusters. 

As discussed in Section II.B there are no experimental data, which would allow for a reliable extrapolation of $U_{opt}(p)$  to large $p$, see Fig. \ref{fig:uopt}. 
However, a momentum $p>2$ GeV$/c$ is achievable already at the beam energies of $\sqrt{s}$ = 3 GeV - as follows from the lower plot in Fig.  \ref{fig:uopt}.
In our model study we explore the 3 parametrizations of $U_{opt}(p)$, shown in Fig. \ref{fig:uopt}, and investigate whether the different parametrization lead to different values of the observables for heavy-ion reactions at $\sqrt{s_{\rm{NN}}}=3$ GeV as well and at 5 GeV, which allow to explore even larger $p$. 

 The result of this study is shown in Fig. \ref{fig:dndy_optPotential} for the rapidity distributions of protons, deuterons, tritons, in  Fig. \ref{fig:v1pt_optPotential} for the in-plane flow $v_1(p_T)$ and in Fig. \ref{fig:v1pt_optPotential} for the elliptic flow $v_2(p_T)$ of protons and deuterons at midrapidity. In these figures we present the observables for  $\sqrt{s_{\rm{NN}}}=3$ GeV (left panels) and 5 GeV (right panels) for all three parameterizations of $U_{opt}$ of Fig. \ref{fig:uopt}. 
 
 Fig. \ref{fig:dndy_optPotential} shows that the rapidity distributions of protons and deuterons and tritons  is practically identical for the three parametrizations. The same is almost true for the in-plane flow, $v_1$. The elliptic flow, $v_2$, however depends on the parametrization. Especially at high $p_T$ the different parametrizations change the elliptic flow by 30\%, for protons as well as for the deuterons. This systematic error of the results due the unknown optical potential at this energy has to be taken into account, when we discuss later the results. 
 
 We note that the calculations, presented in the following Sections, are performed using the parametrization I.

\section{Centrality selection within PHQMD}\label{sec:compare}

The main purpose of this paper is to compare the predictions of the PHQMD transport approach with the best experimental data set, which is, for the relevant energies around  $\sqrt{s_{\rm{NN}}} \le 3$ GeV, presently and for the near future available. This will allow to elucidate how presently transport approaches compare with high statistics results and to indicate where disagreement needs future development.
We compare the Au+Au collisions at $\sqrt{s_{\rm{NN}}}=3$ GeV from the STAR experiment with PHQMD calculations for the three different EoS, S, H, and SM, discussed in the previous section. 

\begin{figure}[h!]             
    \centering    
    \includegraphics[width=0.8\linewidth]{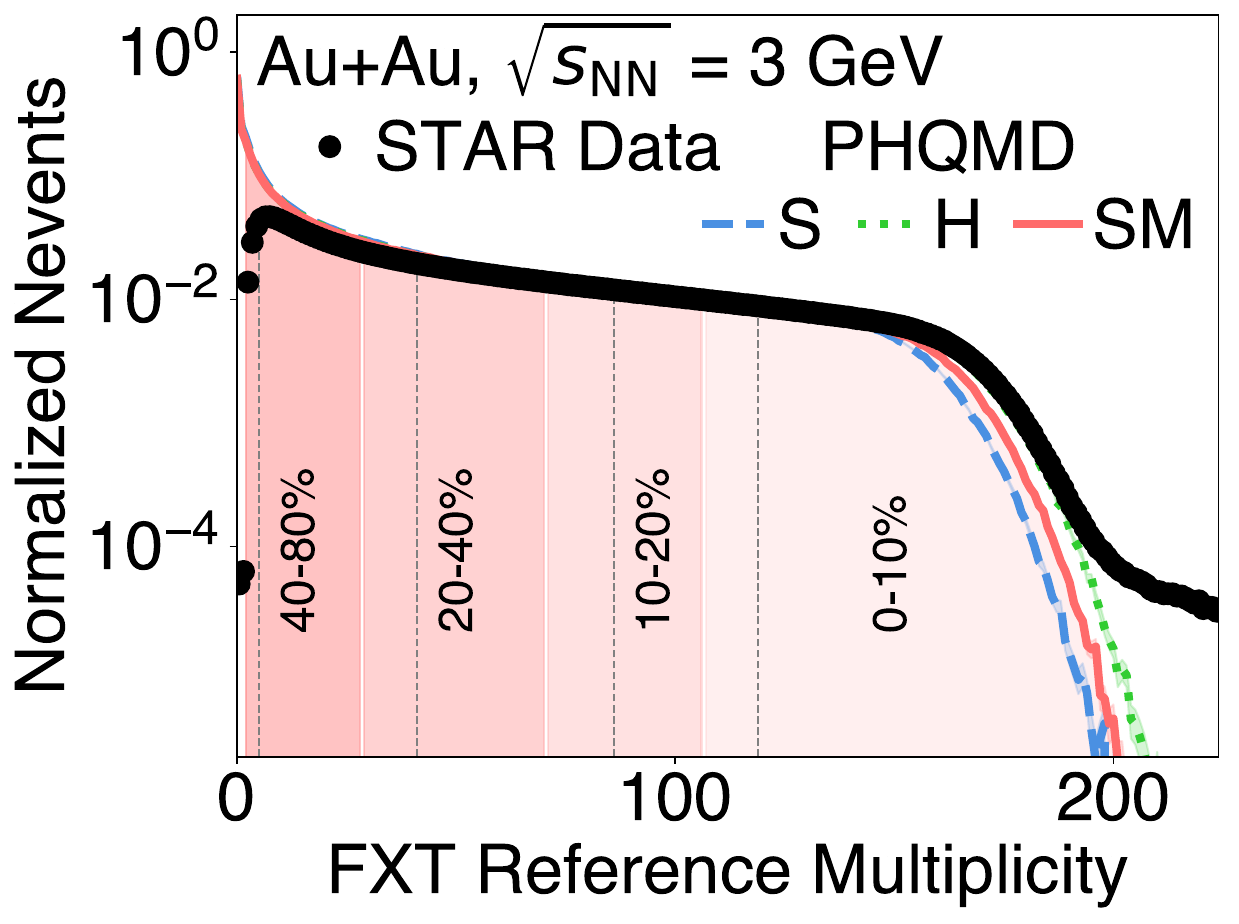}    
    \caption{The multiplicity distribution of charged particles in Au+Au collisions at $\sqrt{s_{\rm{NN}}}=3$ GeV of STAR data (black dots) and of the hadronic transport model PHQMD using a hard (green band), a soft (blue band) and a soft momentum dependent (red) equation of state (EoS). Vertical lines indicate the minimum number of tracks required for an event to be in the corresponding centrality bin for the soft EoS with momentum dependence.
    \label{fig:cent}}
\end{figure}

To compare data and model calculations, we determine the centrality in PHQMD in the same way as done in the experimental analysis. In experimental data, the centrality class is defined using the measured charged particle multiplicity (FXTMult) within the acceptance of the STAR Time Projection Chamber (TPC) ($-2.0<\eta<0$). The relationship between FXTMult and collision centrality is determined by fitting Glauber model calculations~\cite{Miller:2007ri} to FXTMult. In PHQMD, charged particles ($\pi,~K,~p$) are selected within the same $\eta$ range as in data, and with 0.2 $<p_{\rm T}<$ 10.0 GeV/c and $p_{\rm T}>$ 0.35 GeV/c for protons, which mimics the acceptance of the STAR detector. Subsequently, the PHQMD multiplicity distribution is scaled to match the $\sqrt{s_{\rm{NN}}}=3$ GeV STAR data, as shown in Fig.~\ref{fig:cent}. As can be seen, the distribution for the different EoS is rather similar to each other and to the data. The tail in the data at FXTMult$>200$ is due to pileup events, i.e.  multiple collision events, which occur within a short period of time or within the same detector readout window. These events are not included in the experimental analysis. The FXTMult values, corresponding to different collision centralities, are determined using the following formula:
\begin{equation}
    C_{M}=\frac{1}{\sigma^{AA}}\int^{\infty}_{M}dM'\frac{d\sigma}{dM'}
\end{equation}
where $M$ is the FXTMult value, which correspond to a given centrality $C_M$. In Fig.~\ref{fig:cent}. we show the centrality windows in data as well as the event distribution for the three EoS. 

\section{Results on Particle Yield}

\subsection{Transverse Momentum Spectra}
\begin{figure}[h!]             
    \centering \includegraphics[width=0.99\linewidth]{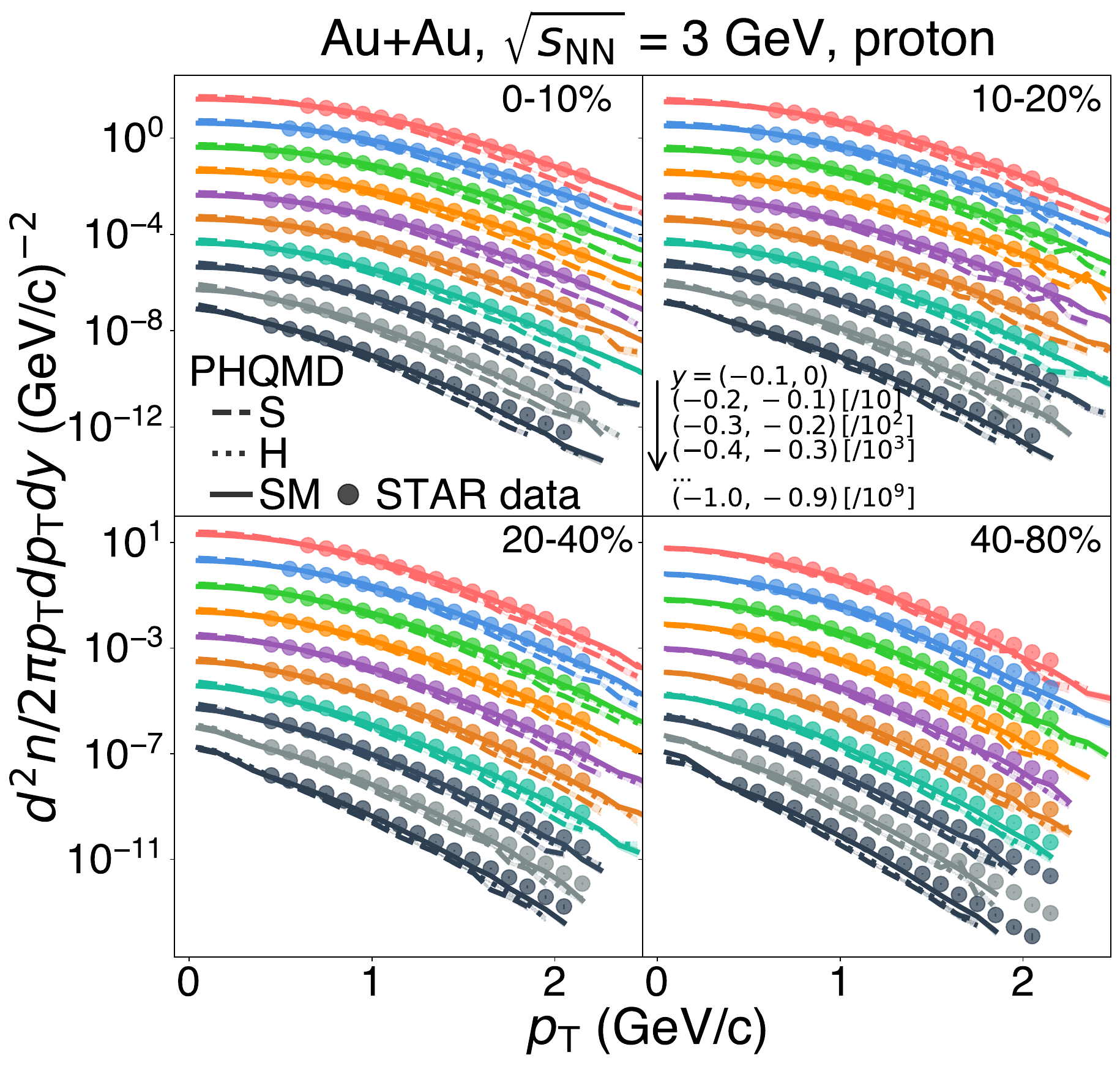}
    \caption{The transverse-momentum spectra of proton for different rapidities and centralities in Au+Au collisions at $\sqrt{s_{\rm{NN}}}=3$ GeV. The proton spectra have been corrected for the feed-down from $\Lambda$ decays. The measured data points from STAR~\cite{STAR:2023uxk} are shown as colored markers and the calculations from PHQMD using soft EOS, hard EOS, and soft EOS with momentum dependence are shown as dotted, dashed, and solid lines respectively. The data points and model calculations are scaled by factors of 1/10 from mid- to forward rapidities as indicated in the legend. 
    \label{fig:pTspectrap}}
\end{figure}
We start out with the comparison of measured differential yields of various particles to PHQMD calculations for several centrality bins with different EoS. Figures~\ref{fig:pTspectrap} \Yj{show the proton $p_T$ spectra at for different rapidities and centralities in Au+Au collisions at $\sqrt{s_{\rm{NN}}}=3$ GeV}. The PHQMD calculations using H and SM EoS describe the data fairly well. The agreement between data and theory increases with increasing centrality and approaching midrapidity. Very good agreement is found near mid-rapidiy, and in central collisions. The predictions underestimate the data near target rapidity in more peripheral collisions. There the emission of light clusters happens by fragmentation or by thermal emission on a much longer time scale than the passing of the two nuclei in heavy ion reactions.  The time scale for cluster production at the rapidity of the residue is given the time to regain the spatial form of the residue and by deexcitation of excited residue states. This time scale is much longer than the time we can pursue the simulations.  Below $p_T=1$ GeV$/c$  calculations using a S EoS are similar to those using a H and SM EoS.  For $p_T>1$ GeV$/c$, the spectra with a S EoS are softer than those with  a H and SM EoS, and the S EoS calculations underestimate the data in all rapidity and centrality bins. \Yj{The deuteron, triton, $^{3}\rm{He}$, $^{4}\rm{He}$ and $\Lambda$ are shown in Appendix Fig.~\ref{fig:pTspectrapdt}, Fig.~\ref{fig:pTspectralambda}, respectively, and exhibit similar overall trends.}

\begin{figure}[h!]             
    \centering 
    \includegraphics[width=0.95\linewidth]{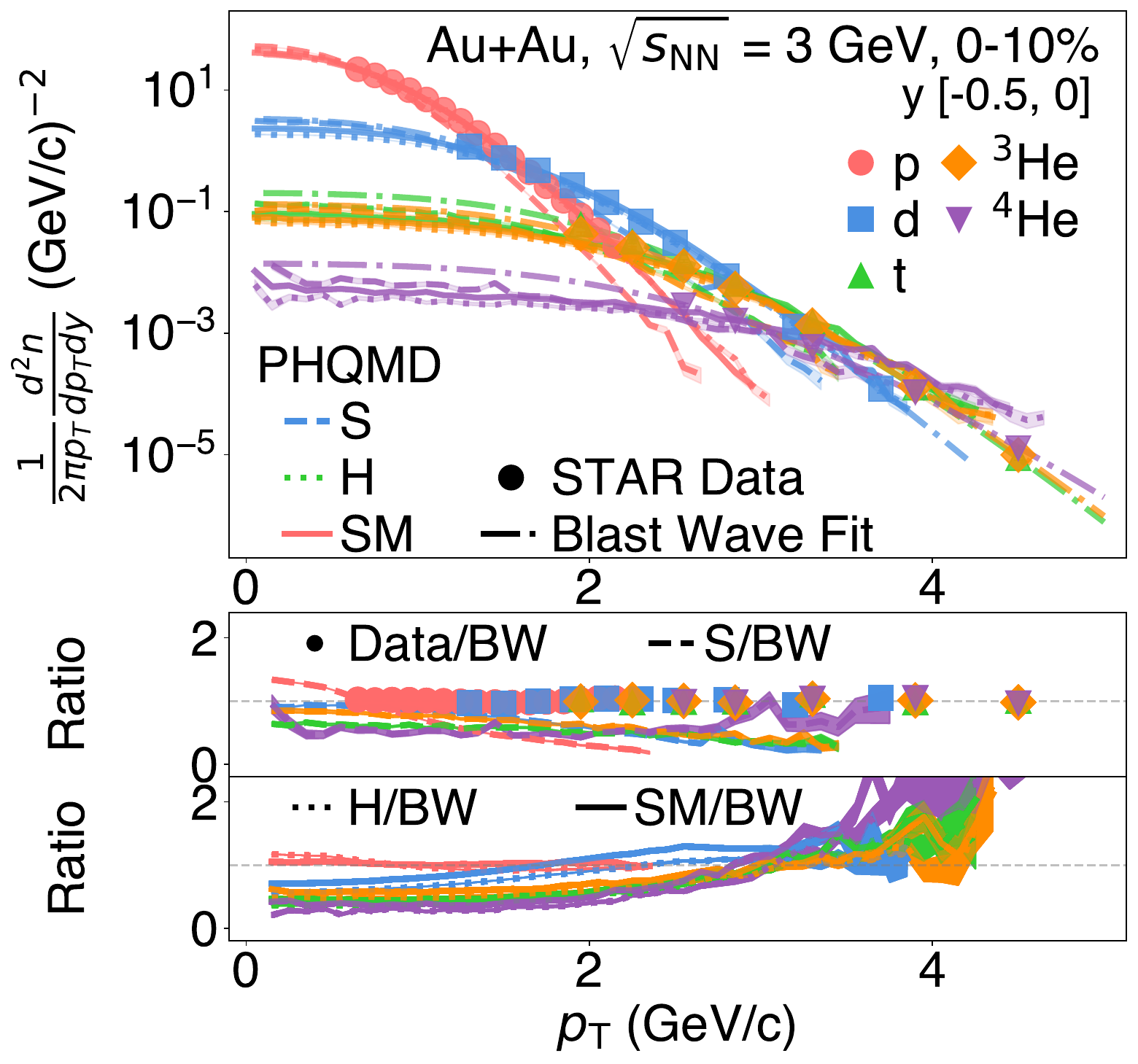}
    \caption{Upper panel: The transverse-momentum spectra of proton, deuteron, triton, $^{3}\rm{He}$ and $^{4}\rm{He}$ for rapidities window between -0.1 and 0,  in $0-10\%$ Au+Au collisions at $\sqrt{s_{\rm{NN}}}=3$ GeV. The measured data points from STAR~\cite{STAR:2024znc} are shown as colored markers and the calculations from PHQMD using soft, hard and soft with momentum dependence EoS  are shown as dotted, dashed, and solid lines respectively. The blast-wave fit to the data is shown as dash-dotted lines. Lower panels: The ratio of the data and PHQMD calculations with soft (upper), hard and soft with momentum dependence EoS (lower) to the blast-wave fit. 
    \label{fig:lightnucleipTspectraratio}}
\end{figure}

To emphasize the cluster size dependence of the $p_T$ spectra we compare in  Fig.~\ref{fig:lightnucleipTspectraratio} the $p_T$ spectra of $p$, $d$, $t$, $^3\rm{He}$, and $^4\rm{He}$ at midrapidty for the most central collisions (0-10\% centrality) with the calculations for the three EoS and with the blast wave fit. It is evident that the spectra show for the different clusters a quite different form. To better visualize the difference between models and data, we calculate also the ratio of the model calculations and the blast-wave fit to the data. The ratios are shown in the lower panels of Fig.~\ref{fig:lightnucleipTspectraratio}. The data-to-fit ratios, represented by the colored markers, are consistent with unity, indicating that the fits are of good quality. The soft EoS predictions are systematically lower than the blast-wave fit for all $p_{T}$ for all clusters. For protons, the H and SM EoS calculations closely follow the blast-wave fit, remaining near unity. In other words, the calculations with H and SM EoS describe the protons at mid-rapidity well, in both the measured region and the extrapolated region. This also implies that the choice of using the blast-wave function for the extrapolation to low $p_T$ values is reasonable. 

For deuterons, the H and SM EoS predictions are mostly consistent with the blast-wave fit in the measured data region. However, some tension can be observed for the slope of the data and the H and SM predictions. Due to this tension, both calculations fall below the blast-wave fit at low $p_{T}$ in the extrapolated region. For tritons, $^{3}\rm{He}$, and $^{4}\rm{He}$, the situation is similar to that of deuterons; although the difference in the slope of the $p_T$ spectra is more significant. This leads to an even larger deviation between the H and SM calculations and the blast-wave fit at low $p_T$.
\begin{figure}[h!]             
    \centering    
    \includegraphics[width=0.49\linewidth]{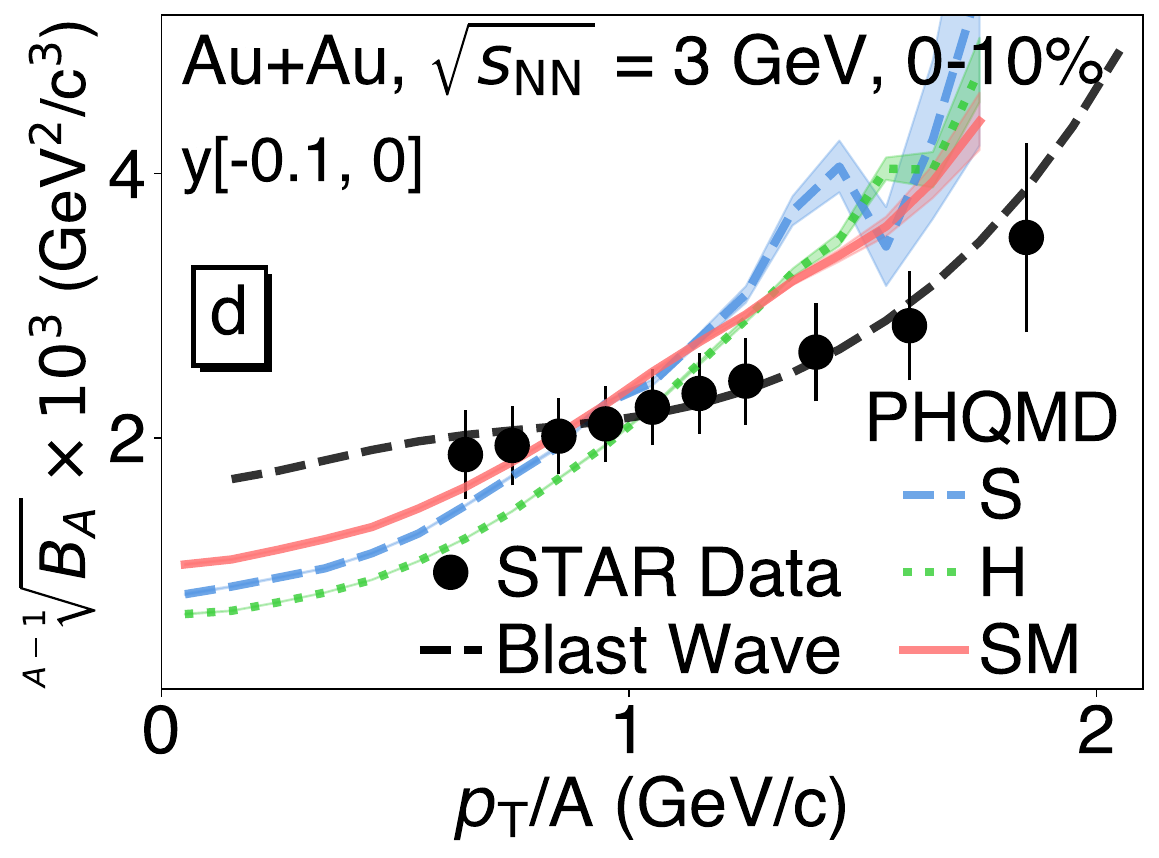}    
    \includegraphics[width=0.49\linewidth]{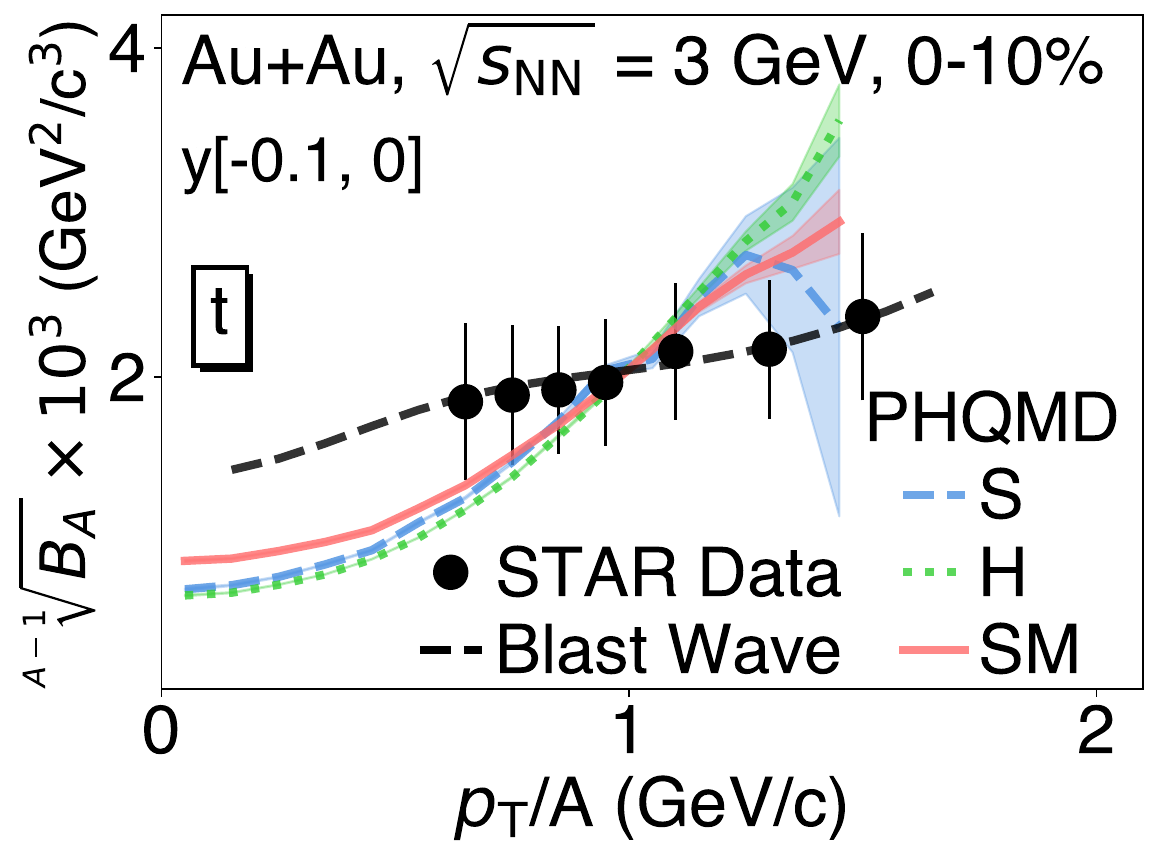}  
    \includegraphics[width=0.49\linewidth]{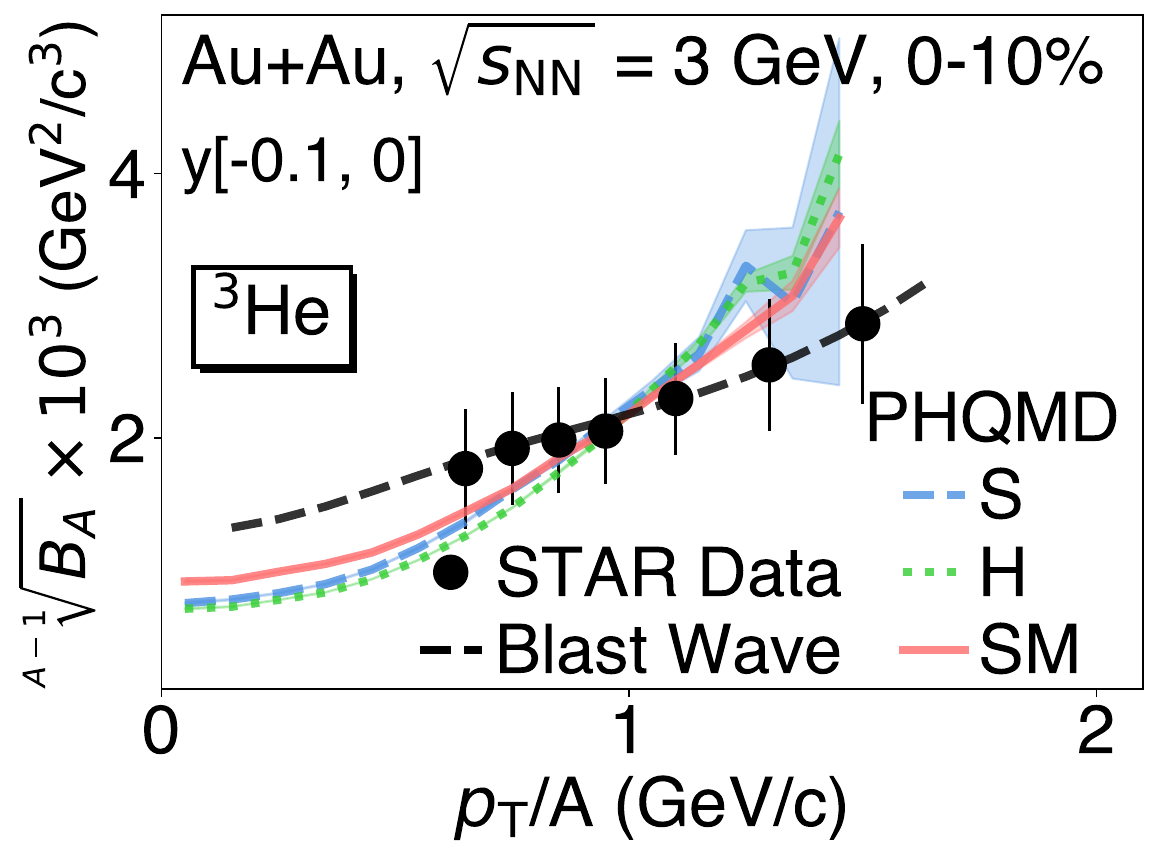}  
    \caption{The coalescence parameters $B_A$ to the power $(A-1)^{-1}$ for deuterons (upper panel), tritons (middle panel) and $^{3}\rm{He}$ (bottom panel) as a function of the $p_T$ divided by $A$ in $0-10\%$ Au+Au collisions at $\sqrt{s_{\rm{NN}}}=3$ GeV. The data are shown as black markers and the blast-wave fit to the data is shown as dash-dotted lines. The calculations from PHQMD using soft, hard and soft with momentum dependence EoS are shown as blue, green and red lines respectively.
    \label{fig:coalescenceparameters}}
\end{figure}

In order to compare the $p_T$ spectra of clusters with those of free nucleons one can introduced the covariant coalescence function, $B_A$:
\begin{eqnarray}    
E_A \frac{d^3N_A}{d^3p_A} = B_A \left( E_p \frac{d^3N_p}{d^3p_p} \right)^Z \left( E_n \frac{d^3N_n}{d^3p_n} \right)^{A-Z}\\
    \approx (1.3)^{A-Z} B_A \left( E_p \frac{d^3N_p}{d^3p_p} \right)^A \bigg|_{p_p = p_n = \frac{p_A}{A}}
\end{eqnarray}
where $A$ and $Z$ are the mass and proton number of the light nuclei, respectively. We adopt the same definition as used in the experimental data~\cite{STAR:2023uxk}, which assumes that protons and neutrons have the same $p_T$, rapidity, and centrality dependence, and the neutron yields is obtained by scaling the proton yields by a factor of $n/p = 1.3 \pm 0.1$. This factor is estimated by assuming the relation $n/p = t/{}^{3}\rm{He}$. Figure~\ref{fig:coalescenceparameters} describes the dependence of the coalescence parameters for $B_2$(d) and $B_3$(t) on the scaled transverse momentum $(p_T/A)$, and $B_3$($^{3}\rm{He}$) at mid-rapidity in $0-10\%$. Data show increasing $B_A$ with increasing $p_T$. These trends are qualitatively reproduced by the PHQMD calculations. The model approximately describes the data within the experimental uncertainties in the measured region, although there is some tension at low $p_T/A$, particularly for $d$ and $t$. Because the tension is in the lower $p_T/A$, the extrapolation to the full $p_T$ range using the blast wave fit is affected as well; as one can see, the dashed line, which represents the blast wave fit, also lies above the PHQMD calculations. The effects of extrapolation and comparison of quantities such as $p_T$-integrated yields and mean transverse momentum will be discussed in more detail in the next section. Finally, we note that differences between different EoS are less pronounced in $B_A$\(\), as the ratio cancels out some of the EoS dependencies.

\subsection{$p_T$-Integrated Yields And Yield Ratios \label{sec:dndy}}
In the experimental data analysis, \Yj{due to detector acceptance and statistics, the low- and high-$p_T$ regions cannot be measured directly}. The $p_T$-integrated yields are therefore estimated by extrapolating the spectra to the unmeasured regions using functions such as the blast-wave formula~\cite{Schnedermann:1993ws},
\begin{align}
\frac{1}{2\pi p_{T}}\frac{d^{2}N}{dp_{T}dy} \propto & \nonumber\\
\int_{0}^{R} r dr m_T & I_0 (\frac{p_T sinh \rho(r)}{T_{kin}}) K_1 (\frac{m_T cosh \rho(r)}{ T_{kin}}) 
\label{eq:blast}
\end{align}

\noindent where $m_T$ is the transverse mass of the particle, $I_0$ and $K_1$ are the modified Bessel functions,
and $\rho(r) = tanh^{-1} (\beta_{T})$, where $\beta_T$ is the radial flow velocity. In the region $0 \le r \le R$, $\beta_T$ can be expressed as $\beta_{T} = \beta_{S}(r/R)^n$, where $\beta_S$ is the surface velocity, and $n$ reflects the form of the flow velocity profile. \Yj{The average transverse flow velocity of the source is then defined as $\langle \beta_{T} \rangle=\frac{2}{n+2}\beta_{S}$, which represents the spatially averaged transverse velocity of the fireball over the radial profile at kinetic freeze-out.} For the experimental data reported in this manuscript, $n$ is fixed to 1, i.e. the transverse velocity profile is chosen to depend linearly on the radius, while $T_{kin}$ and the overall normalization are free parameters in the fit. The fitted values for various particles observed in the  $0-10\%$ centrality bin at mid-rapidity $y=(-0.1,0)$ are tabulated in Tab.~\ref{tab:bwparameters}, \Yj{together with the fractions of the yield within the measured $p_T$ range relative to the total yield~\cite{STAR:2023uxk}}.
 
\begin{table}[h]
    \centering
    \begin{tabular}{lccc}
        \toprule
        Particles & $T_{\text{kin}}$ (GeV) & $\langle \beta_{T} \rangle$ (c) & Measured percentage (\%) \\
        \midrule
        $p$       &  $0.062\pm0.004$ & $0.45\pm0.01$ & $55\pm4$\\
        $\Lambda$ &  $0.062\pm0.006$ & $0.39\pm0.02$ & $55\pm1$\\
        $d$       &  $0.070\pm0.004$ & $0.40\pm0.01$ & $36\pm5$\\
        $t$       &  $0.078\pm0.005$ & $0.37\pm0.01$ & $23\pm2$\\
        $^3$He    &  $0.077\pm0.005$ & $0.38\pm0.01$ & $26\pm2$\\
        $^4$He    &  $0.080\pm0.010$ & $0.36\pm0.02$ & $15\pm1$\\
        \bottomrule        
    \end{tabular}
    \caption{Fitted kinetic freeze-out parameters for various particles and \Yj{the percentage (\%) of the measured $p_T$ range relative to the total $dN/dy$}, measured in $0-10\%$ centrality bin at mid-rapidity $y=(-0.1,0)$ for $\sqrt{s_{\rm{NN}}}=3$ GeV Au+Au collisions. }
    \label{tab:bwparameters}
\end{table}

The blast-wave function is used for all particles except ${}^{3}_{\Lambda}\rm{H}$ and ${}^{4}_{\Lambda}\rm{H}$, because the number of data points is too small for a stable fit. For the hypernuclei, the $m_T$-exponential function, 
\begin{align}
&\frac{dN}{m_T dm_T} \propto e^{-m_{T}/T_{m_{T}}}
\end{align}
\noindent is used instead, where $m_T = \sqrt{m_0^2 + p_T^2}$ is the transverse mass of the particle, and $T_{m_{T}}$ is a fit parameter. Based on the difference of the estimated yields using different extrapolation functions, a systematic uncertainty is assigned. It depends both on the statistical accuracy of the $p_T$ spectra and the fraction of the unmeasured region. For the majority of cases, this uncertainty ranges from \Yj{$3-6\%$ for protons and increases for ${}^{4}_{\Lambda}\rm{H}$ with $10-30\%$}.
\begin{figure}[!htbp]             
    \centering    
    \includegraphics[width=0.7\linewidth]{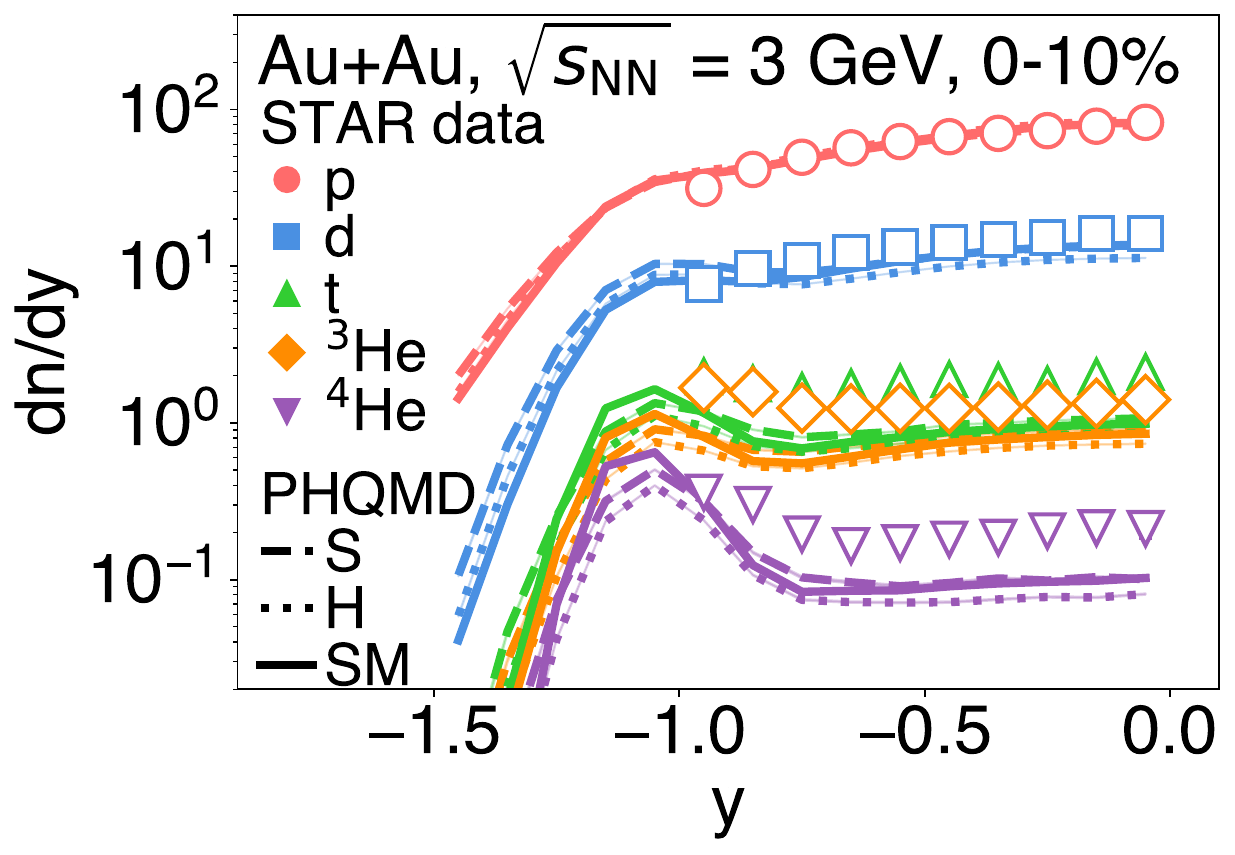}    
    \includegraphics[width=0.7\linewidth]{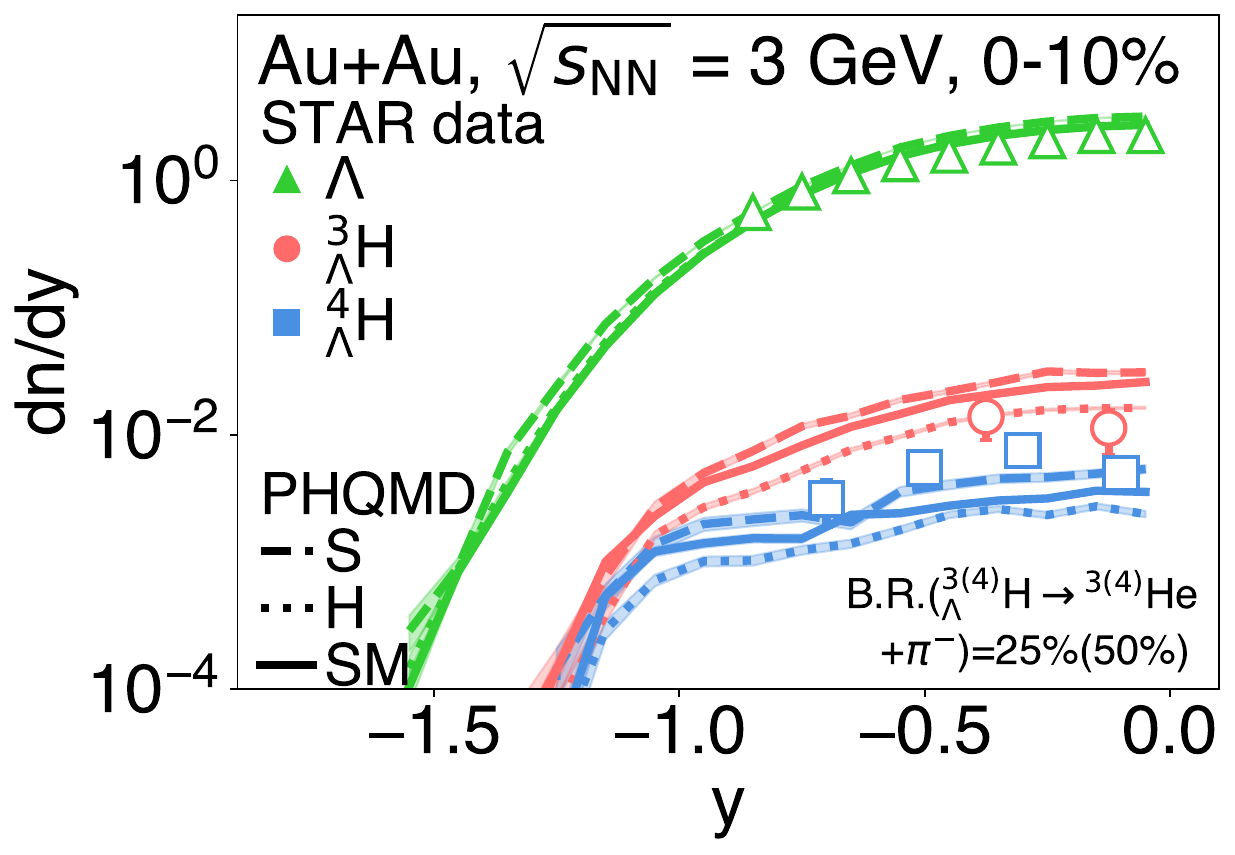} 
    \caption{The $p_T$ integrated yield of proton, deuteron, triton, ${^{3}\rm{He}}$, ${^{4}\rm{He}}$ (up), $\Lambda$ ${}^{3}_{\Lambda}\rm{H}$ and ${}^{4}_{\Lambda}\rm{H}$ (down) as a function of rapidity in $0-10\%$ Au+Au collisions at $\sqrt{s_{\rm{NN}}}=3$ GeV. The measured data points from STAR~\cite{STAR:2024znc, STAR:2023uxk, STAR:2021orx} are shown as colored markers, while the calculations from PHQMD using a soft, hard and soft momentum dependent EoS are shown as dotted, dashed and solid lines, respectively.
    \label{fig:dndynuclei}}
\end{figure}

Fig.~\ref{fig:dndynuclei} shows the $p_T$-integrated $p$, $d$, $t$, ${}^{3}\rm{He}$, ${}^{4}\rm{He}$, $\Lambda$, ${}^{3}_{\Lambda}\rm{H}$ and ${}^{4}_{\Lambda}\rm{H}$ yields as a function of rapidity for $0-10\%$ central Au+Au collisions at $\sqrt{s_{\rm{NN}}}=3$ GeV. As in the previous section, the data~\cite{STAR:2024znc, STAR:2023uxk, STAR:2021orx} are compared to PHQMD with a S, H, and SM EoS. It should be noted that in PHQMD, \Yj{yields are obtained over the full $p_T$ range, so no extrapolation is needed}. Thus, differences of the rapidity distribution between PHQMD calculations and the data can arise from two factors: 1) a difference of the integrated yield in the measured region 2) a difference of the extrapolation of the experimental data into the unmeasured $p_T$ regions by a fit function because the extrapolation function is not uniquely determined. The latter may be a consequence of a different shape of the calculations and data in the measured region, even if both agree in between the error bars.

The proton rapidity distribution is rather well reproduced by all three PHQMD calculations. For deuterons, we observe that PHQMD with H EoS predicts a lower yield compared to those with S and SM EoS, which indicates that the light nuclei yields are sensitive to the stiffness of the EoS. The data favors the S and SM EoS, although both calculations still underestimate the data by $\sim20\%$ at mid-rapidity. Near target rapidity, all three PHQMD calculations predict a prominent peak-like structure due to fragments from the target and projectile residues, but this structure is wider in the data for $t$, ${}^{3}\rm{He}$, and ${}^{4}\rm{He}$ and is not visible for $p$ and $d$. We observe that the calculations using a H EoS are consistently lower than those using a S and SM EoS. However, the calculations underestimate the experimental data by approximately a factor of 2 for $t$ and ${}^{3}\rm{He}$, and the factor of 3 for ${}^{4}\rm{He}$.

As for the strange particles, we observe that the $\Lambda$ yield has a mild sensitivity to the EoS, with calculations with S EoS predicting the largest yields, and SM predicting the lowest yields. This is due to the different densities and hence of the mean free paths, which are obtained in these calculations. Within uncertainties, all three calculations describe the data. As for ${}^{3}_{\Lambda}\rm{H}$ and ${}^{4}_{\Lambda}\rm{H}$, we observe that a S EoS predicts the largest yield, followed by a SM and a H EoS. The difference between the predictions is rather significant in a wide range of rapidities, e.g. predictions from S and H EoS differ by approximately a factor of 2. This indicates a higher sensitivity of the hypernuclei yields to the EoS compared to the yields of other particles. The calculations with H EoS give a good description of the ${}^{3}_{\Lambda}\rm{H}$ data, while calculations with S EoS describes ${}^{4}_{\Lambda}\rm{H}$. 

It is important to note that the experimental yields of 
${}^{3}_{\Lambda}\rm{H}$ and ${}^{4}_{\Lambda}\rm{H}$ were obtained assuming branching ratios (B.R.) of $25\%$ and $50\%$ for their two-body decays, respectively~\cite{STAR:2021orx}. 
The decay B.R. of ${}^{3}_{\Lambda}\rm{H}\rightarrow{{}^{3}He + \pi^{-}}$ was not directly measured. A variation in the range by 15-35\% for the B.R. was considered when calculating the total $dN/dy$. 
For ${}^{4}_{\Lambda}\rm{H}\rightarrow{{}^{4}He + \pi^{-}}$ a variation of 40-60\% was considered in the analysis. PHQMD cannot describe excited states of nuclei.
Therefore the $\gamma$-decay of the excited state of ${}^{4}_{\Lambda}\rm{H}^{*}(1+)$ to the ground state is not taken into account by the PHQMD calculations.

\begin{figure}[h!]             
    \centering    
\includegraphics[width=0.99\linewidth]{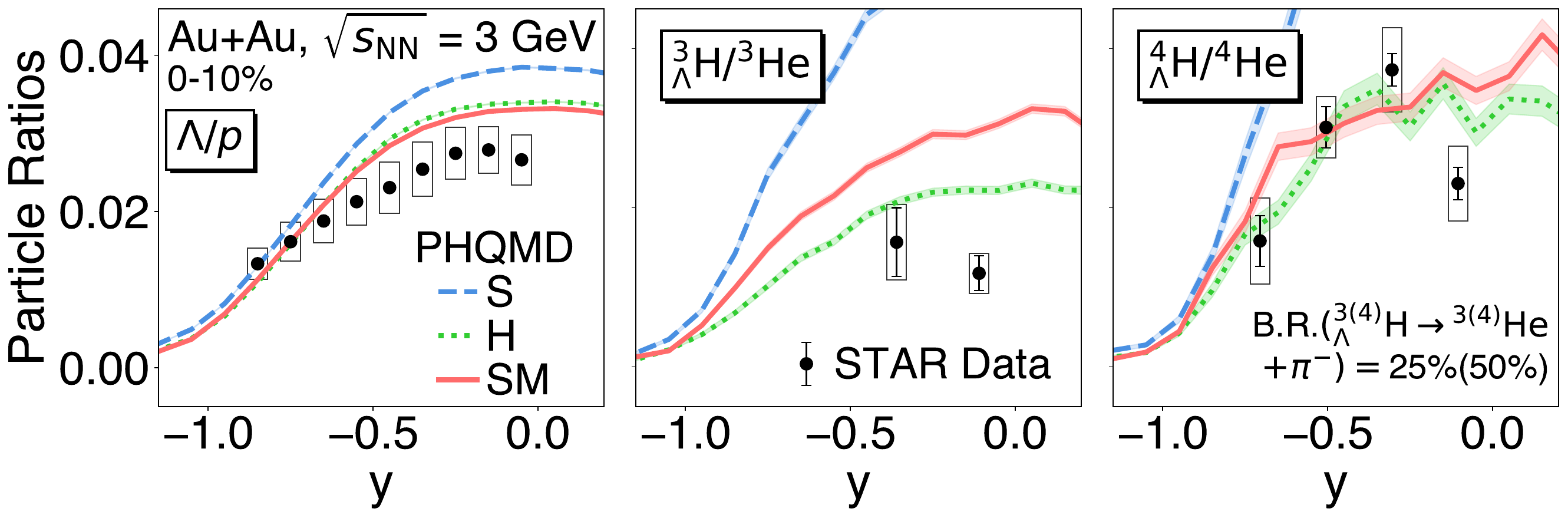}    
    \caption{The yield ratios $\Lambda/p$, ${}^{3}_{\Lambda}\rm{H}/^{3}_{}\rm{He}$ and ${}^{4}_{\Lambda}\rm{H}/^{4}_{}\rm{He}$~\cite{STAR:2024znc,STAR:2021orx,STAR:2023uxk} as a function of rapidity in $0-10\%$ central Au+Au collisions at $\sqrt{s_{\rm{NN}}}=3$ GeV. The measured data points are shown as black markers, while the calculations from PHQMD using soft, hard, and soft momentum dependent EoS are shown as blue, green, and red bands, respectively.    
    \label{fig:dndyhypernucleiyieldratio}}
\end{figure}
\begin{figure}[h!]          
    \centering \includegraphics[width=0.8\linewidth]{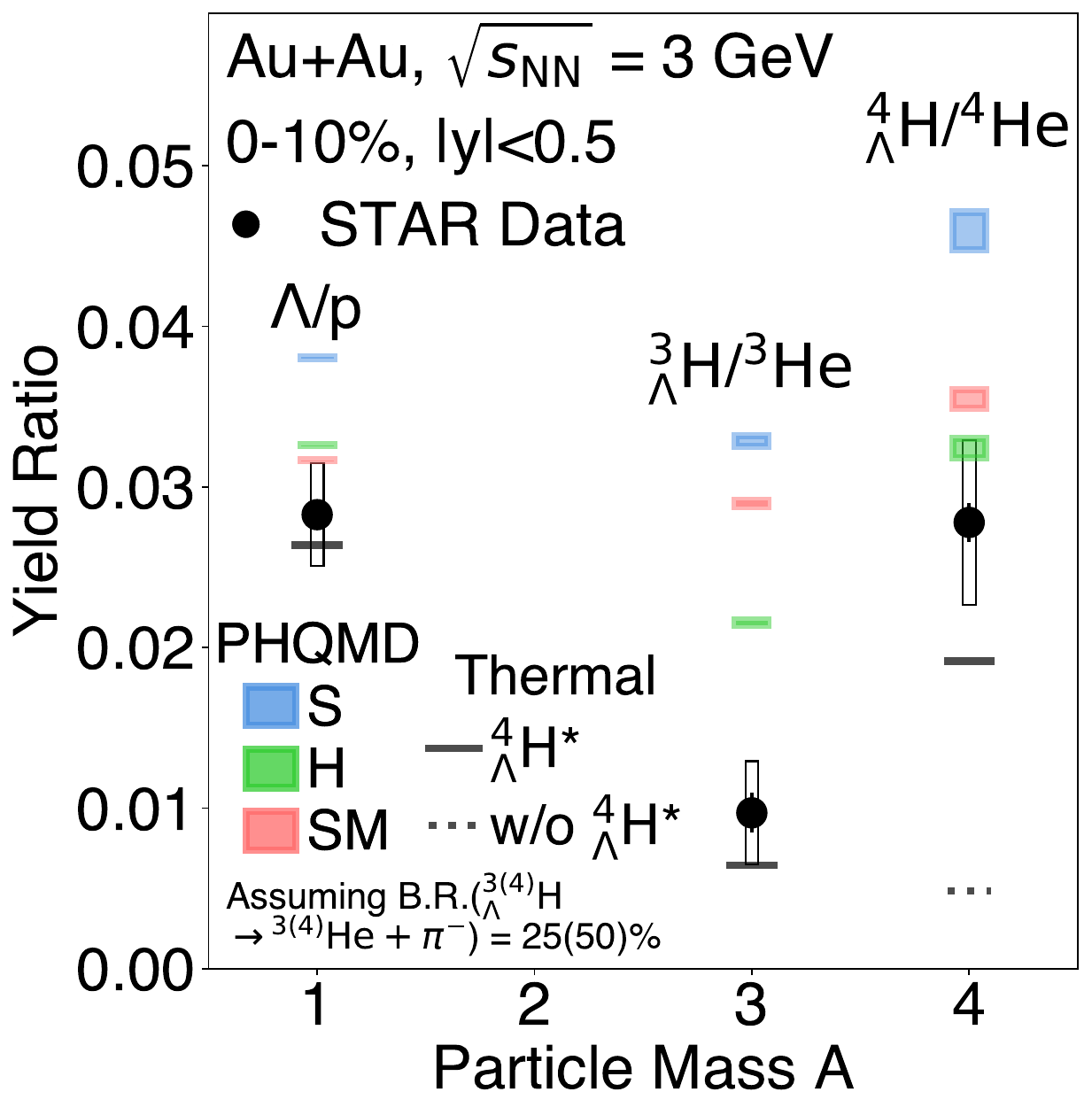}    
    \caption{The yield ratios of hypernuclei to nuclei as a function of the mass number in $0-10\%$ Au+Au collisions at $\sqrt{s_{\rm{NN}}}=3$ GeV~\cite{STAR:2024znc,STAR:2021orx,STAR:2023uxk}. The measured data points are shown as black markers, while the calculations from PHQMD using soft, hard, and soft with momentum dependence EoS are shown as blue, green, and red bands respectively. Expectations from the Thermal-Fist model~\cite{Vovchenko:2018fiy} are also shown. 
\label{fig:dndyhypernucleiyieldratiomid}}
\end{figure}
The yield ratios as a function of rapidity, shown in Fig.~\ref{fig:dndyhypernucleiyieldratio}, are calculated by dividing the $\Lambda$ yield by the proton yield and the yield of each hypernucleus by the yield of the nucleus with the same mass number. For all three ratios $(\Lambda/p, {}^{3}_{\Lambda}\rm{H}/^{3}\rm{He}, {}^{4}_{\Lambda}\rm{H}/^{4}\rm{He})$, calculations with a S EoS overestimate the data, while those with H and SM give a fair description of the data, although the SM and H EoS calculations overestimate the ${}^{3}_{\Lambda}\rm{H}/^{3}\rm{He}$ ratio slightly. Figure~\ref{fig:dndyhypernucleiyieldratiomid} shows the hypernuclei-to-nuclei ratio at mid-rapidity ($|y|<0.5$), this time plotted as a function of the mass number. The dependence of this ratio is non-monotonic. This non-monotonicity is also seen in PHQMD calculations, particularly for the H EoS, however to a lesser extent. We also compare with calculations from a thermal model ~\cite{Vovchenko:2018fiy}, with and without the inclusion of the excited ${}^{4}_{\Lambda}\rm{H^*}(1^+)$ decay. The inclusion of this decay in the model reproduces the trend in the data, which implies that the non-monotonicity could be partially arising from feed-down of the excited state ${}^{4}_{\Lambda}\rm{H^*}$. A more complete understanding in the role of feed-down on hypernuclei production may be needed to establish hypernuclei as probes for the equation-of-state in the future.

\begin{figure}[h!]             
    \centering    
    \includegraphics[width=0.99\linewidth]{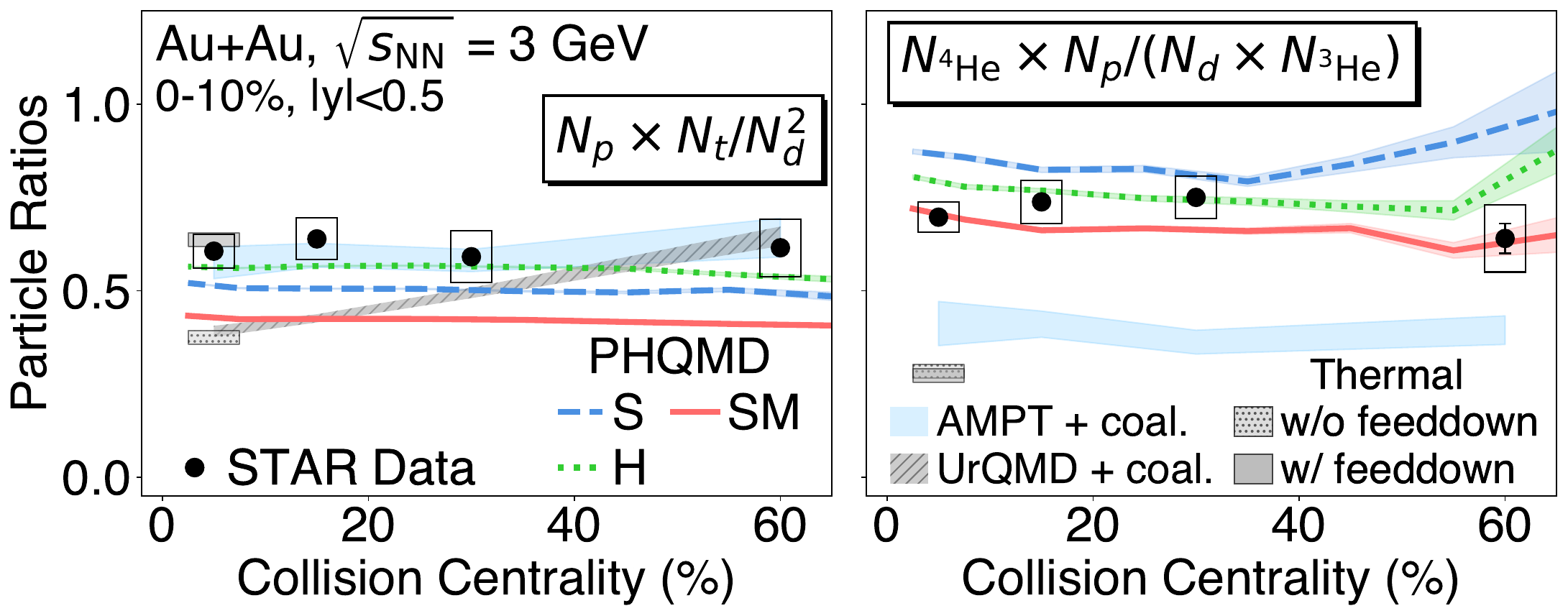}   
    \caption{The yield ratios $N_p\times N_t / N_d^2$ (left) and $N_{^{4}\rm{He}} \times N_p / (N_{^{3}\rm{He}} \times N_d)$ (right) in the rapidity region $y=(-0.5,0)$ as a function of centrality in Au+Au collisions at $\sqrt{s_{\rm{NN}}}=3$ GeV~\cite{STAR:2023uxk}. The calculations from PHQMD using soft, hard, and soft momentum dependent EOS are shown as blue, green, and red bands, respectively. Calculations from transport models with coalescence afterburner~\cite{Sun:2022cxp} and thermal model~\cite{Vovchenko:2018fiy} are also shown for comparison.
    \label{fig:nucleicompoundratios}}
\end{figure}

The light nuclei ratios $N_p\times N_t / N_d^2$ and $N_{^{4}\rm{He}} \times N_t / (N_{^{3}\rm{He}} \times N_d)$, shown in Fig.~\ref{fig:nucleicompoundratios}, have been suggested to be sensitive to baryon density inhomogeneities generated by a possible critical end point or by a first-order phase transition~\cite{Sun:2022cxp}, and have been proposed as sensitive observables for the equation of state~\cite{Sun:2022cxp, Sun:2018jhg, Shuryak:2019ikv} Although at this energy we do not expect to reach the point where a first order phase transition may occur, we think it is meaningful to confront the  data with the predictions. In particular, the observed flat centrality dependence of $N_p\times N_t / N_d^2$ at $\sqrt{s_{\rm{NN}}}=3$ GeV is consistent with coalescence calculations including a first-order phase transition \cite{Sun:2022cxp} (AMPT+coalescence), while those without (UrQMD+coalescence) exhibit an increasing trend which is inconsistent with the data. We note here that the MST clusterization algorithm for PHQMD produces a flat centrality dependence for all employed EoS. The results are qualitatively consistent with the data, without the need of inquiring a first-order phase transition. This also points to the need of a complete understanding of the cluster formation procedure, before one advocates cluster formation as a probe for a first-order phase transition or for a  critical point.

\section{Results on Collectivity}
\subsection{Mean Transverse Momentum}

As we have seen in the preceding section, the experimental transverse momentum spectra of all particles can be well described in a blast-wave model, which has two parameters, the temperature of the fireball T and the collective radial expansion $\beta$ of the medium. 
To condense the information we can define the average transverse momentum as: 

\begin{align}
\langle p_T \rangle(T,\beta) = \frac{\int p_T \frac{dN}{dp_T}dp_T}{\int \frac{dN}{dp_T} dp_T}.
\end{align}

This quantity is based, as the rapidity distribution in Sec.~\ref{sec:dndy}, on the assumption that a blast-wave describes the spectra well also in the unmeasured regions. 
Different functions are used for the extrapolation to estimate the systematic uncertainties. 

\begin{figure}[!h]             
    \centering    
    \includegraphics[width=0.49\linewidth]{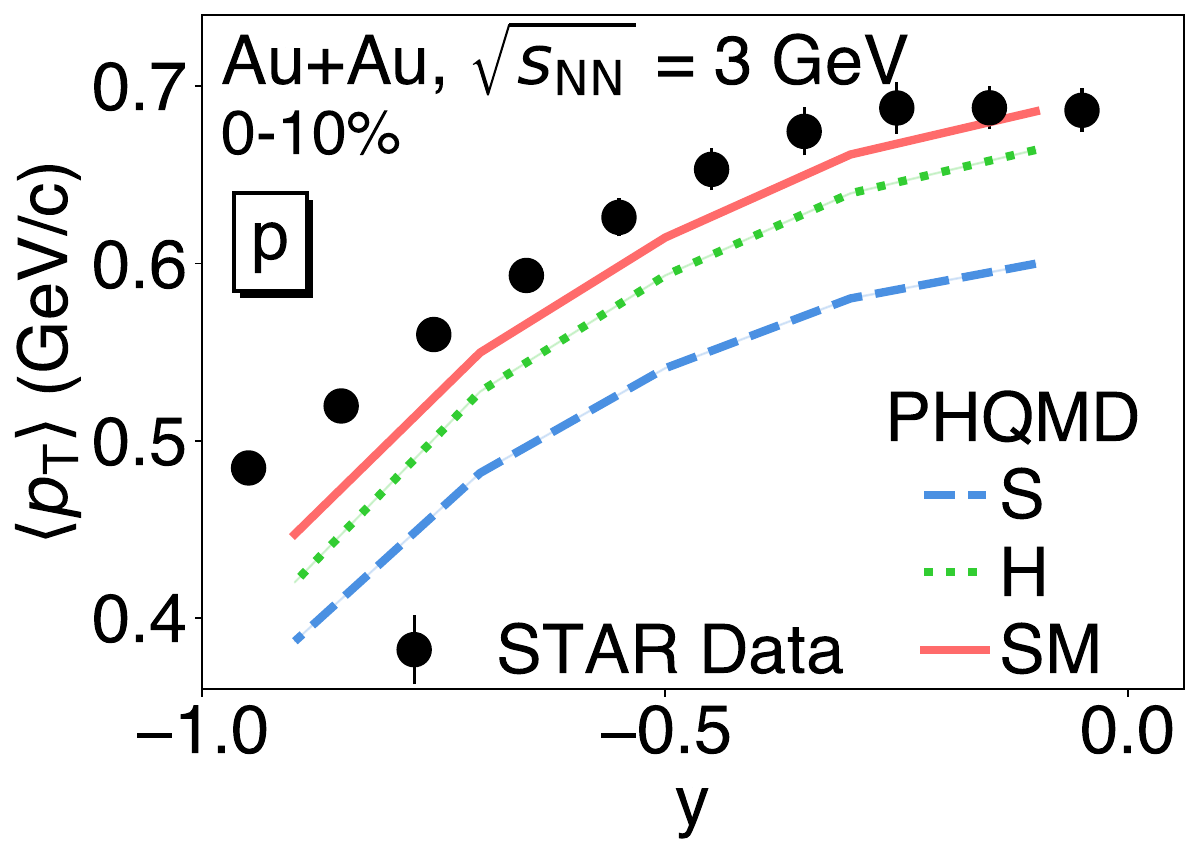}
    \includegraphics[width=0.49\linewidth]{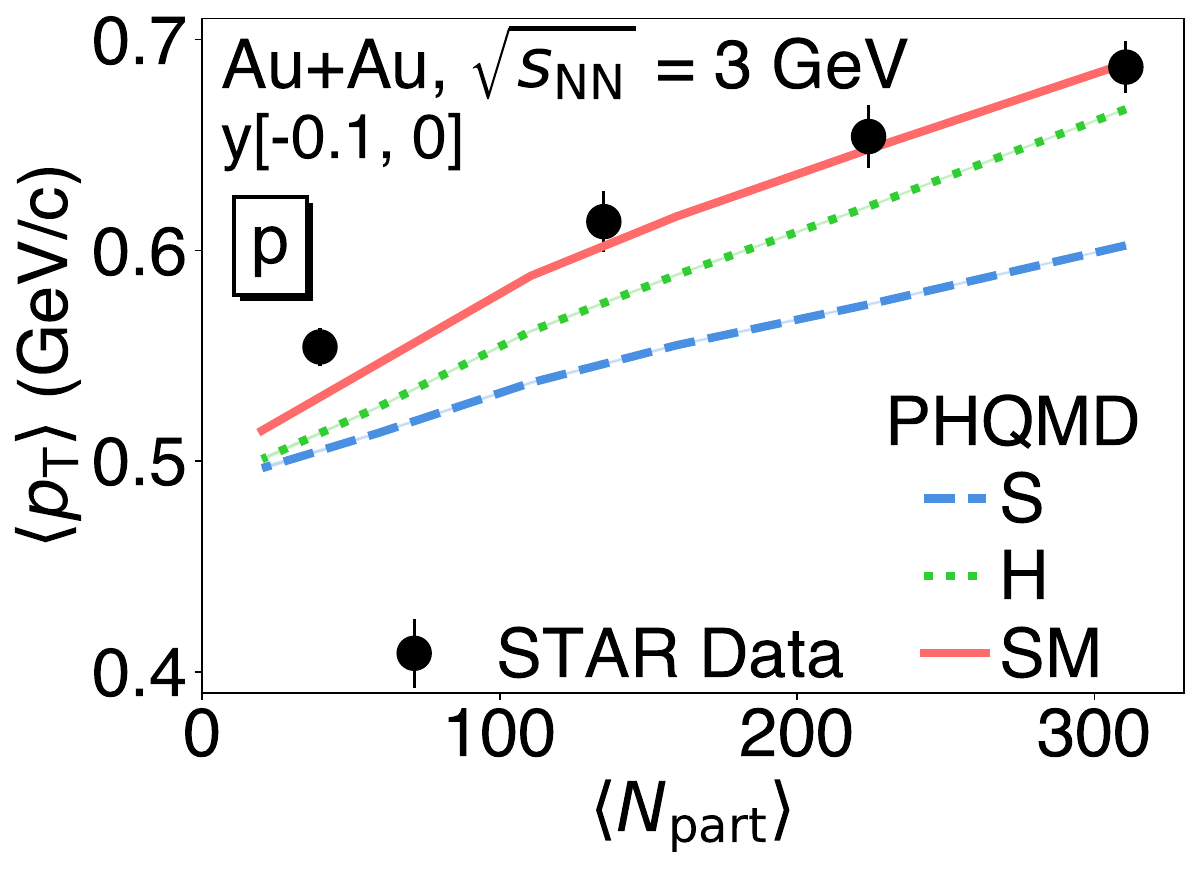}
    \includegraphics[width=0.49\linewidth]{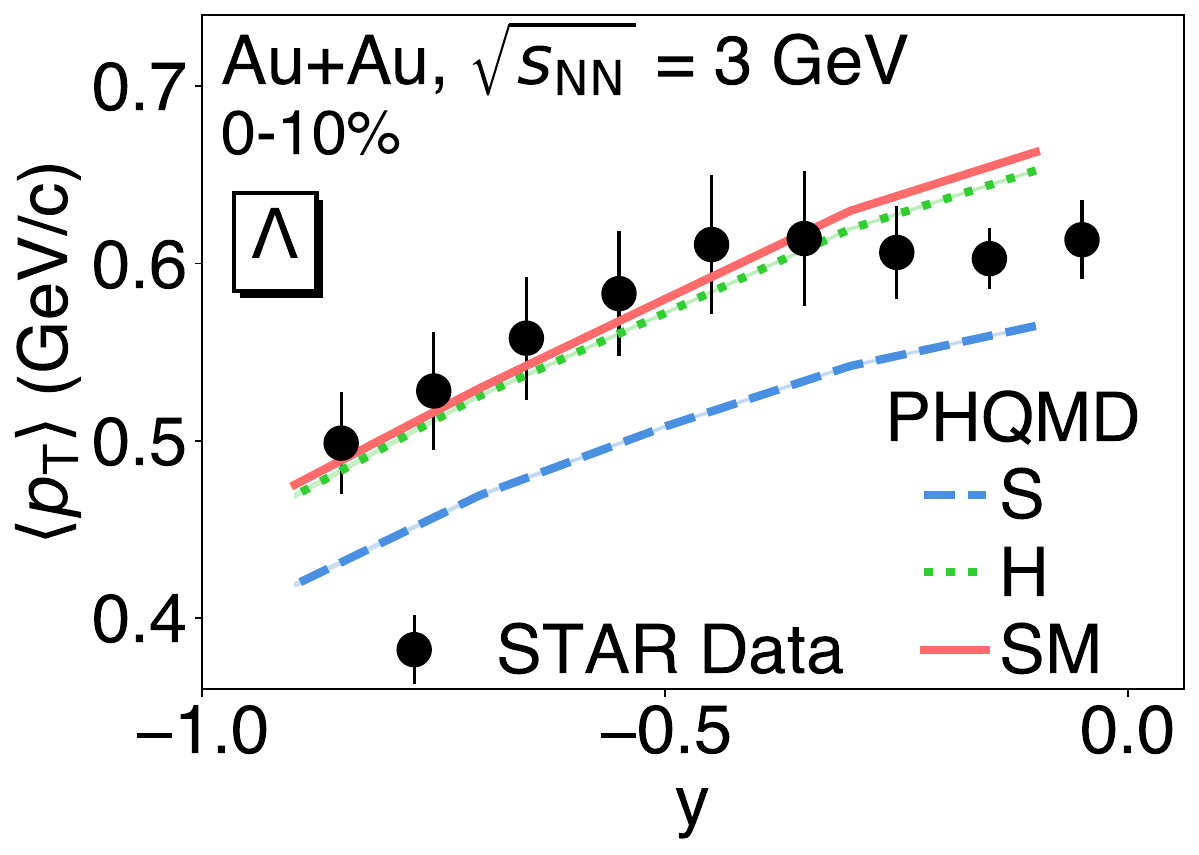}
    \includegraphics[width=0.49\linewidth]{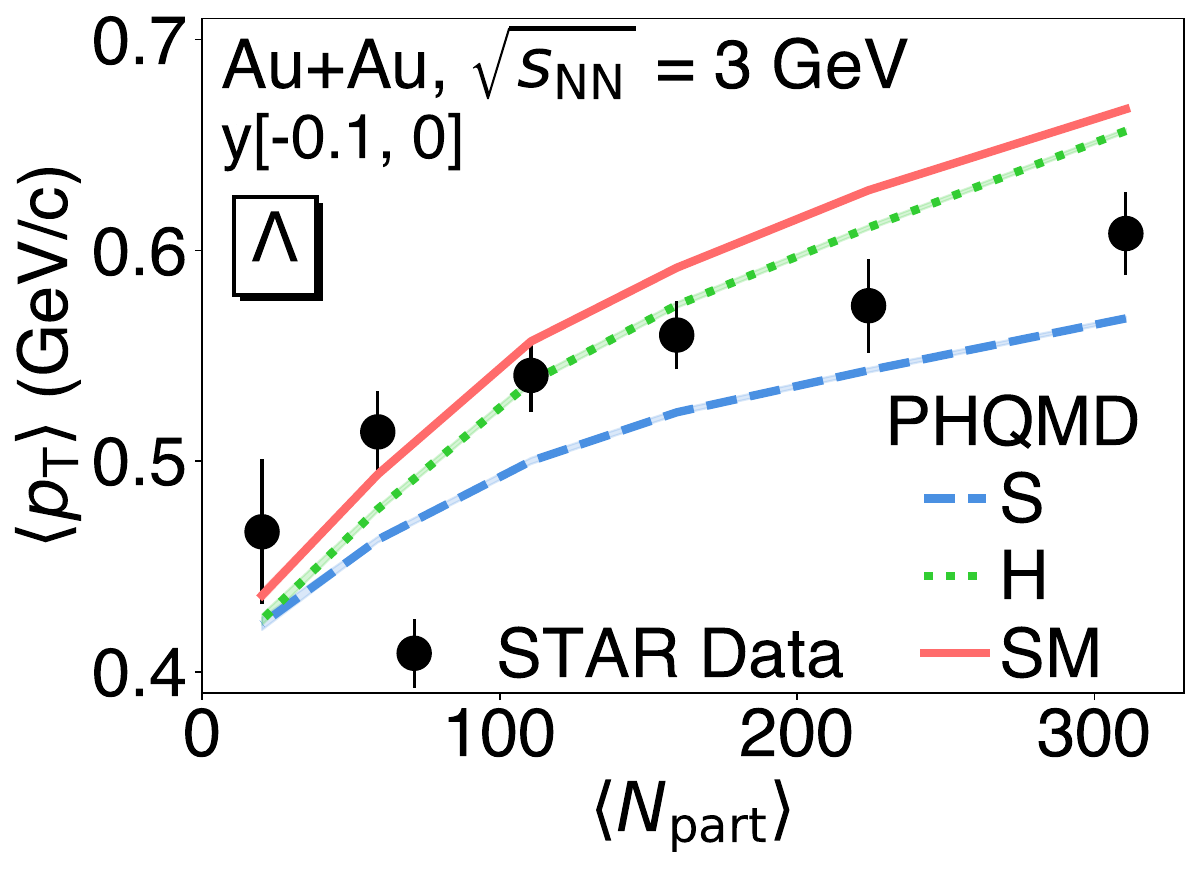}
    \includegraphics[width=0.49\linewidth]{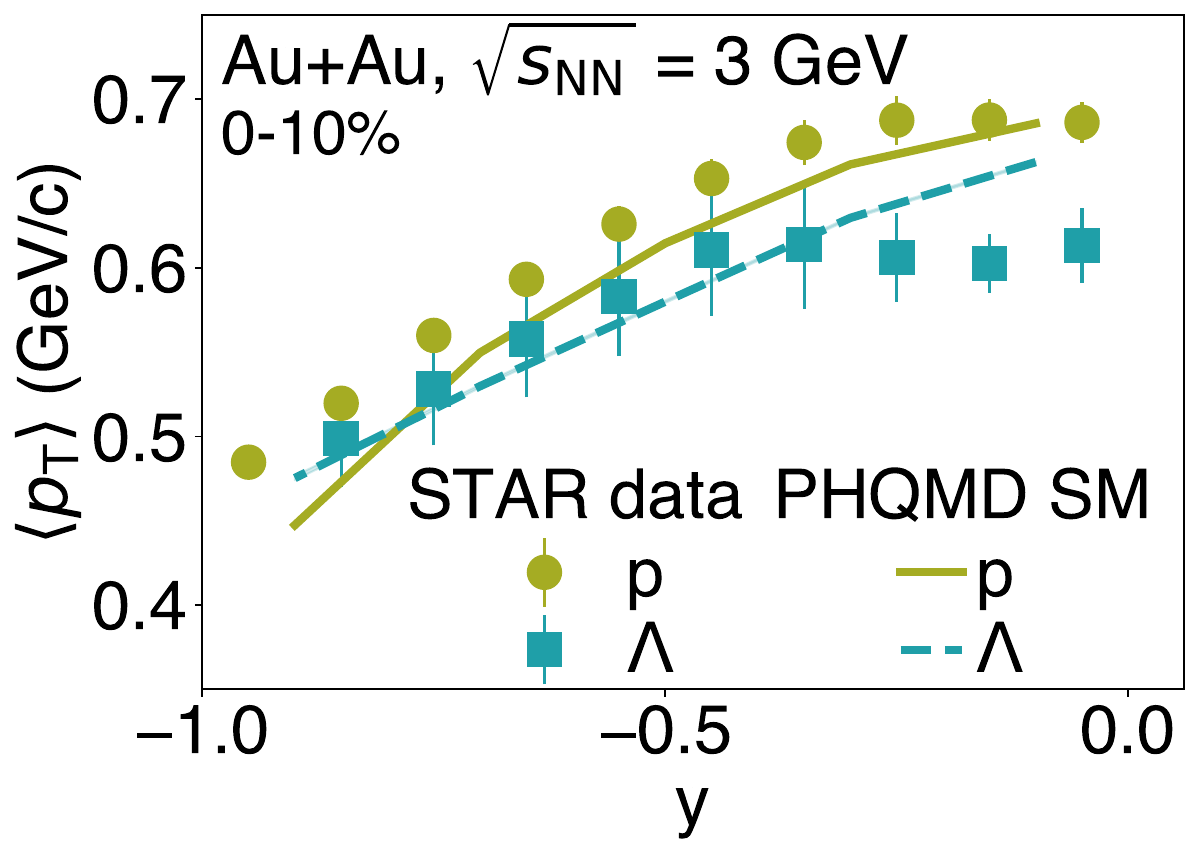}
    \includegraphics[width=0.49\linewidth]{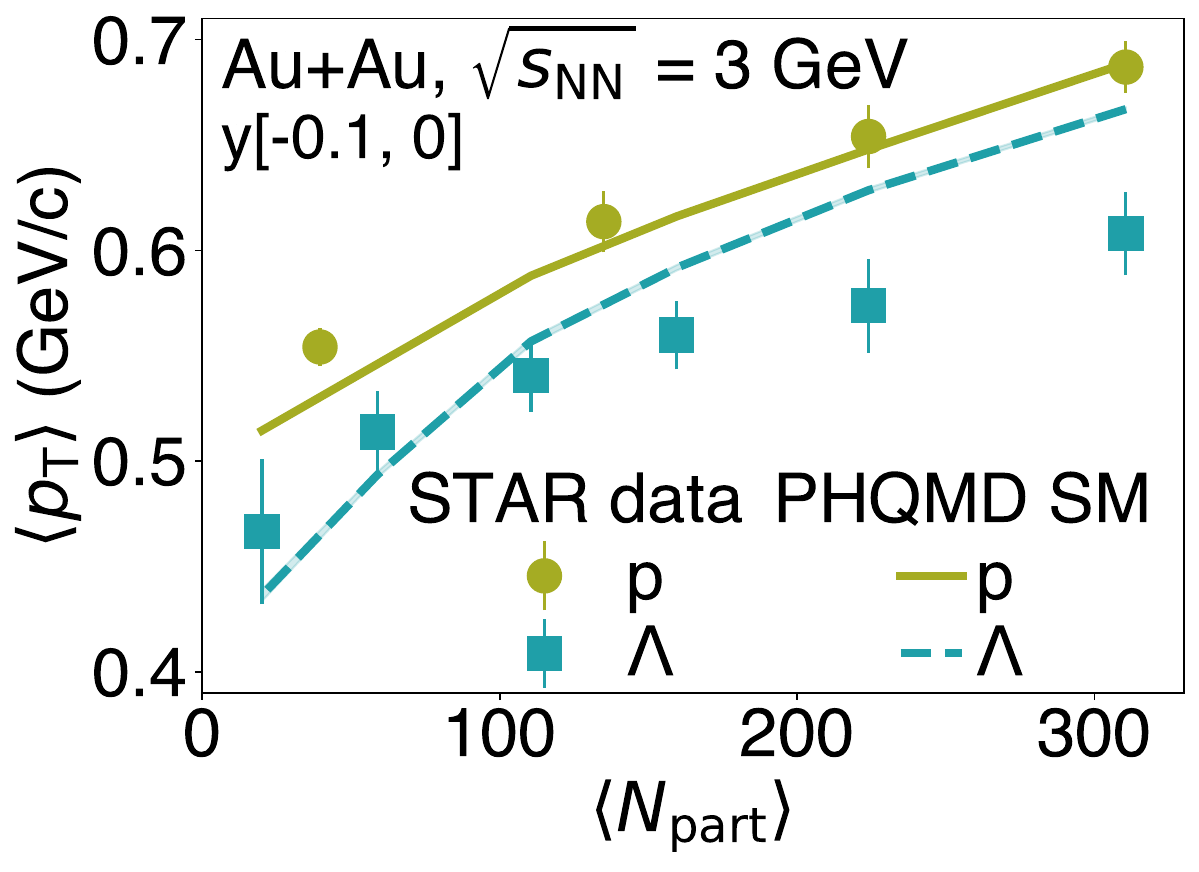}
    \caption{The mean transverse momentum $\langle p_{T} \rangle$ of protons and $\Lambda$s as a function of rapidity (left panels), and at mid-rapidity as a function of the number of participating nucleons (right panels) in $\sqrt{s_{\rm{NN}}}=3$ Au$+$Au GeV collisions~\cite{STAR:2023uxk, STAR:2024znc}. The data are shown as solid markers, while PHQMD calculations are shown as lines.
    \label{fig:meanpt_lambdaproton}}
\end{figure}

Fig.~\ref{fig:meanpt_lambdaproton}  shows $\langle p_T \rangle$ of $p$ and $\Lambda$ as a function of rapidity on the left and for the rapidity interval $-0.1<y<0$ as a function of the number of participants on the right hand side. The number of participants for each centrality is determined by the same Glauber calculations, which we used for the  centrality determination, described in Sec.~\ref{sec:compare}. Calculations with a S EoS underestimate the data for both, $p$ and $\Lambda$. As a function of rapidity, calculations using a S EoS reproduce the same functional form as the data for both $p$ and $\Lambda$. In centrality dependence, however, the discrepancy between data and calculations increases with centrality for $p$, whereas for $\Lambda$, the calculations follow the experimental trend which flattens with increasing centrality.

Calculations with a H and a SM EoS come much closer to the data than the calculation with a S EoS and describe the data reasonably well. Calculations with a SM EoS describe the data best but fail to reproduce
the flattening of the rapidity distribution towards midrapidity, which is seen in the data.

It is worth noting, see bottom figures,  that the $\langle p_T \rangle$ of $p$ is larger than that of $\Lambda$ at mid-rapidity in all centrality regions, despite the larger mass of the $\Lambda$. Similar observations have been reported previously at similar collision energies~\cite{FOPI:2007usx, Albergo:2002tn}, and are often attributed to the fact that $\Lambda$ are produced particles while a fraction of protons may be spectator protons at such low energies. 

\begin{figure}[!h]             
    \centering    
    \includegraphics[width=0.8\linewidth]{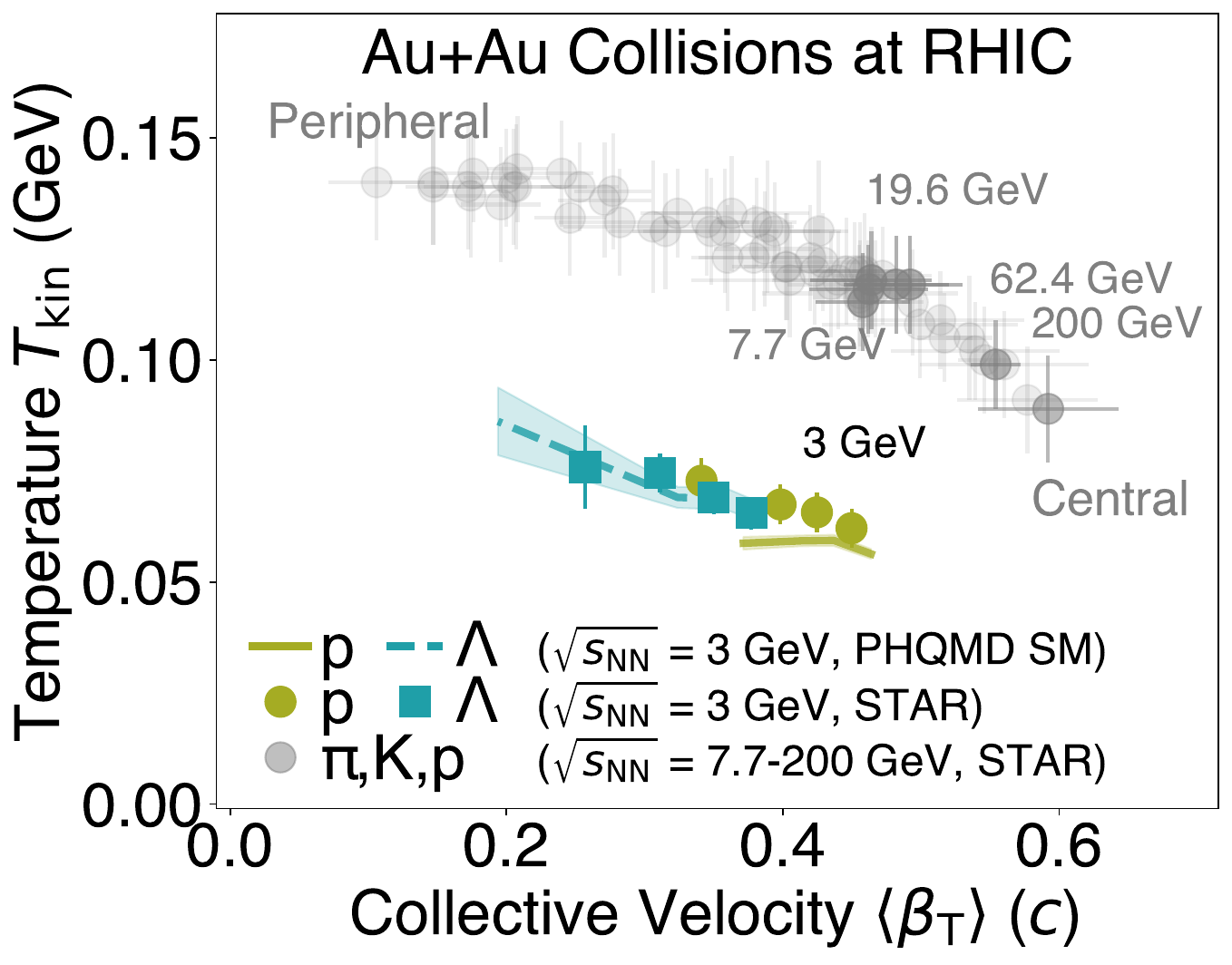}    
    \caption{Kinetic freeze-out temperature $T_{kin}$ as a function of collective radial velocity $\langle \beta_{T} \rangle$ from blast-wave fits to mid-rapidity proton, $\Lambda$ spectra at $\sqrt{s_{\rm{NN}}}=3$ GeV (colored markers), and $\pi^{\pm}, K^{\pm}, p(\bar{p})$ spectra from $\sqrt{s_{\rm{NN}}}=7.7$ to 200 GeV (dark and light gray dots)~\cite{STAR:2017sal}. The different points correspond to different centralities, peripheral to central from left to right. The dark gray dots correspond to the most central bin $0-5\%$ for $\sqrt{s_{\rm{NN}}}=7.7$ to 200 GeV, while the light gray dots correspond to other centralites. Colored bands represent the blast-wave fit results over the full $p_T$ range from PHQMD SM simulations at $\sqrt{s_{\rm{NN}}}=3$ GeV.}
   \label{fig:tkin}
\end{figure}

\begin{figure}[!h]             
    \centering    
    \includegraphics[width=0.8\linewidth]{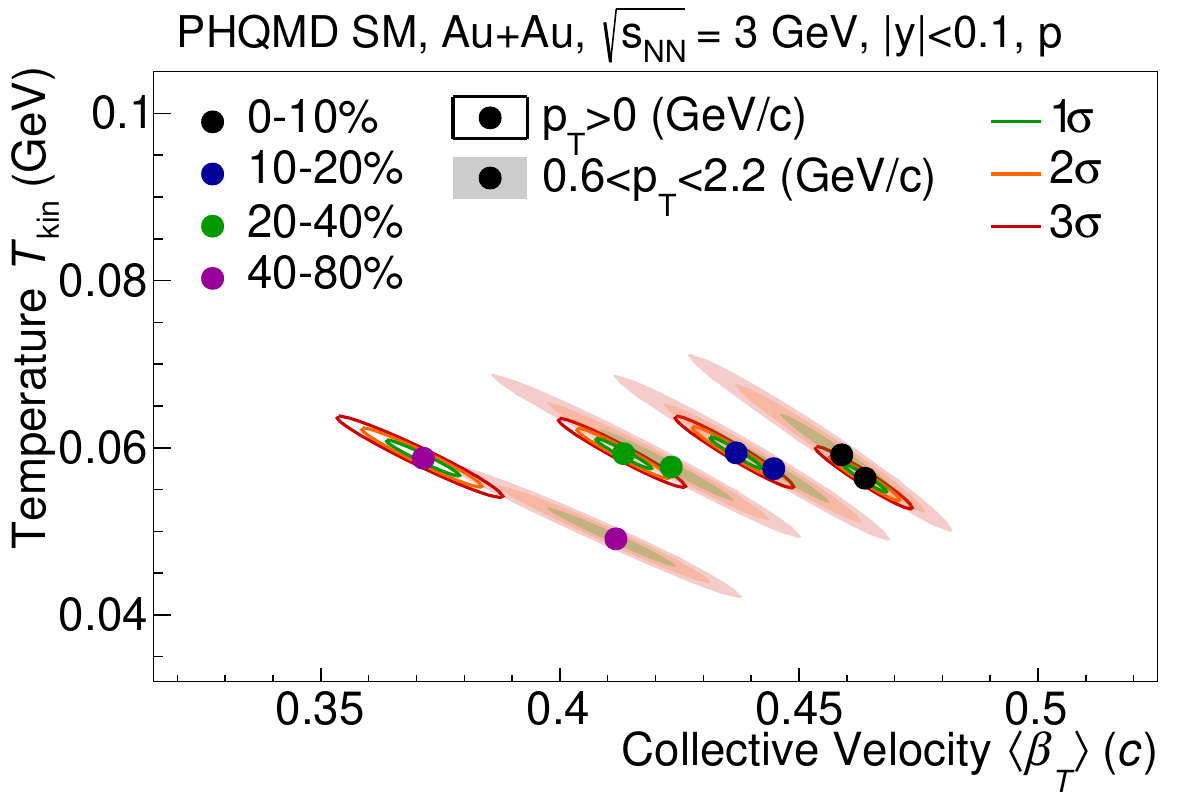}    
    \includegraphics[width=0.8\linewidth]{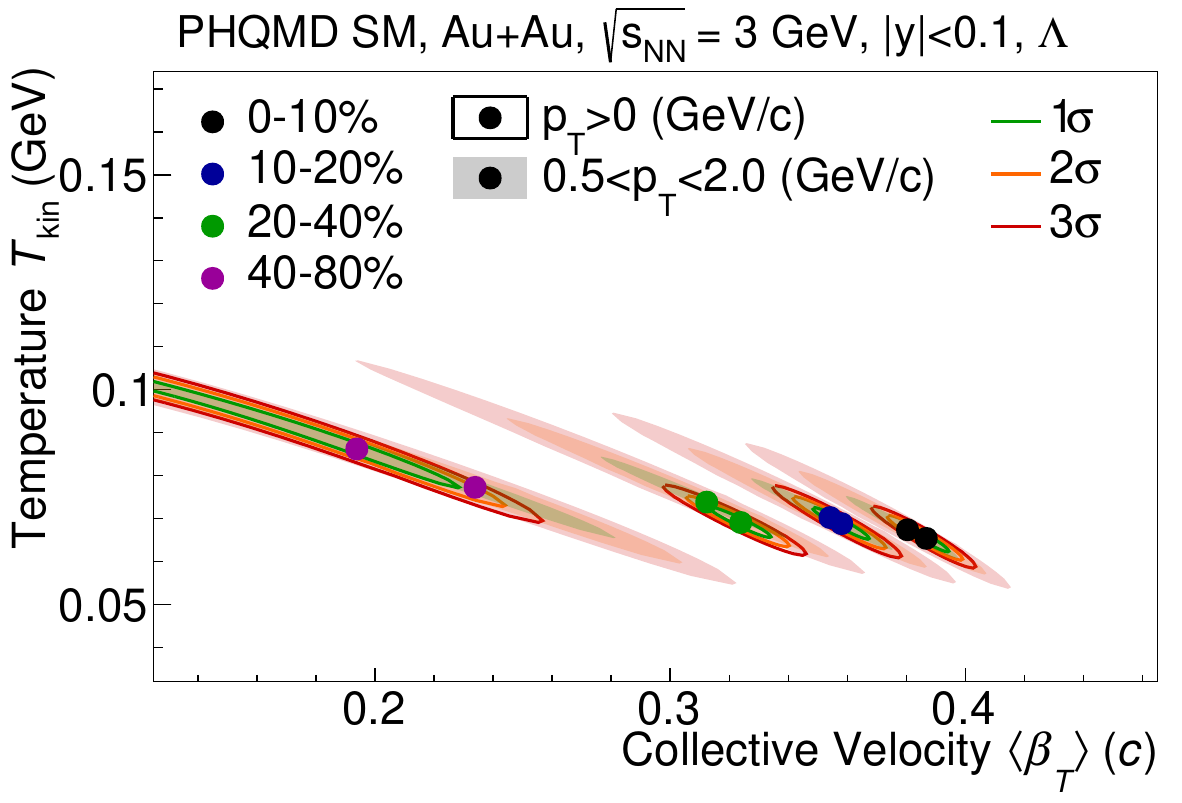}    
    \caption{Blast-wave fit results for the kinetic freeze-out temperature $T_{kin}$ and average transverse collective velocity $\langle \beta_{T} \rangle$ from PHQMD SM simulations of Au+Au collisions at $\sqrt{s_{\rm{NN}}}=3$ GeV at mid-rapidity, $|y| < 0.1$, shown separately for protons (top) and $\Lambda$ hyperons (bottom). Different centrality classes are indicated by different colors. Ellipses indicate $1\sigma$, $2\sigma$, and 3$\sigma$ confidence contours from the fits. 
    Two $p_{T}$ fit ranges are compared: $p_T>0$ $GeV/c$ (contours indicated using solid lines) and data measured $p_{T}$ ranges (contours indicated using shaded regions). }
   \label{fig:tkin_phqmd}
\end{figure}

It is worthwhile to investigate this further.
To separate the influence from the temperature and the collective expansion on the transverse momentum spectra, we perform blast-wave fits to the proton and $\Lambda$ spectra in the rapidity region $y=(-0.1,0)$ in four different centralities, $0-10\%, 10-20\%, 20-40\%$, and $40-80\%$ to extract the temperature at kinetic freeze out $T_{kin}$ and the average collective velocity $\langle \beta_T \rangle$. The radial flow profile parameter $n$ is fixed to 1 in these studies. The results are shown in Fig.~\ref{fig:tkin}. The temperature of protons and $\Lambda$s are identical within uncertainties but the collective radial flow of protons is systematically larger than that of $\Lambda$s. $\Lambda$s are produced particles and the production probability is higher close to the center of the fireball where the nucleon density is high and consequently the mean free path small. In the center of the fireball the collective velocity is small (see eq. \ref{eq:blast}) which results in a lower collective velocity of the $\Lambda$s compared to protons. The blast-wave fit results over the full $p_T$ range from PHQMD SM simulations are shown in the same figure as colored bands. When performing the blast-wave fits to the PHQMD $p$ and $\Lambda$ $p_T$ spectra, an additional 5\% uncertainty (fudge factor) was added to the statistical errors. This adjustment was applied to achieve a reasonable $\chi^2/\mathrm{ndf}$ and to match the level of uncertainty of the experimental data~\cite{STAR:2023uxk, STAR:2024znc}. PHQMD describes the trends observed in the data well, and the deviation seen in peripheral collisions for protons has been discussed in the previous section.

Due to detector acceptance limitations, experimental data cannot cover the entire low- and high-$p_T$ regions. Therefore, we also performed blast-wave fits to the PHQMD spectra within the measured $p_T$ ranges and compared the results to those obtained over the full $p_T$ range, as shown in Fig.~\ref{fig:tkin_phqmd}. For protons, in central and mid-central collisions, the extracted temperature and collective velocity are consistent within uncertainties between the two fit ranges. This suggests that the freeze-out conditions are similar across the entire $p_T$ spectrum. In contrast, in peripheral collisions, the fit within the measured $p_T$ range yields a lower temperature and higher collective velocity compared to the full $p_T$ fit. This difference may be attributed to biases arising from limited low-$p_T$ coverage and could suggest distinct freeze-out dynamics between low- and high-$p_T$ regions in peripheral collisions. For $\Lambda$ hyperons, the results are consistent across all centrality classes and between the two fit ranges, suggesting uniform freeze-out conditions within the $p_T$ coverage for these particles.

We also compare the freezeout parameters to those obtained by fits to $\pi^\pm, K^\pm,$ and $p(\bar{p})$ at $\sqrt{s_{\rm{NN}}}=7.7-200$~\cite{STAR:2017sal}. Focusing on the most central collisions, from $\sqrt{s_{\rm{NN}}}$=200 GeV to 7.7 GeV, the temperature increases, while the collective velocity decreases. However, the freezeout parameters at $\sqrt{s_{\rm{NN}}}=3$ GeV do not follow the same trend. This implies a change in medium properties or expansion dynamics from $\sqrt{s_{\rm{NN}}}=3$ to 7.7 GeV. Future data from STAR beam energy scan-II can help elucidate the nature of the evolution of the freeze out properties.

\begin{figure}[h!]             
    \centering    
    \includegraphics[width=0.75\linewidth]{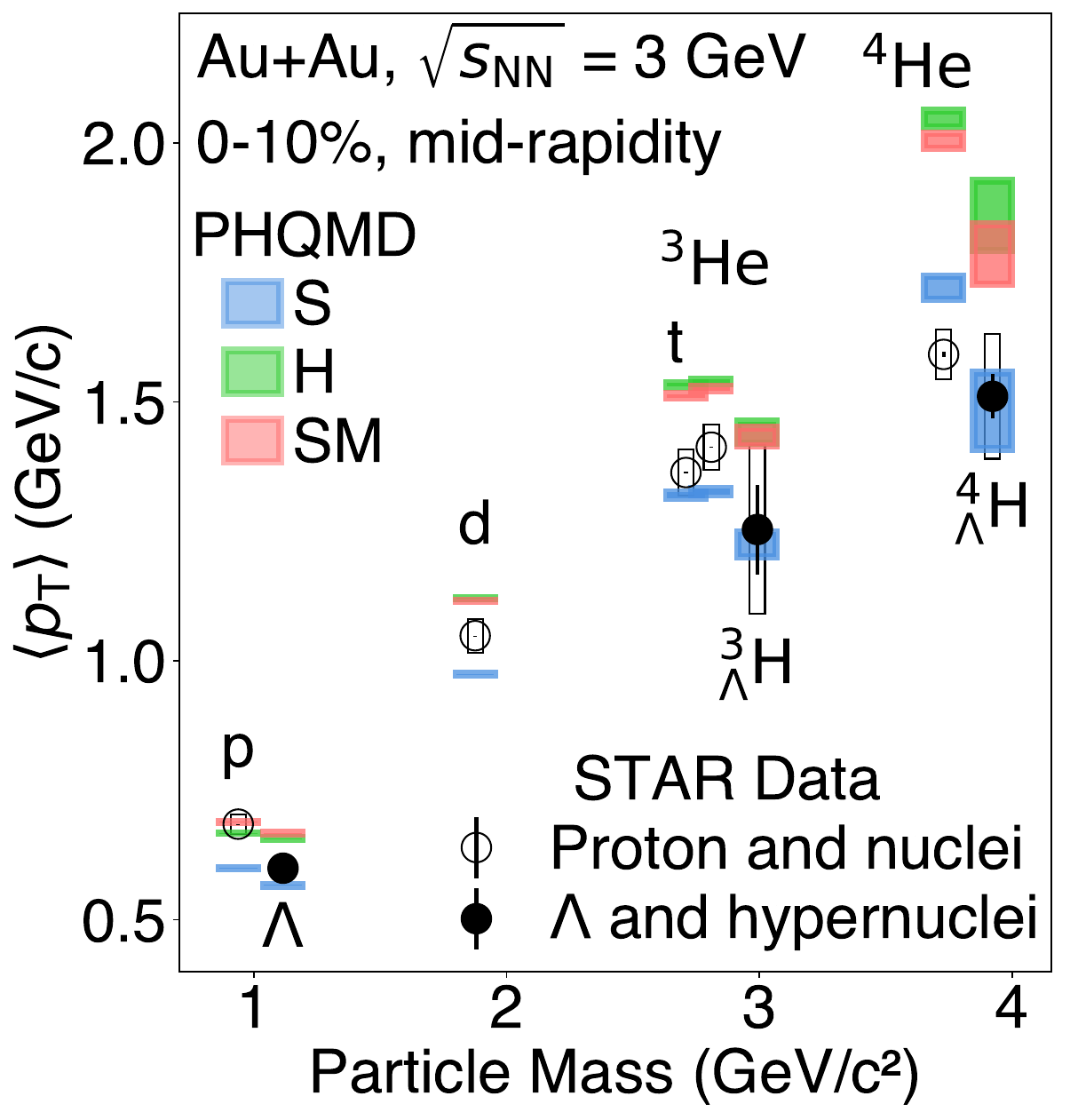}
    \caption{The mean transverse momentum spectra of proton, $\Lambda$, light nuclei and hypernuclei at mid-rapiditiy as a function of mass in $0-10\%$ Au+Au collisions at $\sqrt{s_{\rm{NN}}}=3$ GeV. The measured data points from STAR~\cite{STAR:2023uxk, STAR:2021orx, STAR:2024znc} are shown as black and open markers, and the calculations from PHQMD using soft, hard, and soft with momentum dependence EoS are shown as colored bands. 
    \label{fig:meanpTnuclei}}
\end{figure}
Fig.~\ref{fig:meanpTnuclei} shows $\langle p_T \rangle$ of $p$, $d$, ${}^{3}\rm{He}$, ${}^{4}\rm{He}$, $\Lambda$, ${}^{3}_{\Lambda}\rm{H}$, ${}^{4}_{\Lambda}\rm{H}$ as a function of the particle mass for 0-10\% central collisions at midrapidity. One can see that $\langle p_T \rangle$ of protons and light nuclei increases approximately linearly with mass. Although calculations with H and SM EoS describe the $\langle p_T \rangle$ of protons, they overestimate the $\langle p_T \rangle$ of all the light nuclei. As Fig. 6 shows, this is due to the underprediction of the yield of low $p_T$ clusters and the disagreement of the extrapolation of the experimental data to $p_T\to0$ with respect to the PHQMD calculations. The level of discrepancy increases with the mass; the H and SM EoS overestimates the $\langle p_T \rangle$ of ${}^{4}\rm{He}$ by 500 MeV. On the contrary, calculations with S EoS give a rather fair description of the light nuclei $\langle p_T \rangle$. 

For $\Lambda$ and hypernuclei, as mentioned in the last section, the $\Lambda$ $\langle p_T \rangle$ is lower as compared to that of protons, despite of having a larger mass. We observe that hypernuclei $\langle p_T \rangle$ also tend to be lower than that of the nuclei with the same mass number, although the large uncertainties prevent us from drawing a strong conclusion. Nonetheless, we note that the PHQMD calculations indicate that the  hypernuclei $\langle p_T \rangle$ is smaller than that of non-strange nuclei $\langle p_T \rangle$  when they have the same mass number regardless of the EoS which is used. This is presumably a consequence of the lower collective velocities of $\Lambda$ as compared to nucleons, which has been discussed above. Similar to what is observed for nuclei, calculations with H and SM tend to overestimate the hypernuclei data, while calculations with S EoS provide a good description. 

\subsection{Directed Flow}
\begin{figure}[h!]             
    \centering 
    \includegraphics[width=0.49\linewidth]{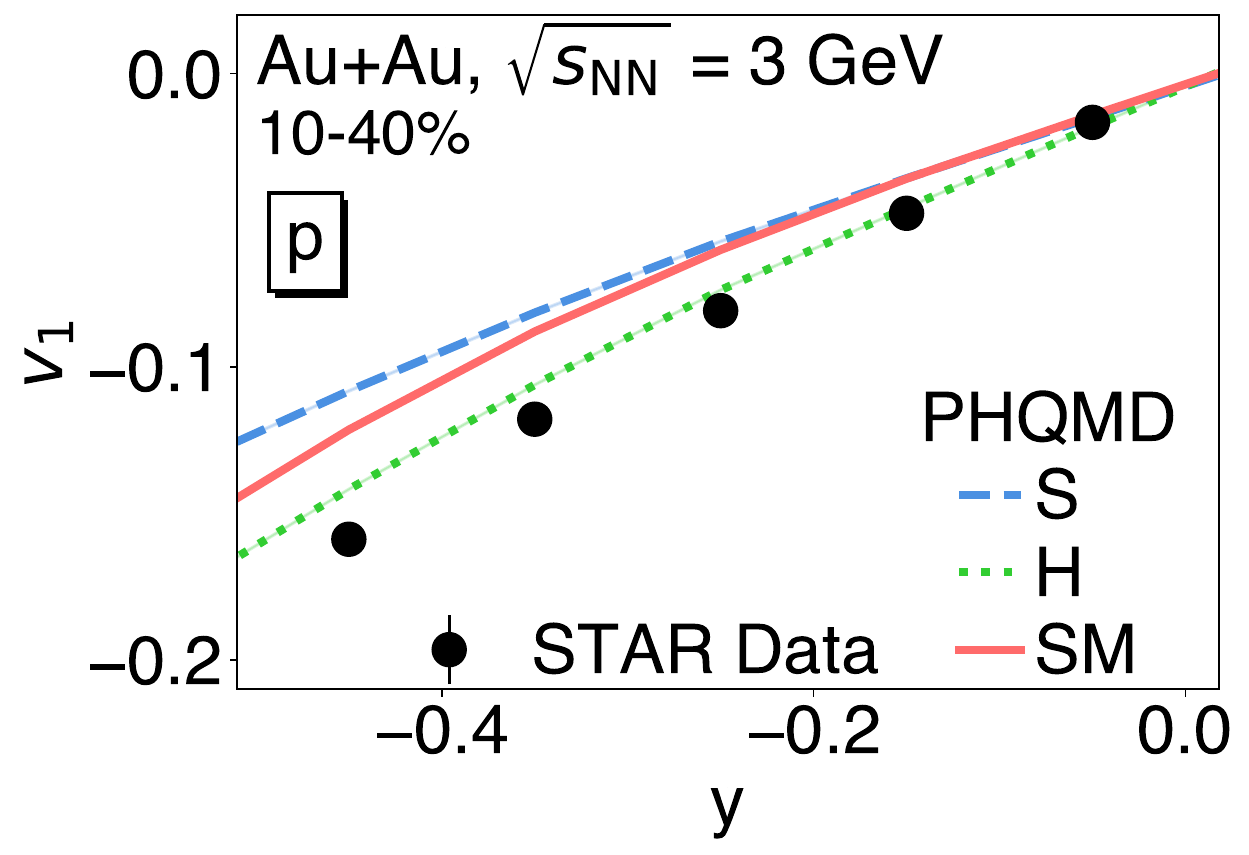}
    \includegraphics[width=0.49\linewidth]{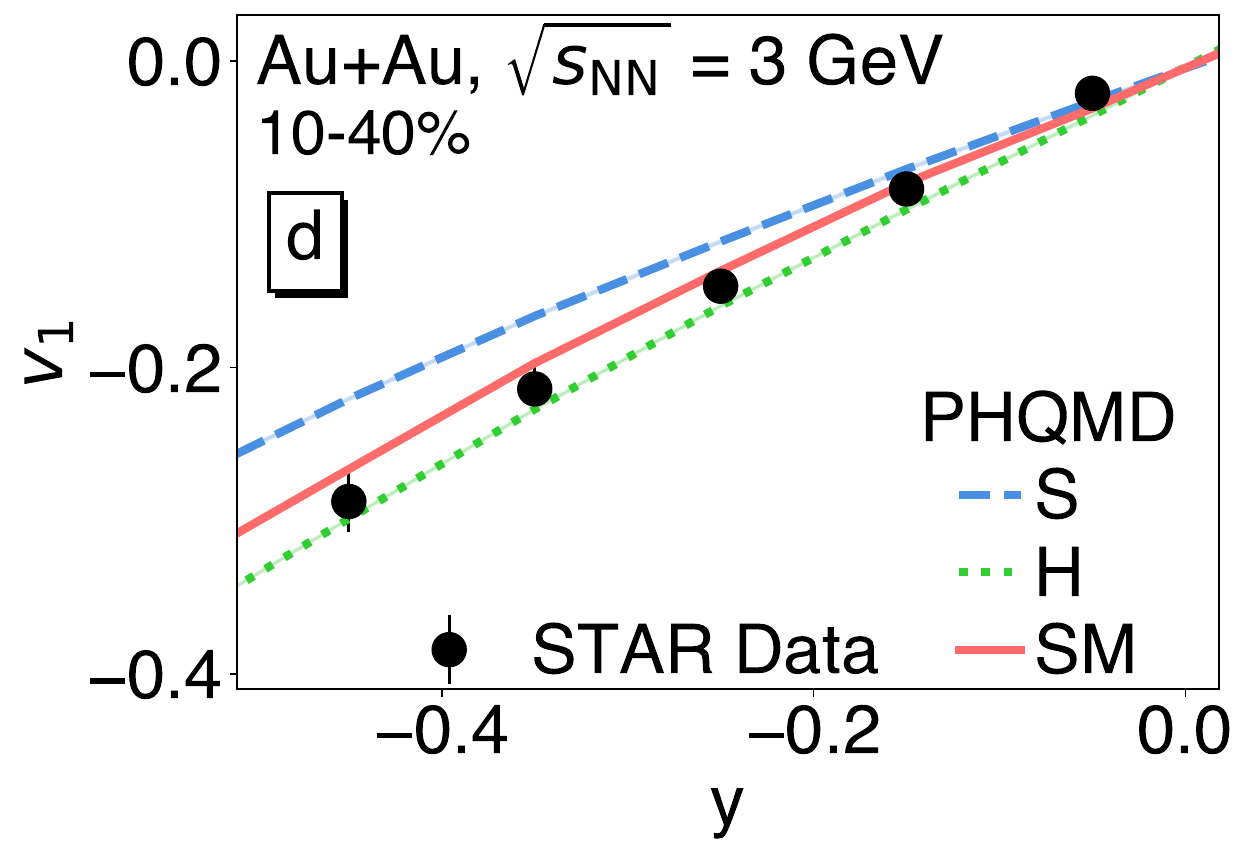}
    \includegraphics[width=0.49\linewidth]{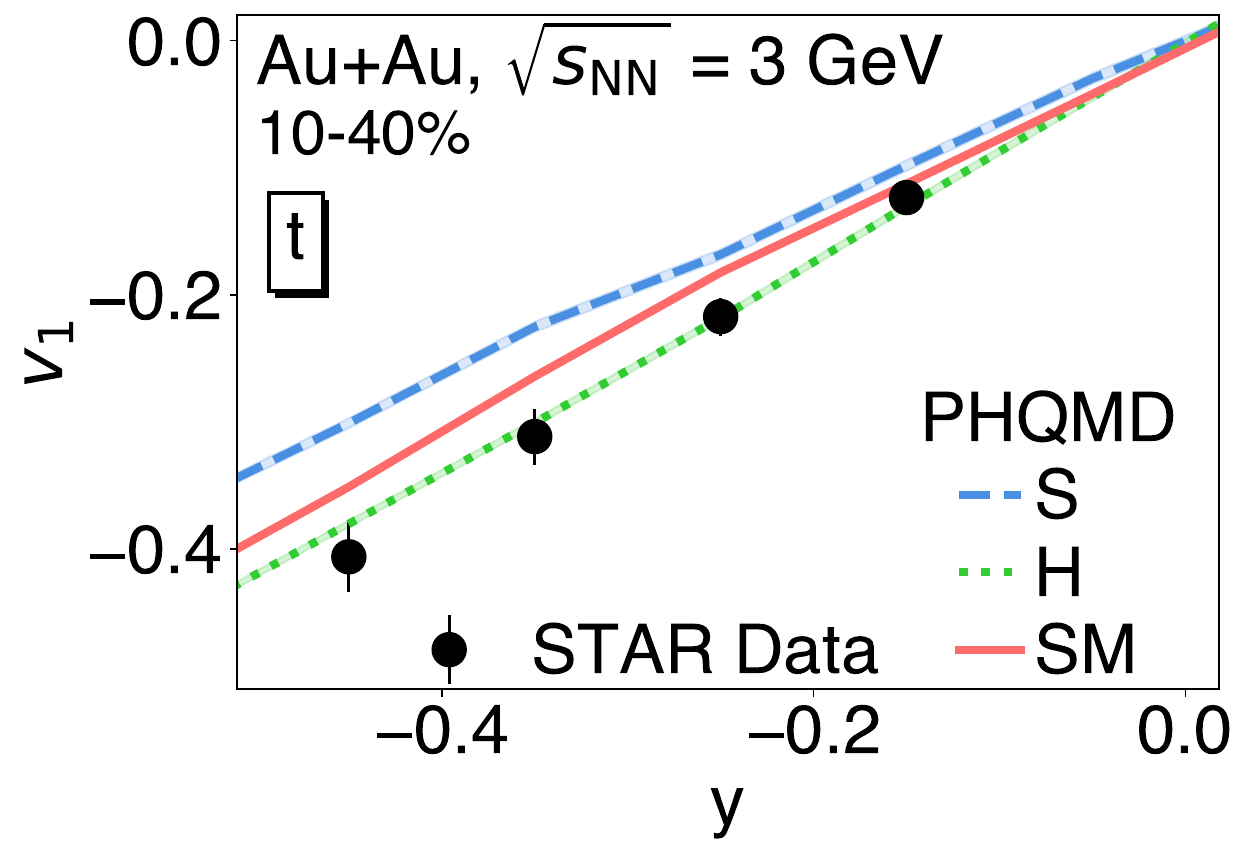}
    \includegraphics[width=0.49\linewidth]{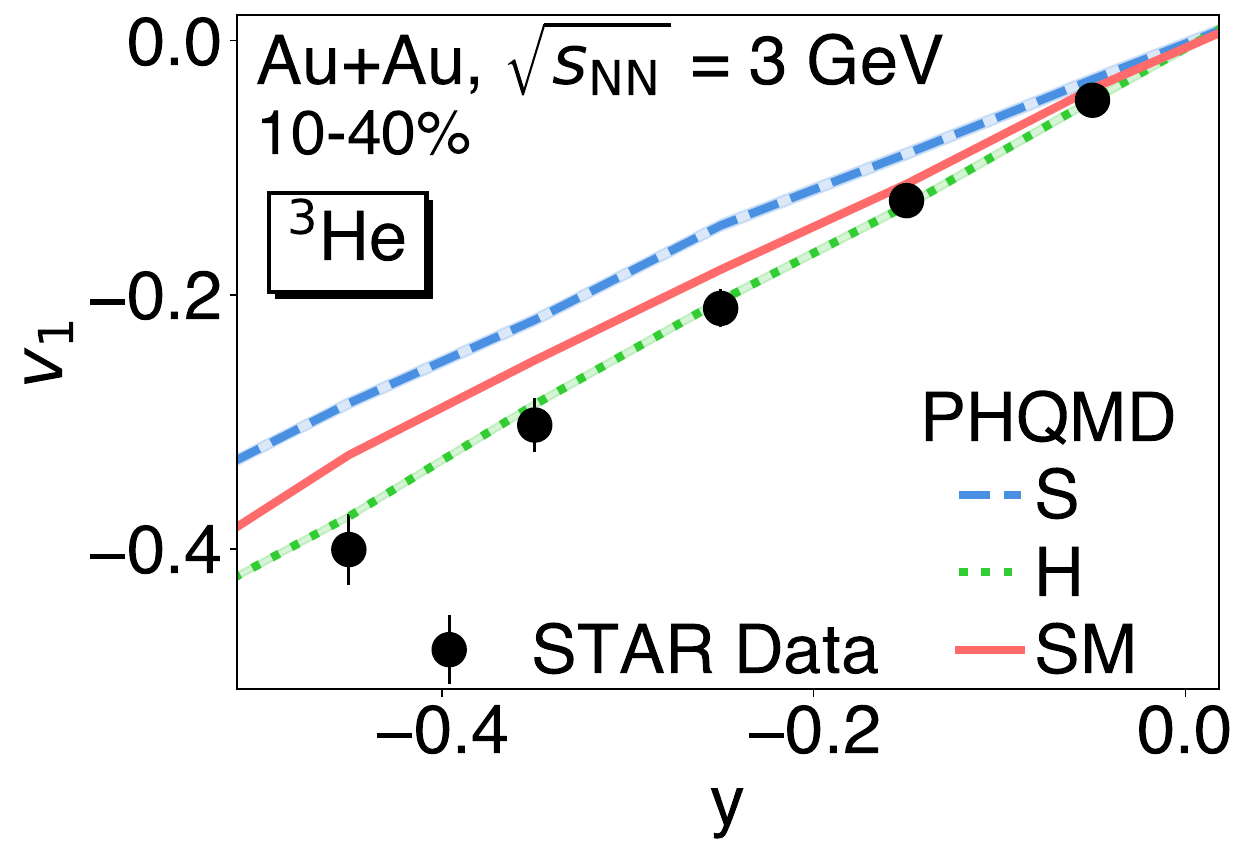}
    \includegraphics[width=0.49\linewidth]{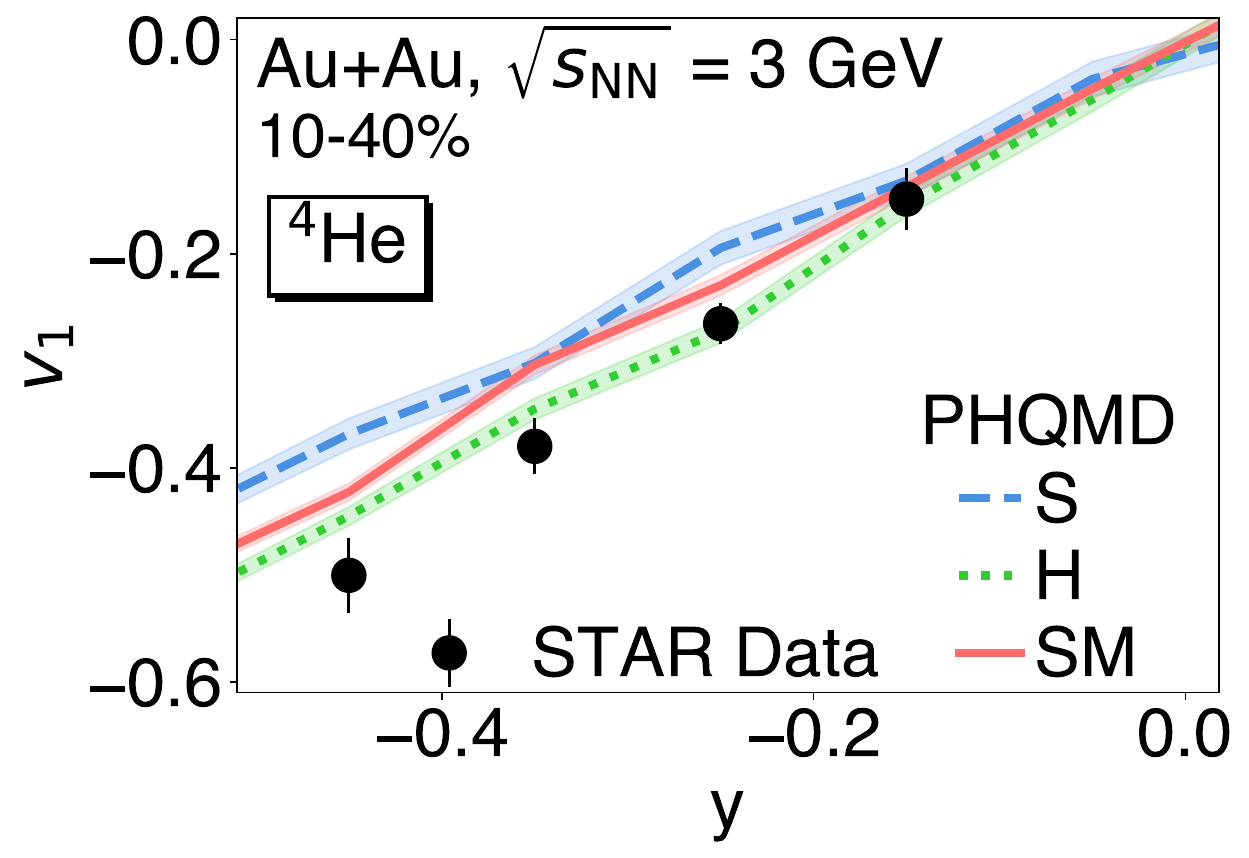}
    \includegraphics[width=0.49\linewidth]{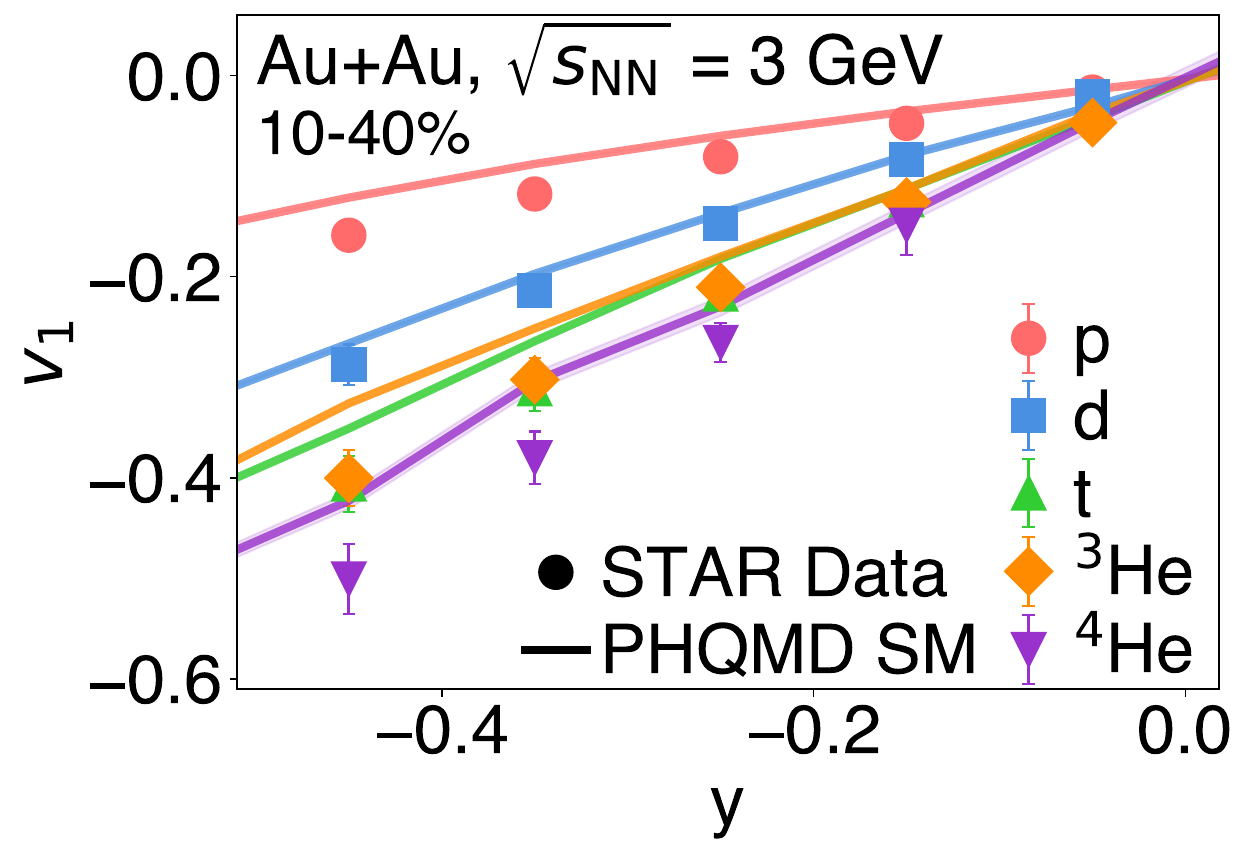}
    \caption{The directed flow $v_1$ of proton, deuteron, triton, $^{3}\rm{He}$, and $^{4}\rm{He}$ as a function of rapidity in $10-40\%$ mid-central Au$+$Au collisions at $\sqrt{s_{\rm{NN}}}=3$ GeV. The STAR data~\cite{STAR:2021ozh} are shown as colored markers while PHQMD calculations are shown as colored lines or bands. 
    \label{fig:v1y_pdth34}}
\end{figure}

Figure~\ref{fig:v1y_pdth34} shows the PHQMD results for the directed flow $v_1$ of protons, deuterons, tritons , ${}^{3}\rm{He}$, and ${}^{4}\rm{He}$ as a function of rapidity integrated in the chosen $p_{T}$ ranges for $\sqrt{s_{\rm{NN}}}=3$ GeV Au+Au collisions in the centrality range $10-40\%$. The calculations are compared to the STAR data. One can see that for all clusters as well as for protons, the $v_1$ is negative. The soft EoS underestimates the data. The difference between the hard and the soft momentum-dependent EoS are smaller; however it is worth to mention  that the SM EoS leads to slightly smaller $v_1$ compared to H, and gives a worse description of the data. The bottom right panel of Fig~\ref{fig:v1y_pdth34} displays $v_1(y)$ for a SM EoS for clusters of different sizes in comparison with the experimental results. There is a clear mass ordering  of $v_1(y)$: The heavier the mass of a nucleus, the stronger is the rapidity dependence of $v_1$, what is indeed seen in the experimental data. It is worth to mention that at the somewhat lower energy of $\sqrt{s_{\rm{NN}}}= 2.4$ GeV we observe the opposite trend: there, the PHQMD results for SM are closest to the data but also slightly underpredict the experimental $v_1$ values \cite{Kireyeu:2024hjo}.

\begin{figure}[h!]             
    \centering    
    \includegraphics[width=0.49\linewidth]{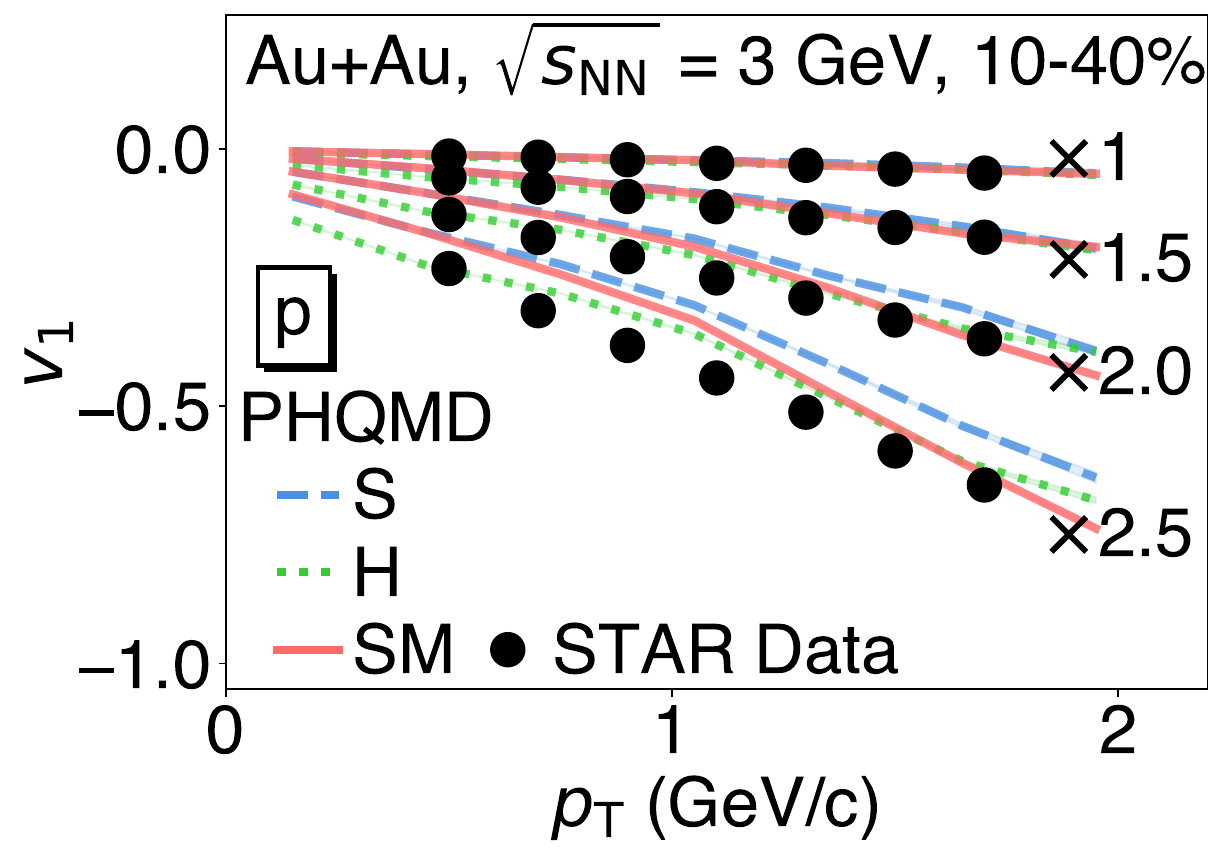}
    \includegraphics[width=0.49\linewidth]{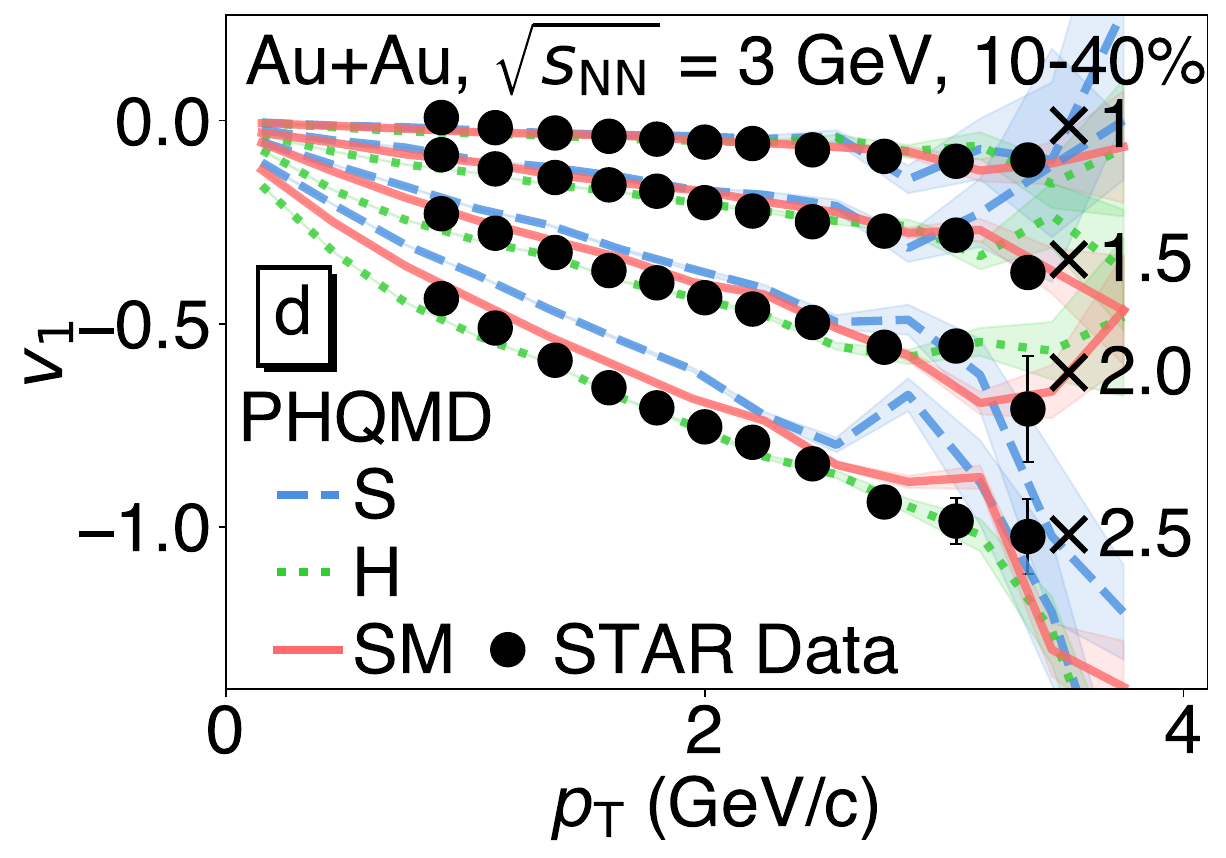}
    \includegraphics[width=0.49\linewidth]{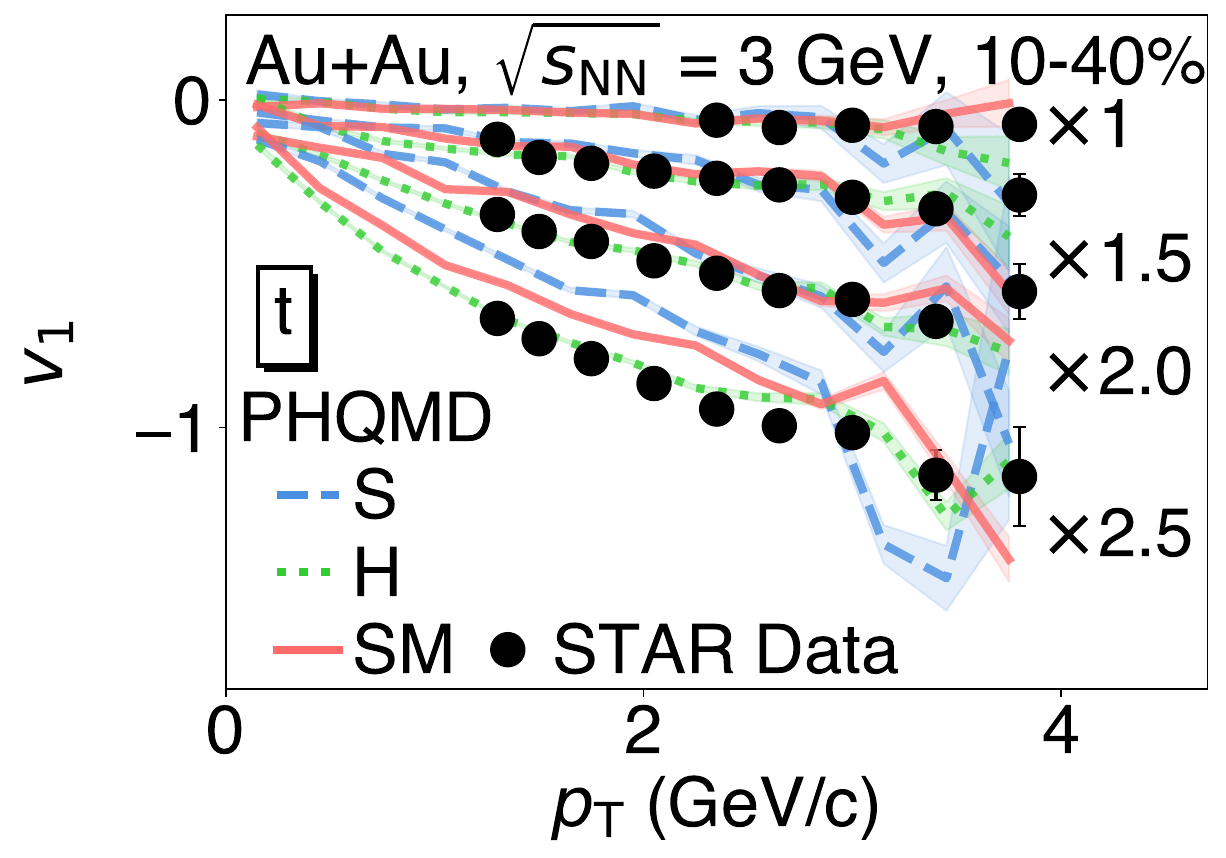}
    \includegraphics[width=0.49\linewidth]{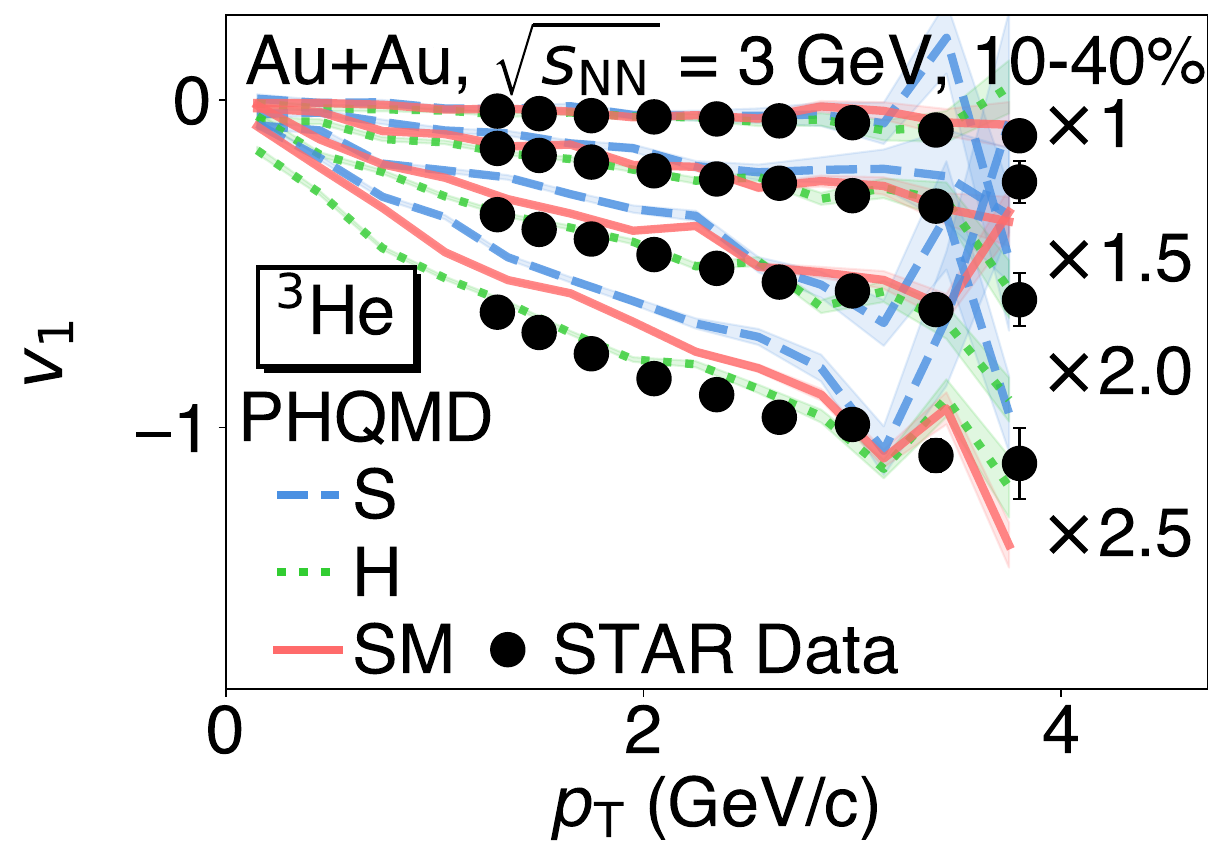}
    \includegraphics[width=0.49\linewidth]{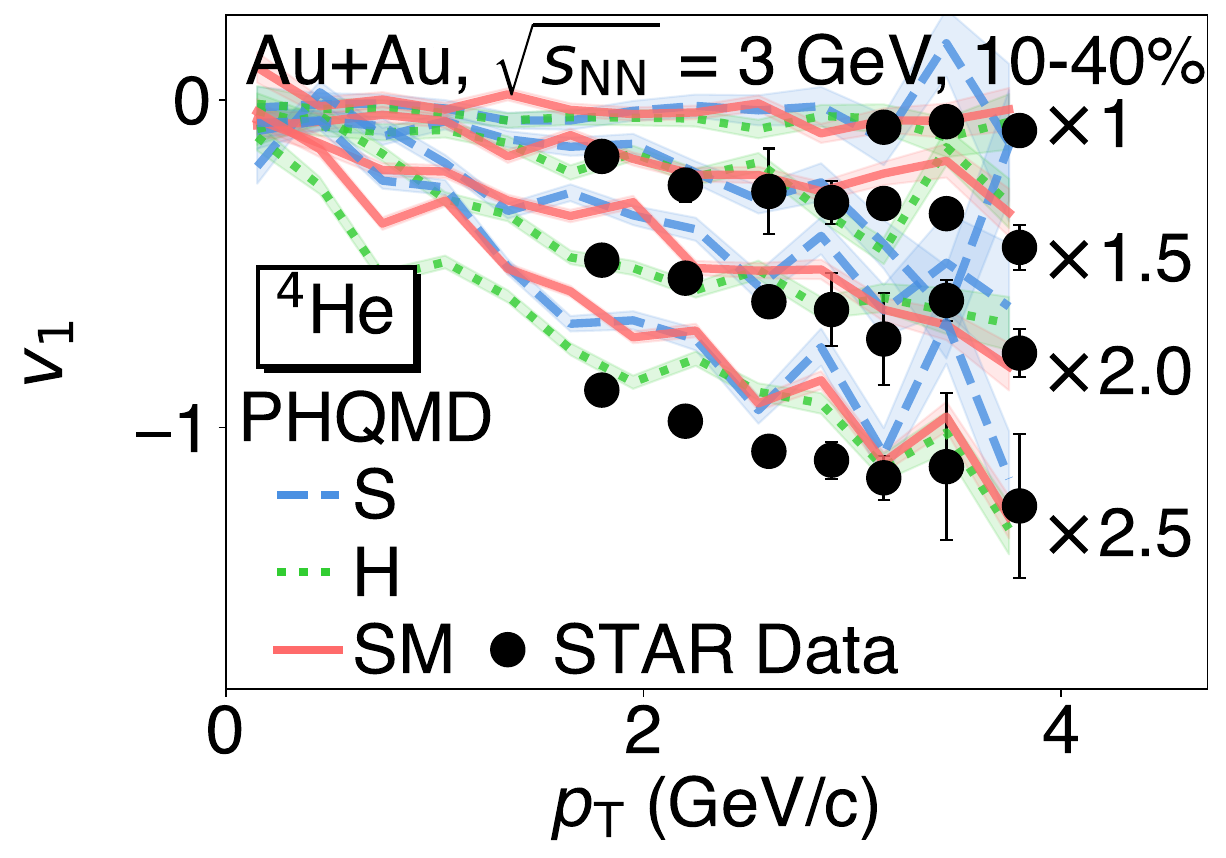}
    \includegraphics[width=0.49\linewidth]{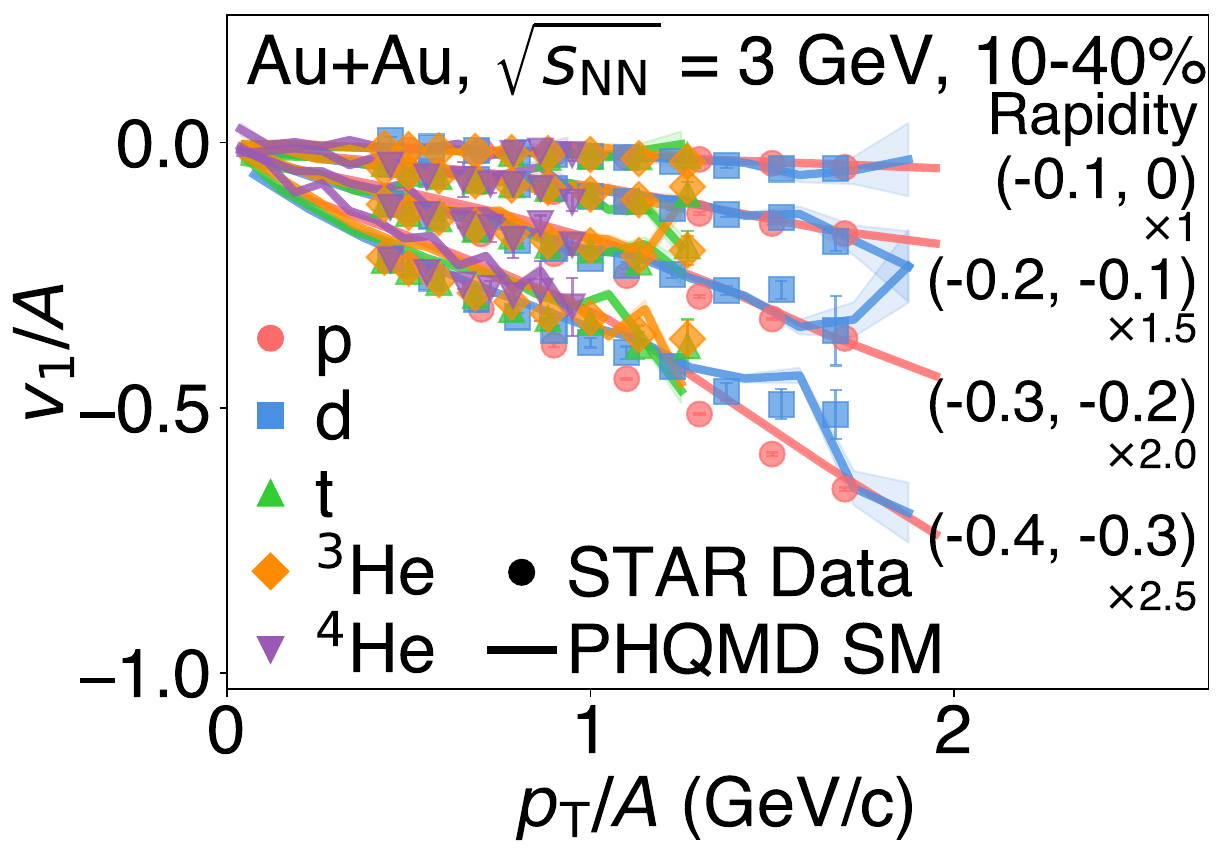}

    \caption{The directed flow $v_1$ of proton, deuteron, triton, $^{3}\rm{He}$, and $^{4}\rm{He}$ as a function of $p_T$ in different rapidity intervals in $10-40\%$ mid-central Au$+$Au collisions at $\sqrt{s_{\rm{NN}}}=3$ GeV. The STAR data~\cite{STAR:2021yiu} are shown as black markers while PHQMD calculations are shown as colored bands. 
    \label{fig:v1pt_pdth34}}
\end{figure}

In Figure~\ref{fig:v1pt_pdth34} we present the PHQMD results for protons, deuterons, tritons, ${}^{3}\rm{He}$, and ${}^{4}\rm{He}$ as a function of $p_T$ in different rapidity intervals for $\sqrt{s_{\rm{NN}}}=3$ GeV in $10-40\%$ central Au+Au collisions, in comparison to the STAR data. One can see that $v_1(p_T)$ depends strongly on the EoS for protons and clusters alike. However, the behaviour is not trivial - for protons, at low $p_T$, $v_1$ calculated with a hard EoS, is significantly more negative than that calculated with a SM EoS, while at larger $p_T$, the SM EoS gives comparable and even larger $v_1$ values than the H EoS. This tendency is also seen for clusters. Similar to what is observed for the rapidity dependence, the SM EoS leads to slightly smaller $v_1$ as compared to H, and gives a less good description of the data, in distinction to the PHQMD results at $\sqrt{s_{\rm{NN}}}= 2.4$ GeV~\cite{Kireyeu:2024hjo}.

Figure~\ref{fig:v1pt_pdth34}, \Yj{bottom right panel}, shows that the values of $v_1/A$ of all light nuclei, including protons, approximately follows an A scaling for $-0.3<y<0$. This is is also seen in PHQMD calculations. For all cluster sizes the theoretical $v_1(y)$ and $v_1(p_T)$ calculations with a H EoS provide the best description of the data. SM slightly underpredicts the data, while a S EOS substantially deviates from the experimental data.

\begin{figure}[!h]             
    \centering    
    \includegraphics[width=0.49\linewidth]{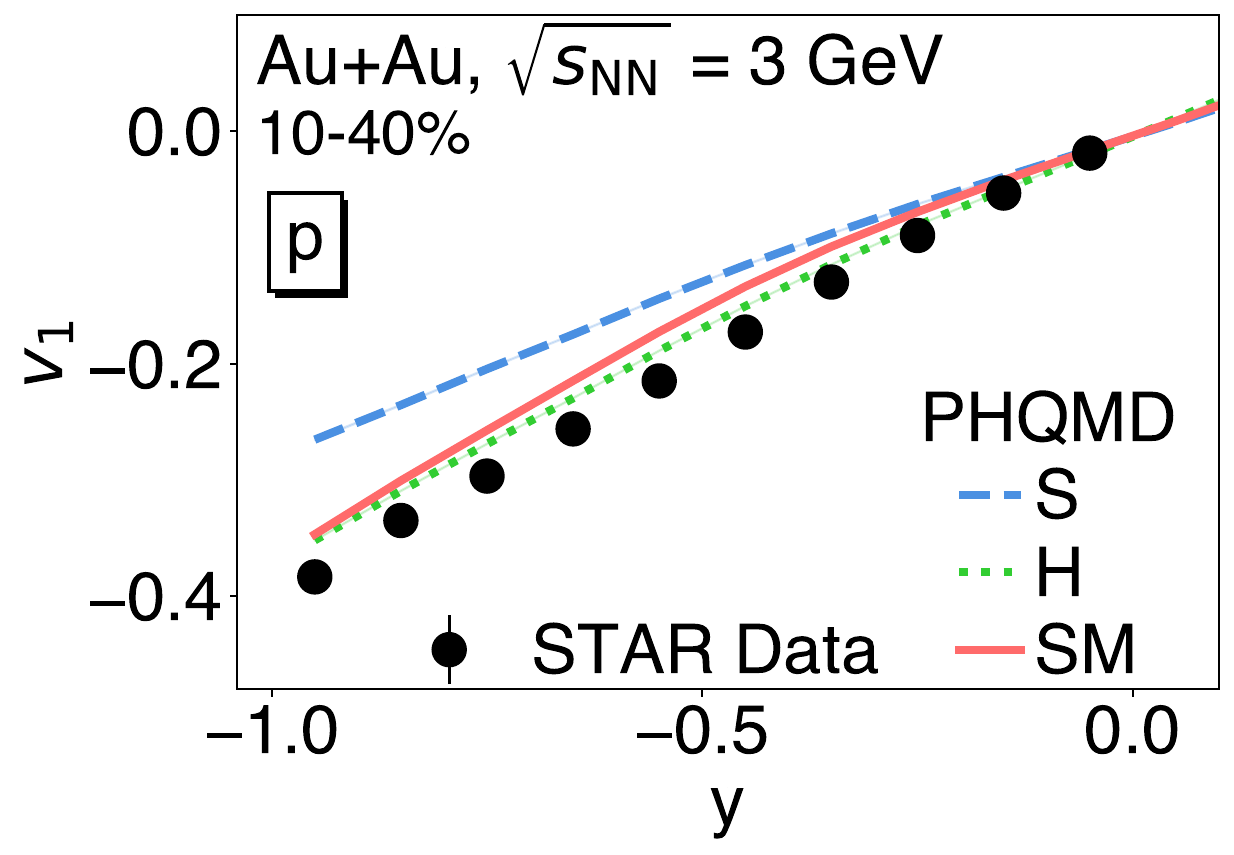}
    \includegraphics[width=0.49\linewidth]{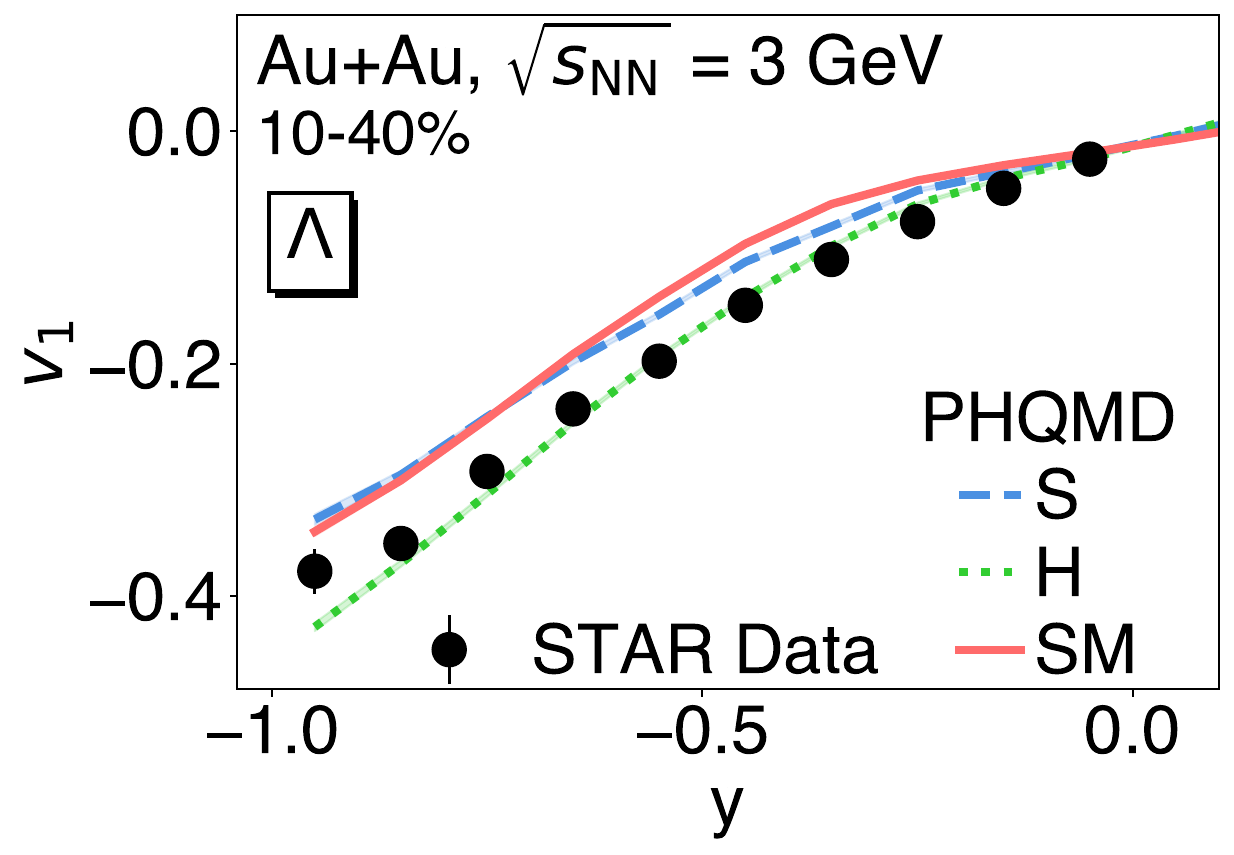}
    \caption{The directed flow $v_1$ of proton and $\Lambda$ as a function of rapidity in $10-40\%$ mid-central $\sqrt{s_{\rm{NN}}}=3$ GeV Au$+$Au collisions. \Yj{The STAR data~\cite{STAR:2021yiu} are shown as black markers while PHQMD calculations are shown as colored bands.}
    \label{fig:v1ylambdaproton}}
\end{figure}

In Fig.~\ref{fig:v1ylambdaproton} we present the PHQMD results for $v_1(y)$ of protons (left), $\Lambda$ (right) in comparison to the STAR data~\cite{STAR:2021ozh} for $\sqrt{s_{\rm{NN}}}=3$ GeV $10-40\%$ central Au+Au collisions and for the $p_T$ interval $0.4<p_T<2.0$ GeV$/c$ for a hard, soft and momentum-dependent EoS. $v_1(y)$ of protons and $\Lambda$s is negative. It is remarkable that the data for both $\Lambda$ and proton are similar despite of the quite different radial expansion velocities of the blast wave fit.

We also note that although $v_1$ of protons in H and SM calculations are similar, it differs quite significantly for $\Lambda$. In PHQMD the results differ because the protons close to the transverse surface of the fireball obtain more directed flow than those close to the center of the fireball  \cite{David:1998qu}. Since $\Lambda$s have a higher chance to be produced in the center of the fireball than close to the surface, their flow (as well as that of $K^+$, the produced associated meson), is smaller. This may indicate that the parametrization of the SM EoS may be too simplistic in being not stiff enough at high densities or that the $ \Lambda N$ scattering is underestimated, although in PHQMD the experimentally measured (free) $\Lambda N$ cross section is implemented.  A detailed investigation of the directed flow of all strange hadrons may elucidate this interesting issue. We observe that calculations with a S EoS, similar to the comparisons of the $\langle p_T \rangle$, do not provide a good description of the data. They underestimate the $v_1$ for both $\Lambda$ and proton in all measured rapidity ranges. However, although calculations with SM, which provides good description of $\langle p_T \rangle$, describe the $v_1$ of $p$ fairly well, they overestimate the $v_1$ of $\Lambda$. The flow  is also influenced by the choice of the $\Lambda$ potential~\cite{Nara:2022kbb}, which we assume here to be identical to the potential of  nucleons.

\subsection{Elliptic Flow}

\begin{figure}[h!]             
    \centering    
    \includegraphics[width=0.49\linewidth]{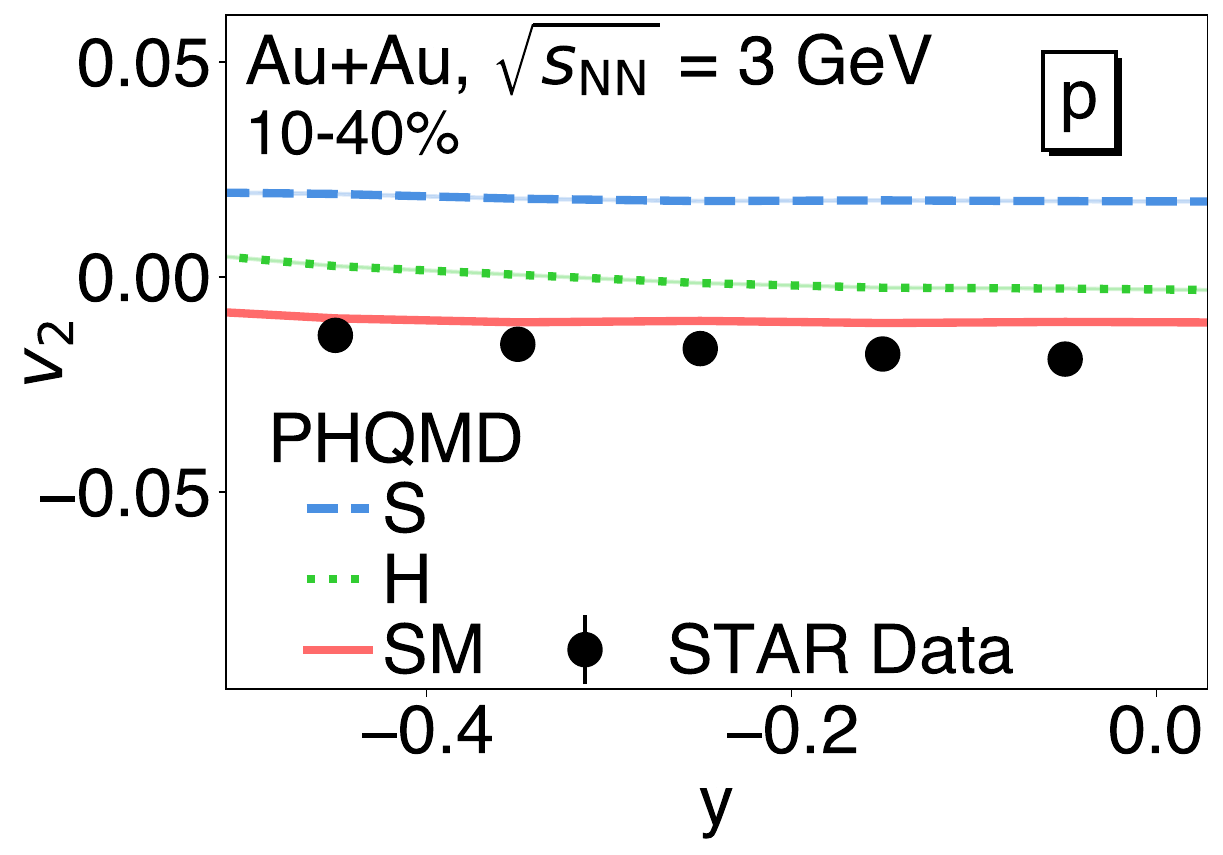}
    \includegraphics[width=0.49\linewidth]{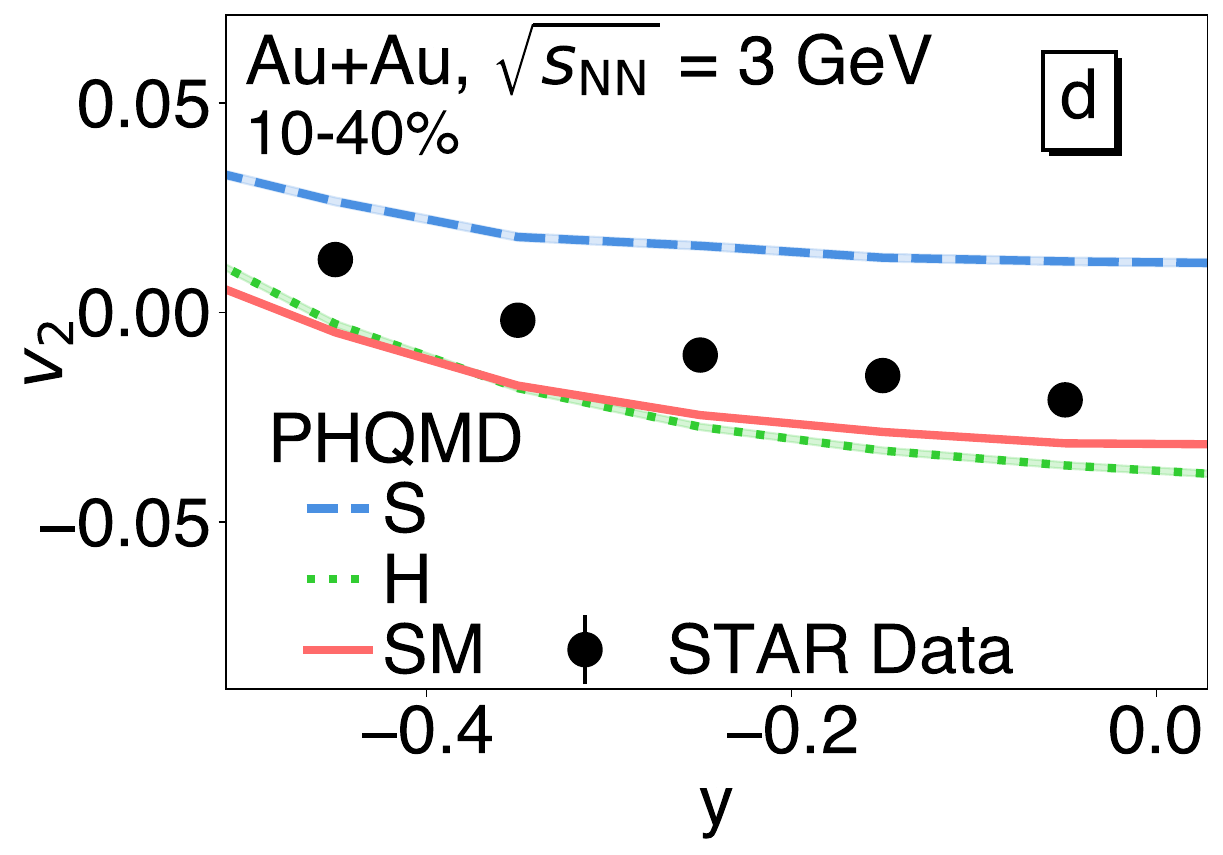}
    \includegraphics[width=0.49\linewidth]{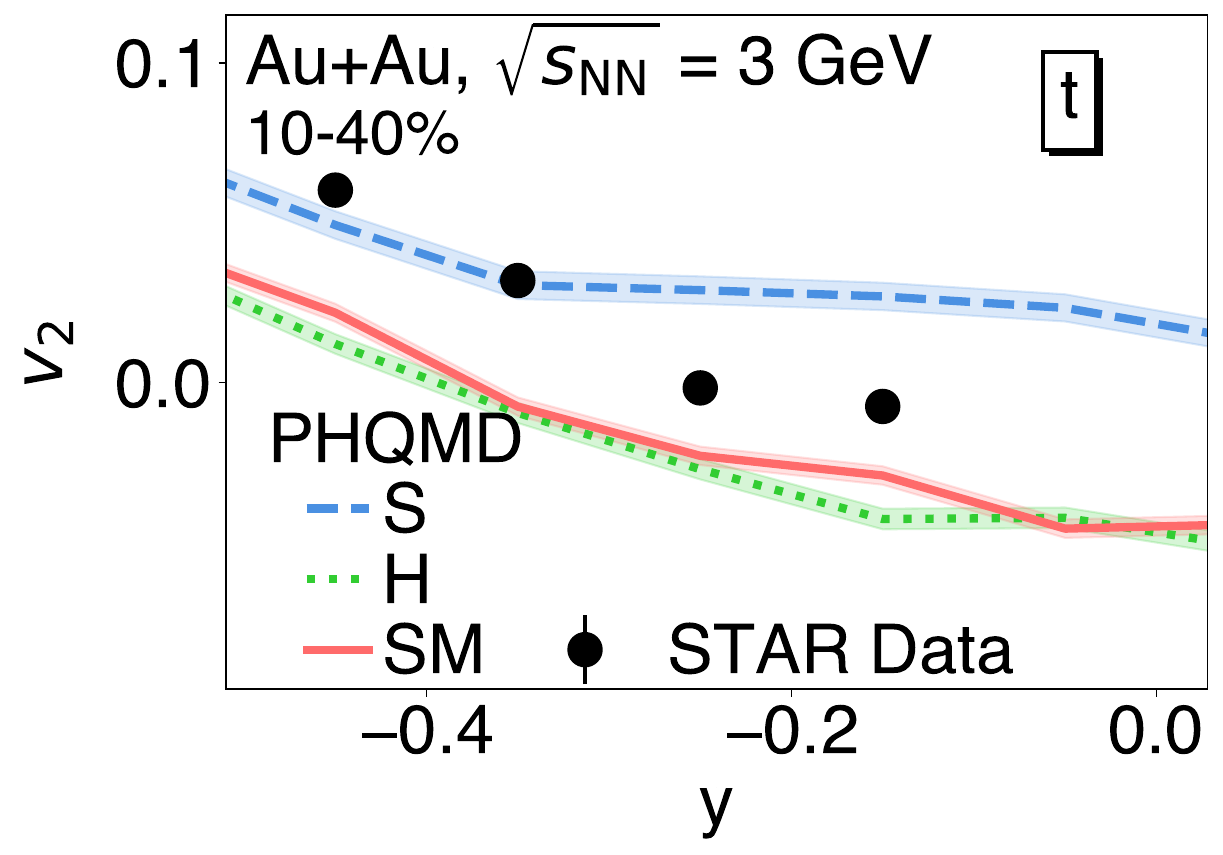}
    \includegraphics[width=0.49\linewidth]{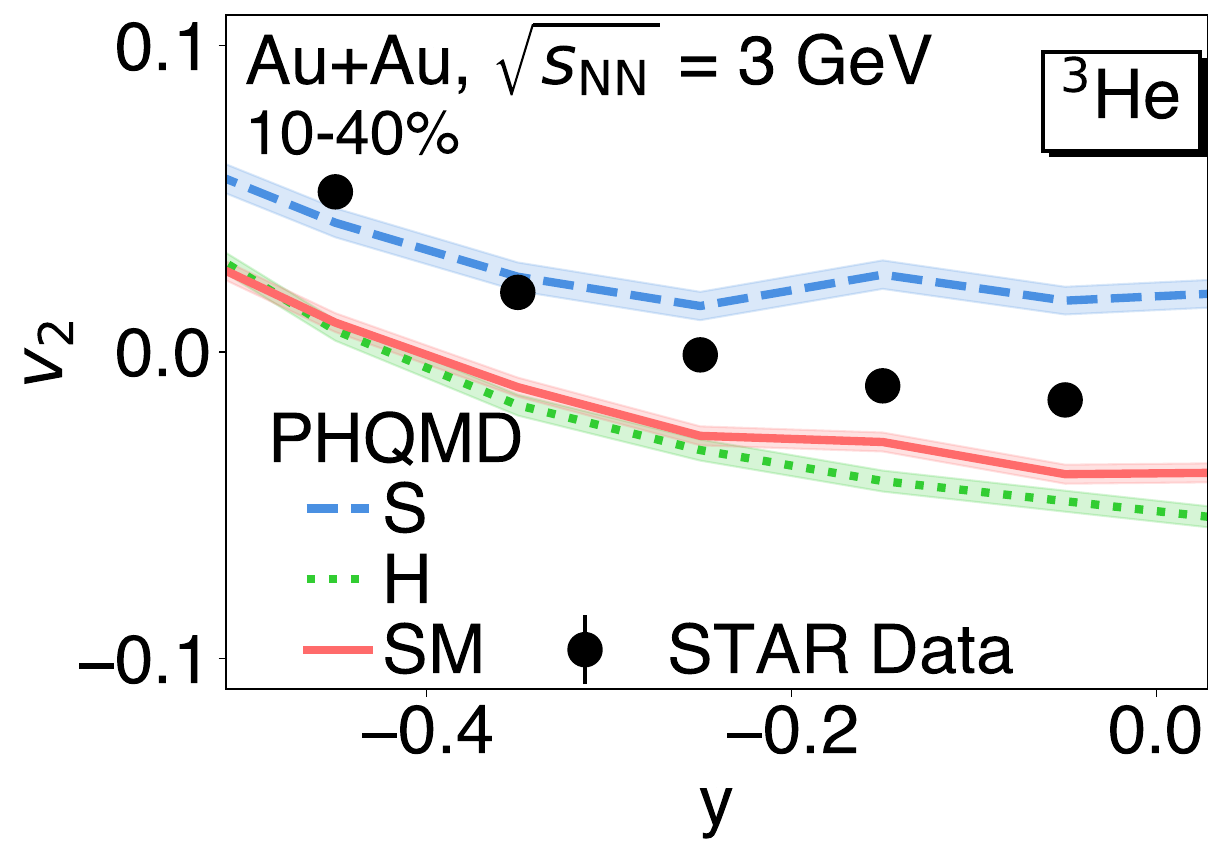}
    \includegraphics[width=0.49\linewidth]{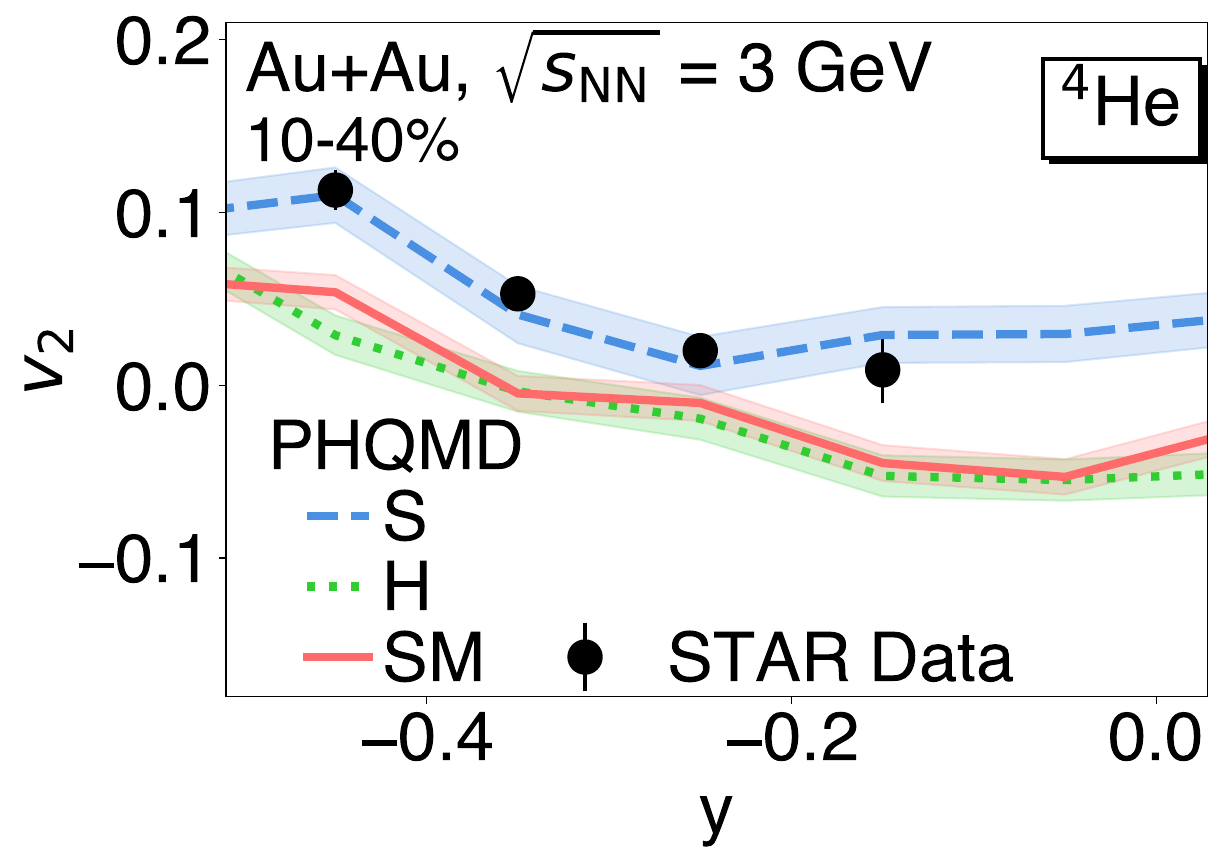}
    \includegraphics[width=0.49\linewidth]{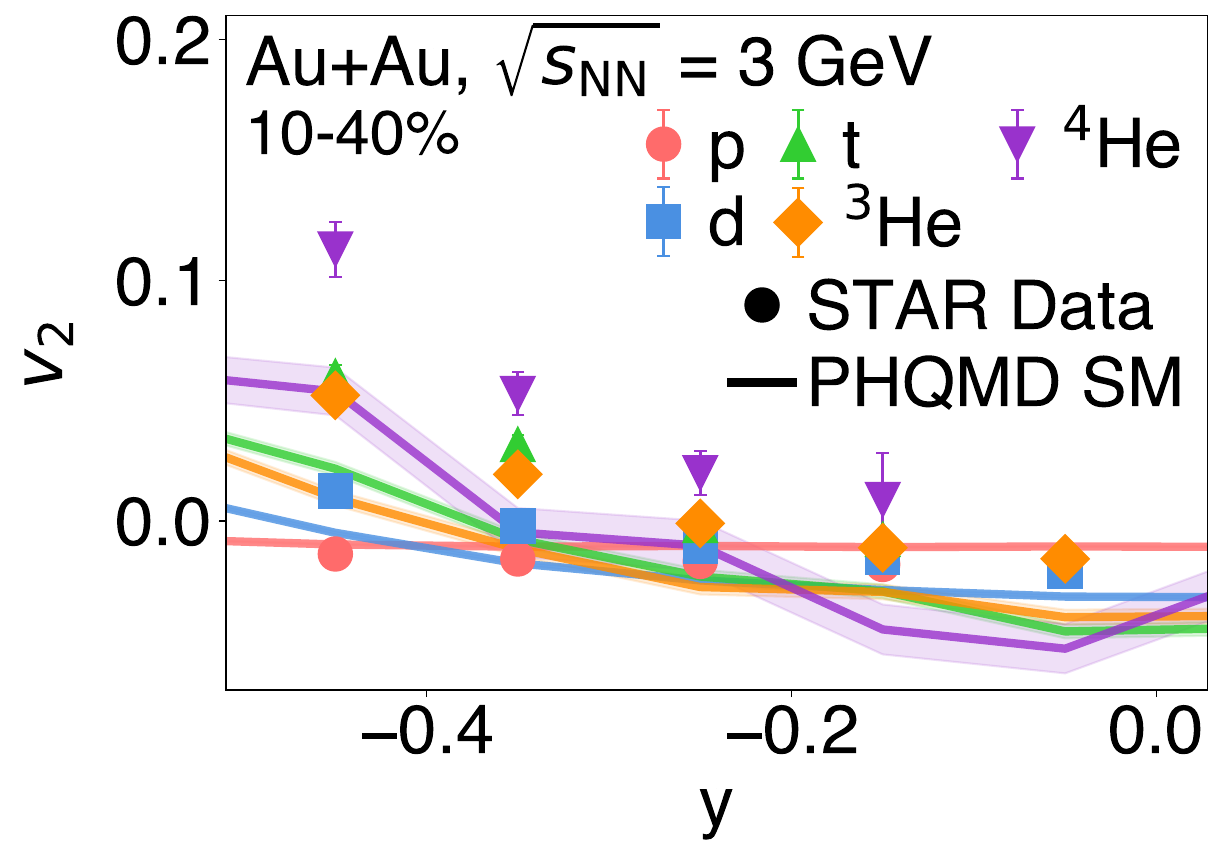}
    \caption{The elliptic flow $v_2$ of proton, deuteron, triton, $^{3}\rm{He}$, and $^{4}\rm{He}$ as a function of rapidity $10-40\%$ $\sqrt{s_{\rm{NN}}}=3$ GeV Au$+$Au collisions. The STAR data~\cite{STAR:2021ozh} are shown as colored markers while PHQMD calculations are shown as colored lines or bands. 
    \label{fig:v2y_pdth34}}
\end{figure}

Now we step to a comparison of the PHQMD results for the elliptic flow for Au+Au collisions at $\sqrt{s_{\rm{NN}}}=3$ GeV with the STAR data~\cite{STAR:2021yiu}. In Fig.~\ref{fig:v2y_pdth34} we show the elliptic flow ($v_2$) of protons, deuterons and tritons, $^{3}\rm{He}$, and $^{4}\rm{He}$ as a function of the rapidity $y$ and integrated in the chosen $p_T$ ranges for the three EoS. For protons, $v_2$ is slightly negative in the considered $p_T$ interval. One can see that the $v_2$ of protons is quite well described by the SM EoS. The H EoS predicts almost zero $v_2$, while the S EoS predicts a positive $v_2$, which is opposite to the data. For deuterons, tritons, $^{3}\rm{He}$, and $^{4}\rm{He}$, the soft EoS yields a positive $v_2(y)$, in contradistinction to the negative $v_2$ of the experimental data. For all clusters SM and H EOS give rather similar results . Both show a more negative $v_2$ than the data in the whole rapidity range in distinction to the PHQMD calculations at $\sqrt{s_{\rm{NN}}}= 2.4$ GeV \cite{Kireyeu:2022qmv}, where the $v_2$ for SM and H is less negative than the data, however in a slightly different $p_T$ and centrality interval. It is remarkable that the sign of $v_2$ depends on the cluster size. For protons, $v_2$ is always negative, for $d$, $t$ and $^3{\rm{He}}$ it changes from positive values at $y=-0.5$ to negative values close to midrapidity, whereas for heavier fragments $v_2$ is always positive. PHQMD calculations with H and SM reproduce the trend of $v_2$ as a function of the rapidity.

In the bottom right plot of Fig.~\ref{fig:v2y_pdth34} we compile $v_2(y)$ of the experimental results and the theoretical predictions with a SM EoS for protons, deuterons, triton, $^{3}\rm{He}$, and $^{4}\rm{He}$. The rapidity distributions of $v_2$ for different clusters exhibit a clear mass ordering, similar to $v_1$. This is qualitatively reproduced by the PHQMD calculations using a H or SM EoS. At mid-rapidity, $-0.1<y<0$, the value of $v_2$ is negative and nearly identical for $p$, $d$, and $^{3}\rm{He}$. A soft EoS give, as said,  positive $v_2$ values for all clusters in $-0.5<y<0$, opposite to the data.

\begin{figure}[h!]             
    \centering 
    \includegraphics[width=0.49\linewidth]{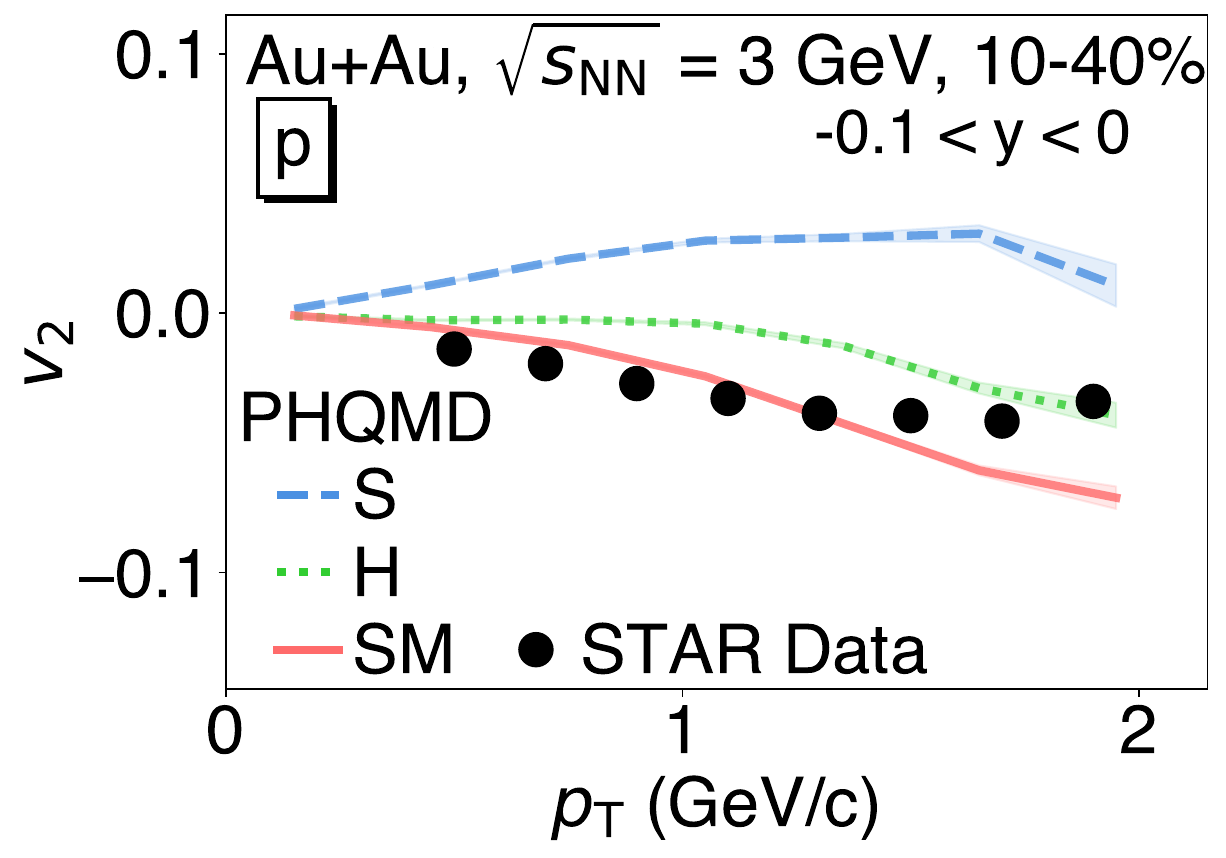}
    \includegraphics[width=0.49\linewidth]{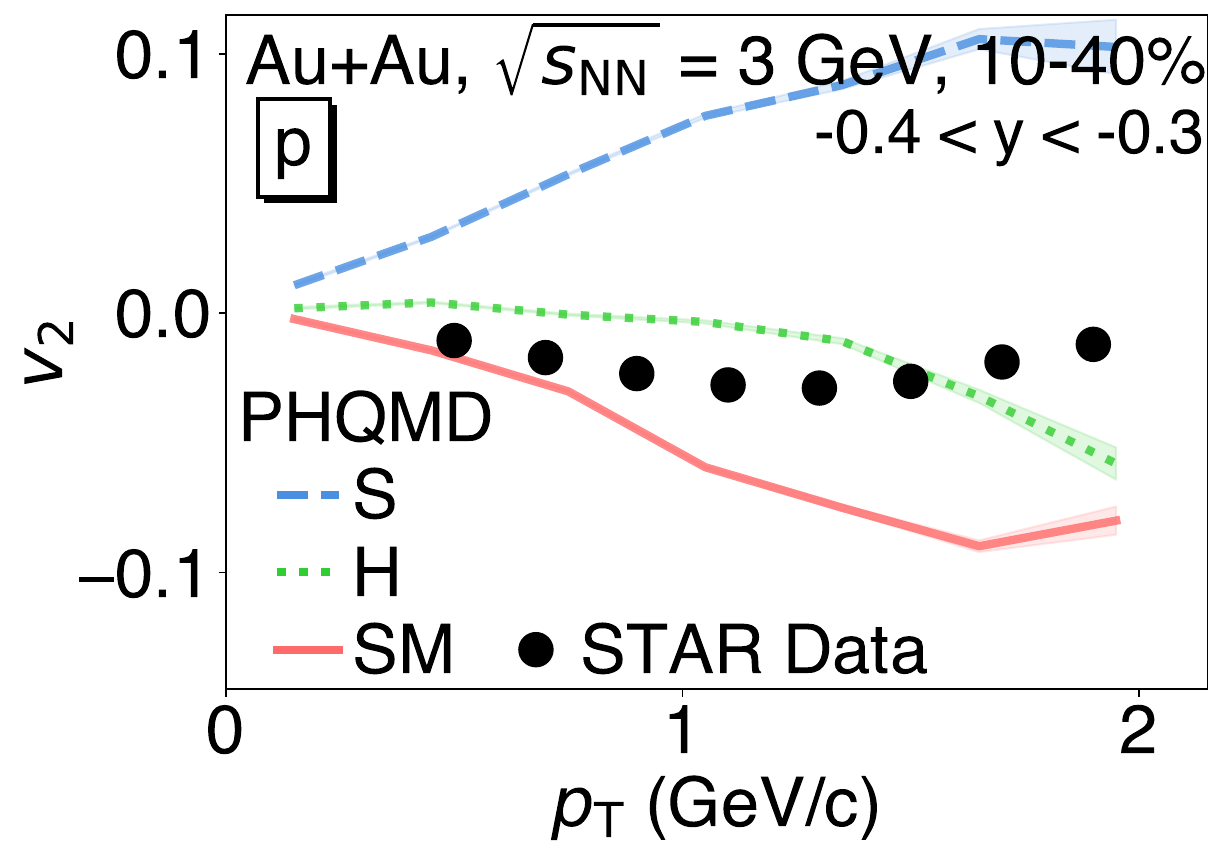}
    \includegraphics[width=0.49\linewidth]{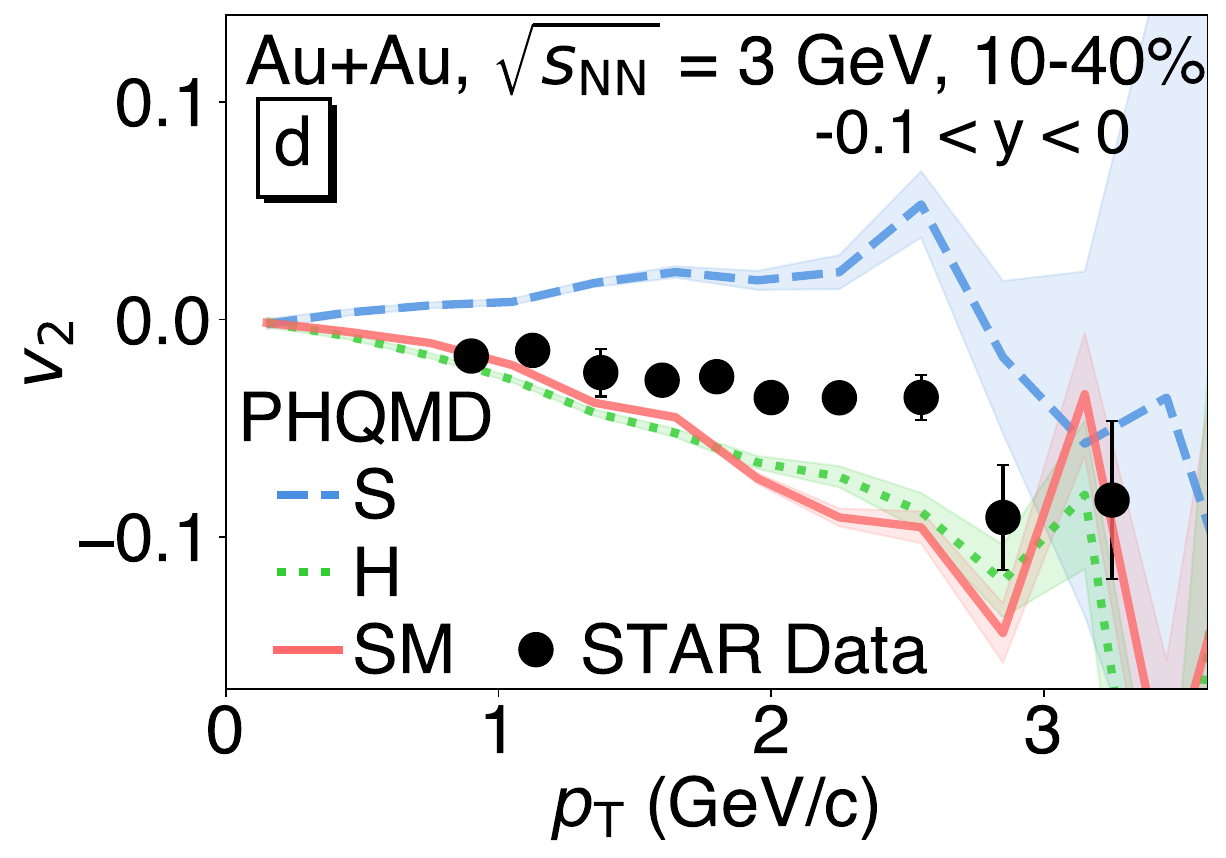}
    \includegraphics[width=0.49\linewidth]{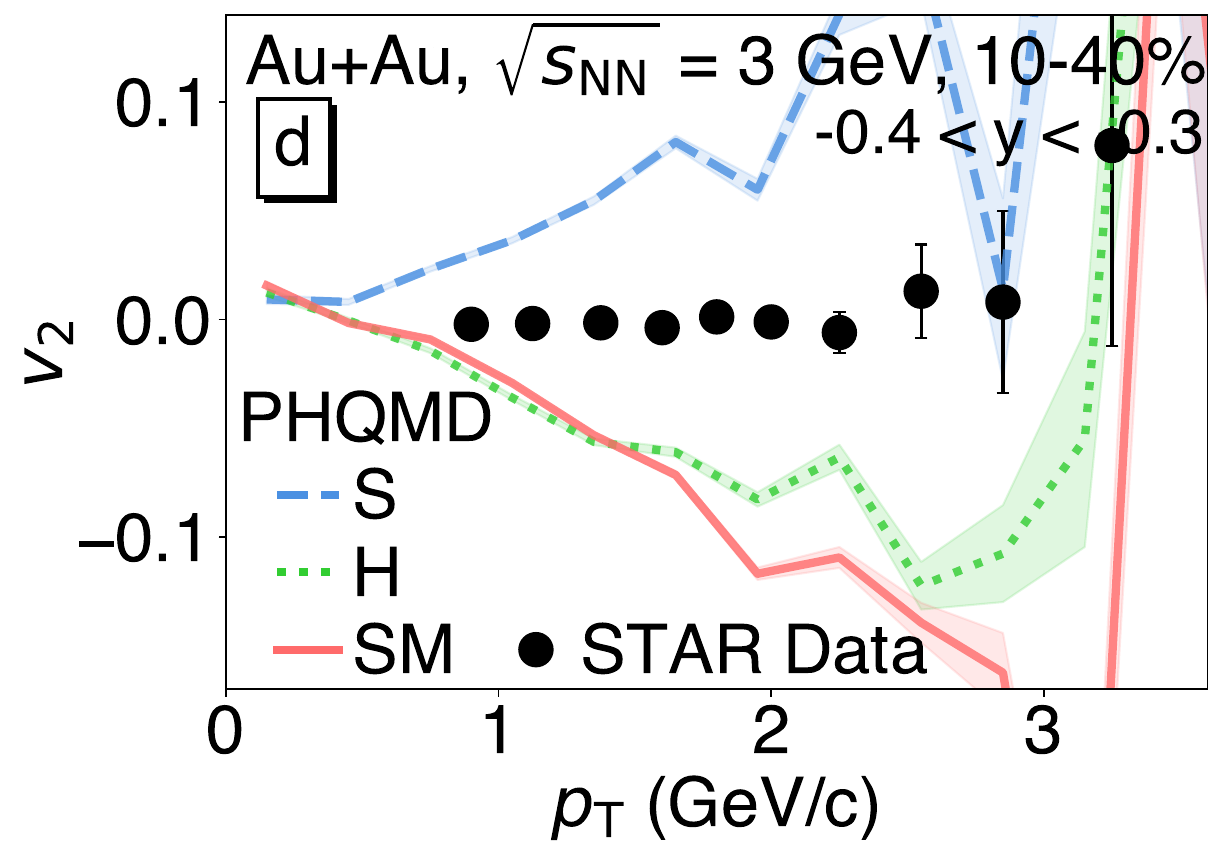}
    \includegraphics[width=0.49\linewidth]{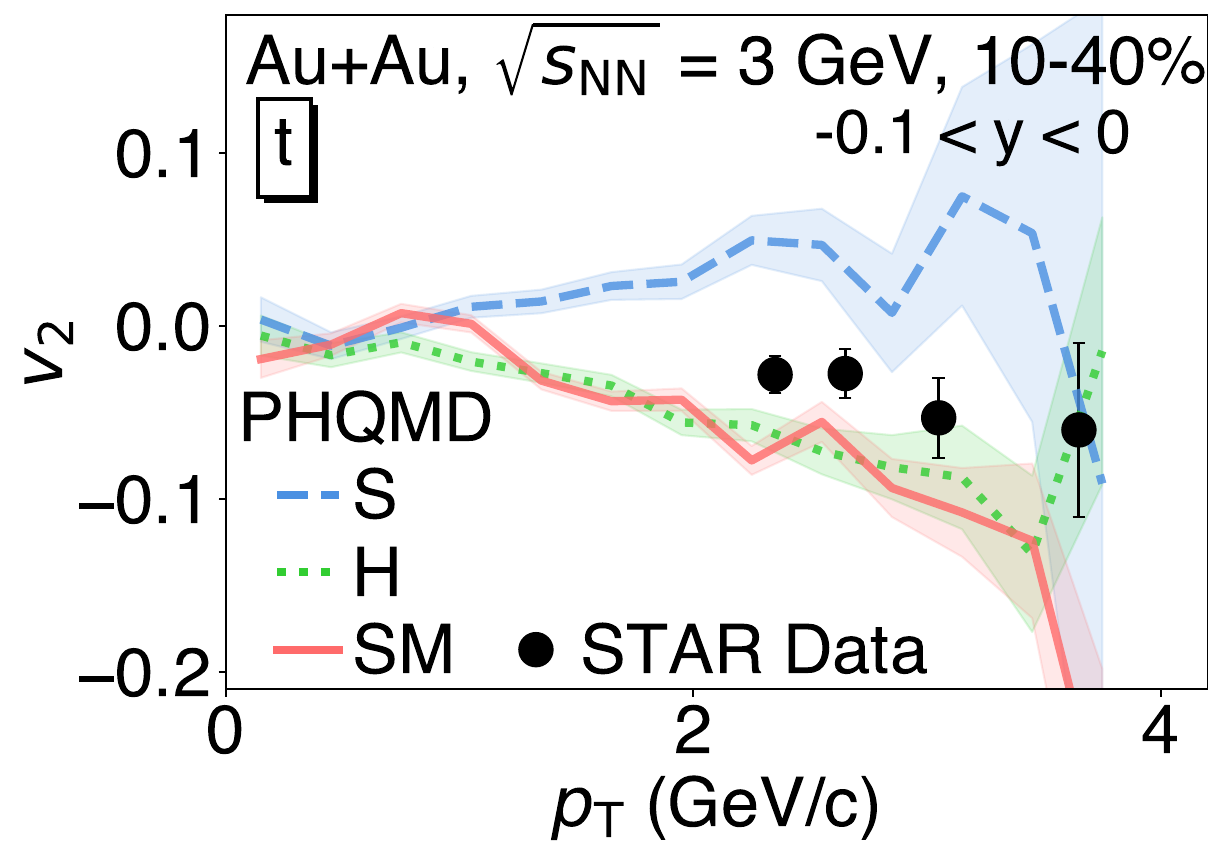}    
    \includegraphics[width=0.49\linewidth]{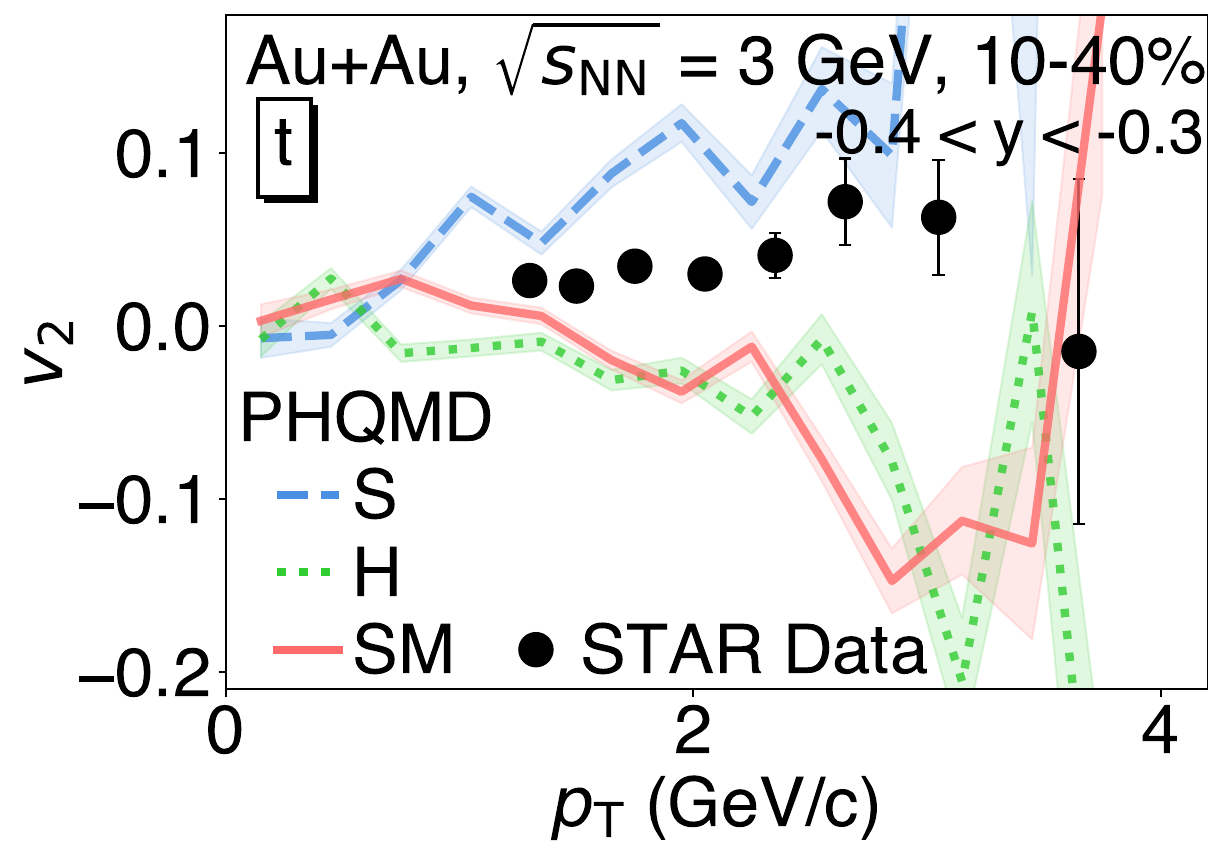}
    \includegraphics[width=0.49\linewidth]{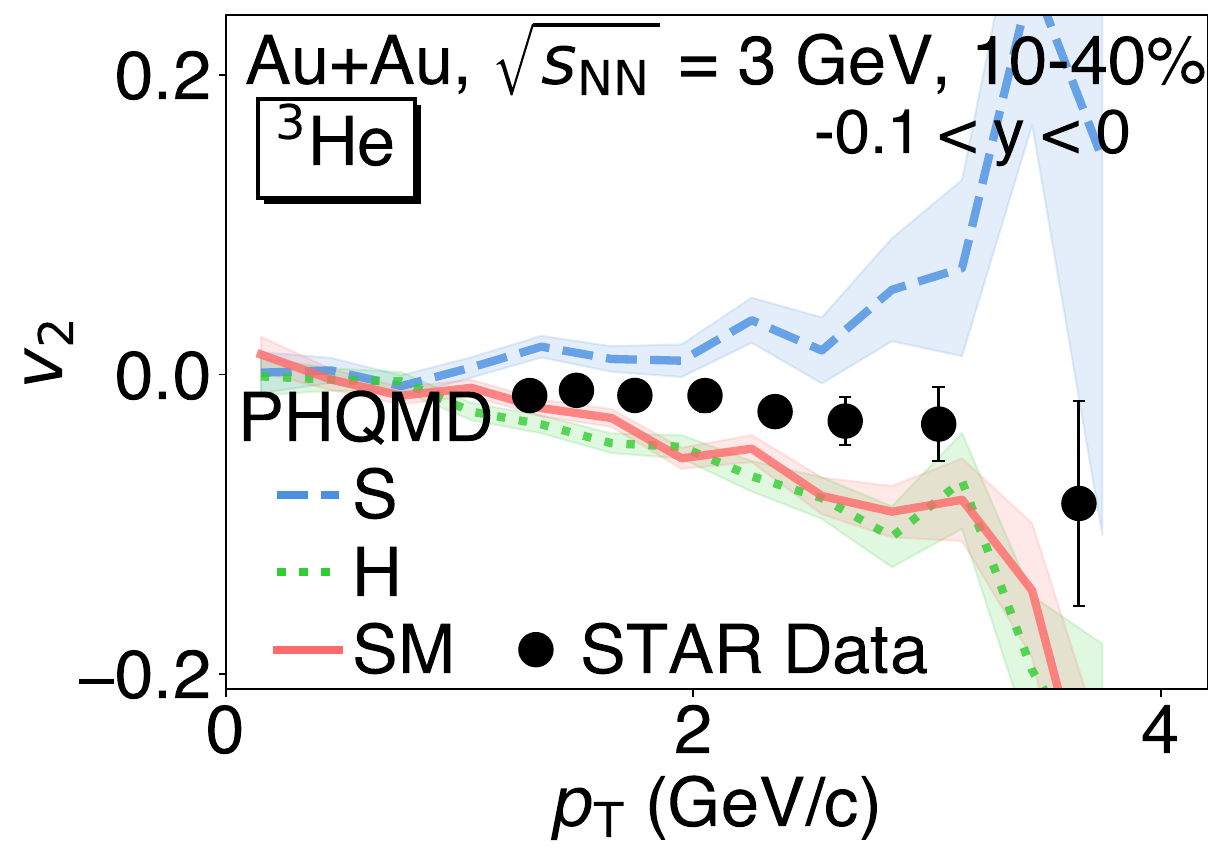}
    \includegraphics[width=0.49\linewidth]{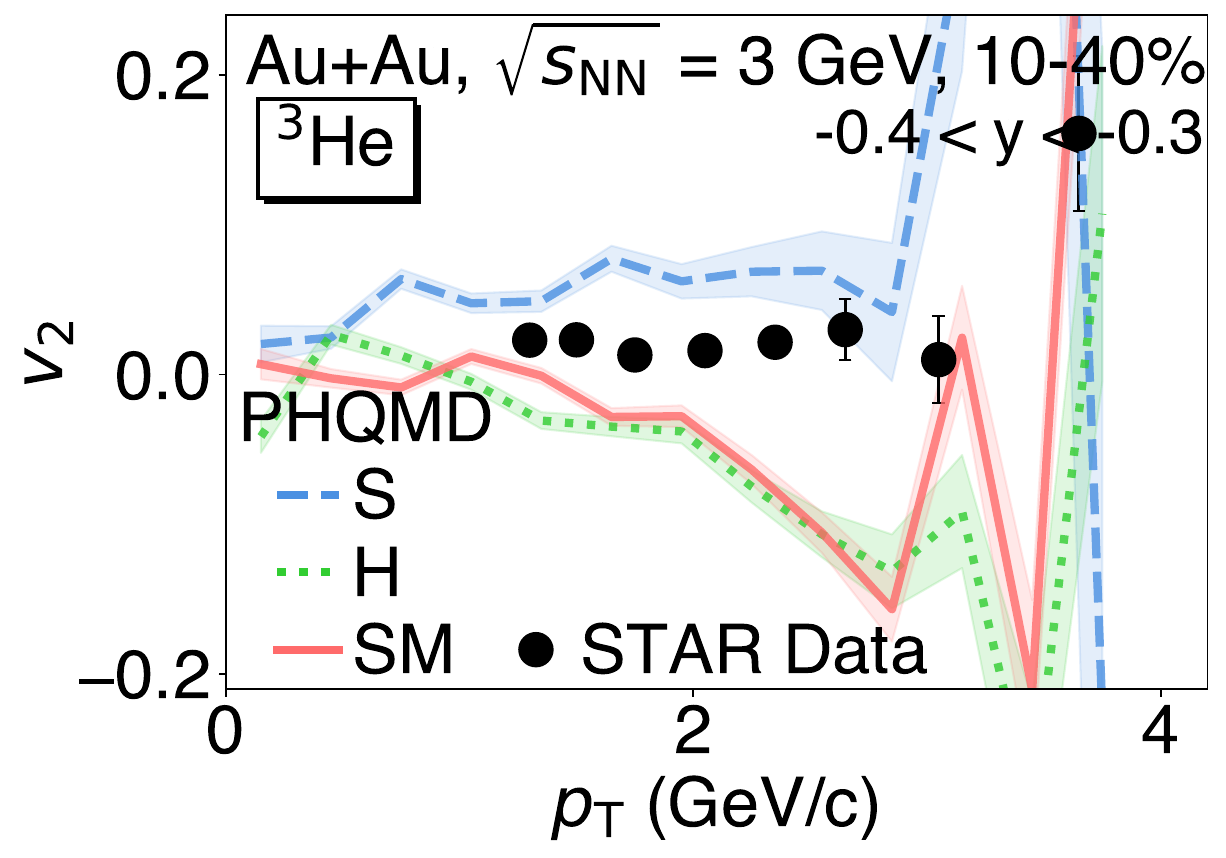}
    \caption{The elliptic flow $v_2$ of protons, deuterons, tritons, $^{3}\rm{He}$, and $^{4}\rm{He}$ in the rapidity regions $(-0.1,0)$ and $(-0.4,-0.3)$ as a function of $p_T$ in $10-40\%$ $\sqrt{s_{\rm{NN}}}=3$ GeV Au$+$Au collisions. The STAR data~\cite{STAR:2021ozh} are shown as black markers while PHQMD calculations are shown as colored bands. 
    \label{fig:v2pt_pdth3}}
\end{figure}
The transverse momentum dependence of $v_2$ is presented in Fig.~\ref{fig:v2pt_pdth3} for protons, deuterons, tritons, and $^{3}\rm{He}$ as a function of $p_T$ for the rapidity intervals $-0.1<y<0$ (left column) and $-0.4<y<-0.3$ (right column) for $10-40\% $ mid-central Au$+$Au collisions at $\sqrt{s_{\rm{NN}}}=3$ GeV. First of all we observe in the calculations a strong $p_T$ dependence of $v_2(p_T)$ and a strong dependence on the EoS for all clusters and in both rapidity intervals. For both rapidity intervals the soft EoS gives mostly a positive $v_2$ for all clusters, whereas the data show mostly a negative $v_2$. Although the calculations using a H and a SM EoS are quite different for the proton $v_2(p_T)$ , for deuterons, tritons and $^{3}\rm{He}$ the two EoS give almost the same result at forward as well as at mid-rapidity. Both give, however, a less positive or more negative $v_2$ compared to the experimental data. 

For protons at midrapidity, the SM EoS gives the largest negative $v_2(p_T)$, followed by the hard EoS, similar to what has been observed in $v_2(y)$. Calculations with SM give a result close to the experiment data up to $p_T=1.4$ GeV$/c$, but above 1.4 GeV$/c$, the calculations continue to decrease while the data increase. For all clusters H and SM calculations predict a similar $v_2(p_T)$. The $v_2$ calculated by  PHQMD  follows qualitatively the experimental data but give a larger negative $v_2$. 

For the backward rapidity interval ($-0.4<y<-0.3$, Fig.~\ref{fig:v2pt_pdth3}, right column) the difference between theory and experiment is more pronounced. 
The proton data show a $v_2(p_T)$, which is slightly less negative than that in the rapidity interval $-0.1<y<0$,  whereas the PHQMD calculations yield a more negative $v_2$, similar to that observed at midrapidity. The cluster data
display a qualitative different behaviour: The PHQMD results are quite similar to those obtained at midrapidity whereas the deuteron data show a $v_2(p_T)$, which is compatible with zero, and larger clusters show a positive $v_2(p_T)$.  Thus the experimental $v_2$ of clusters changes sign when going from $\sqrt{s_{\rm{NN}}}= 2.4$ to $\sqrt{s_{\rm{NN}}} = 3$ GeV \cite{Kireyeu:2024hjo}.  This indicates that in this rapidity interval clusters are more than  an ensemble of randomly picked nucleons.  
The PHQMD calculations yield,  as for $\sqrt{s_{\rm{NN}}}$ = 2.4 GeV,  for $p_T> 1$ GeV a negative $v_2(p_T)$. This discrepancy between theory and experiment points towards a cluster production process, which is not yet correctly modeled in PHQMD and which will be explored in an upcoming publication.  

\begin{figure}[h!]             
    \centering    
    \includegraphics[width=0.49\linewidth]{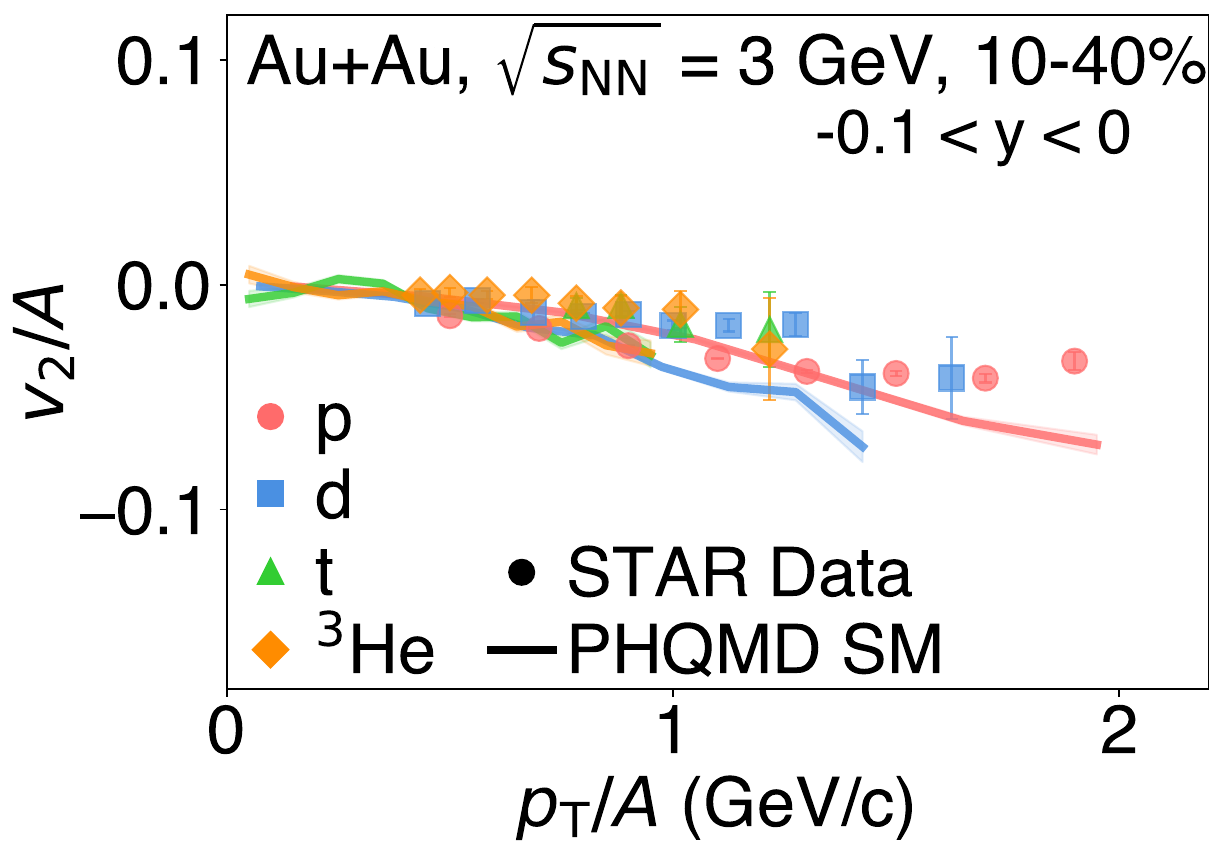}
    \includegraphics[width=0.49\linewidth]{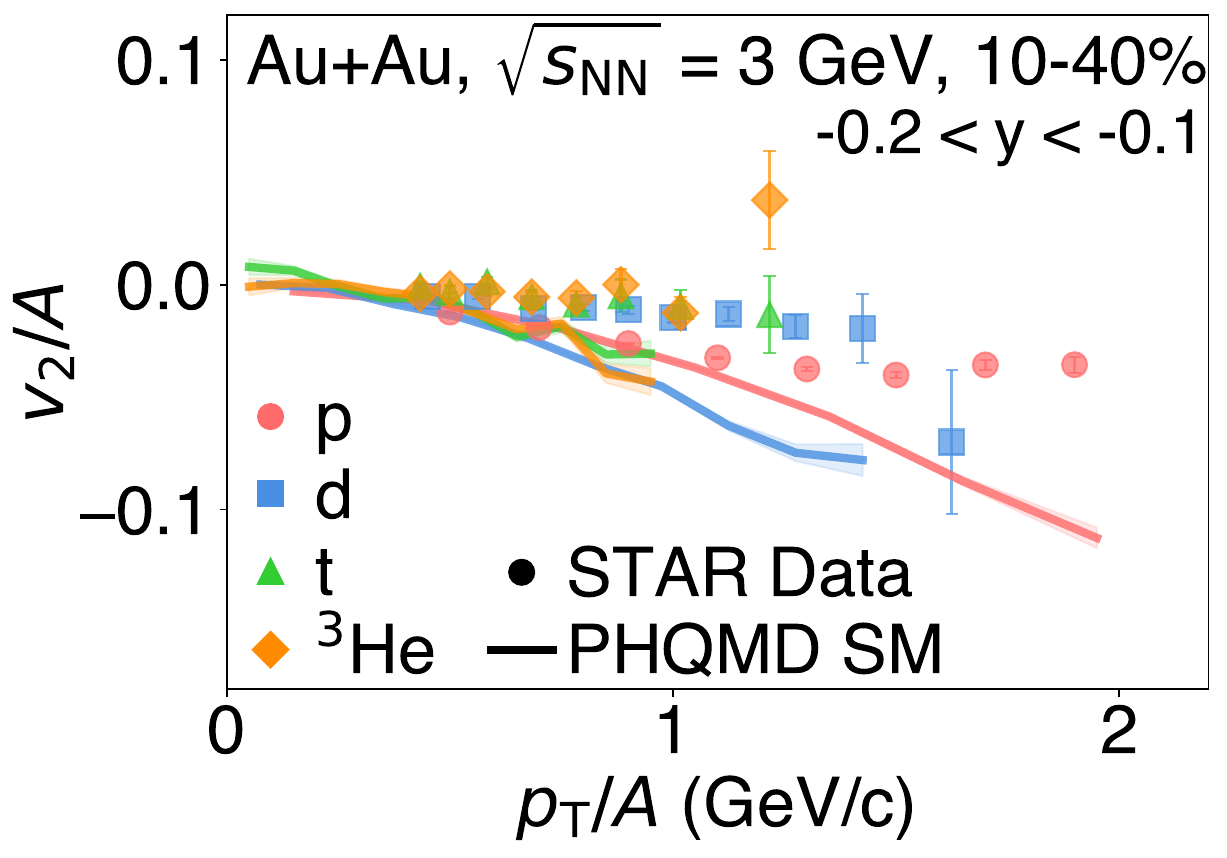}
    \includegraphics[width=0.49\linewidth]{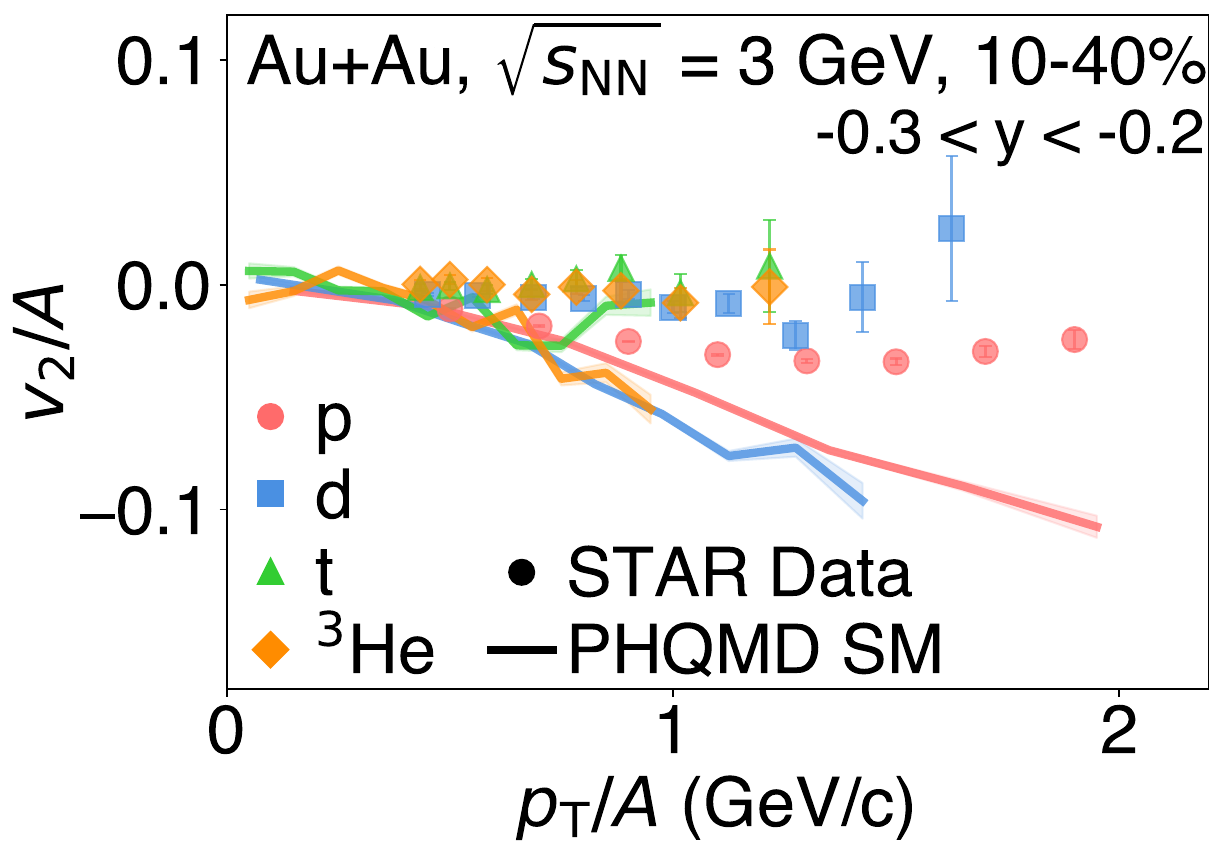}
    \includegraphics[width=0.49\linewidth]{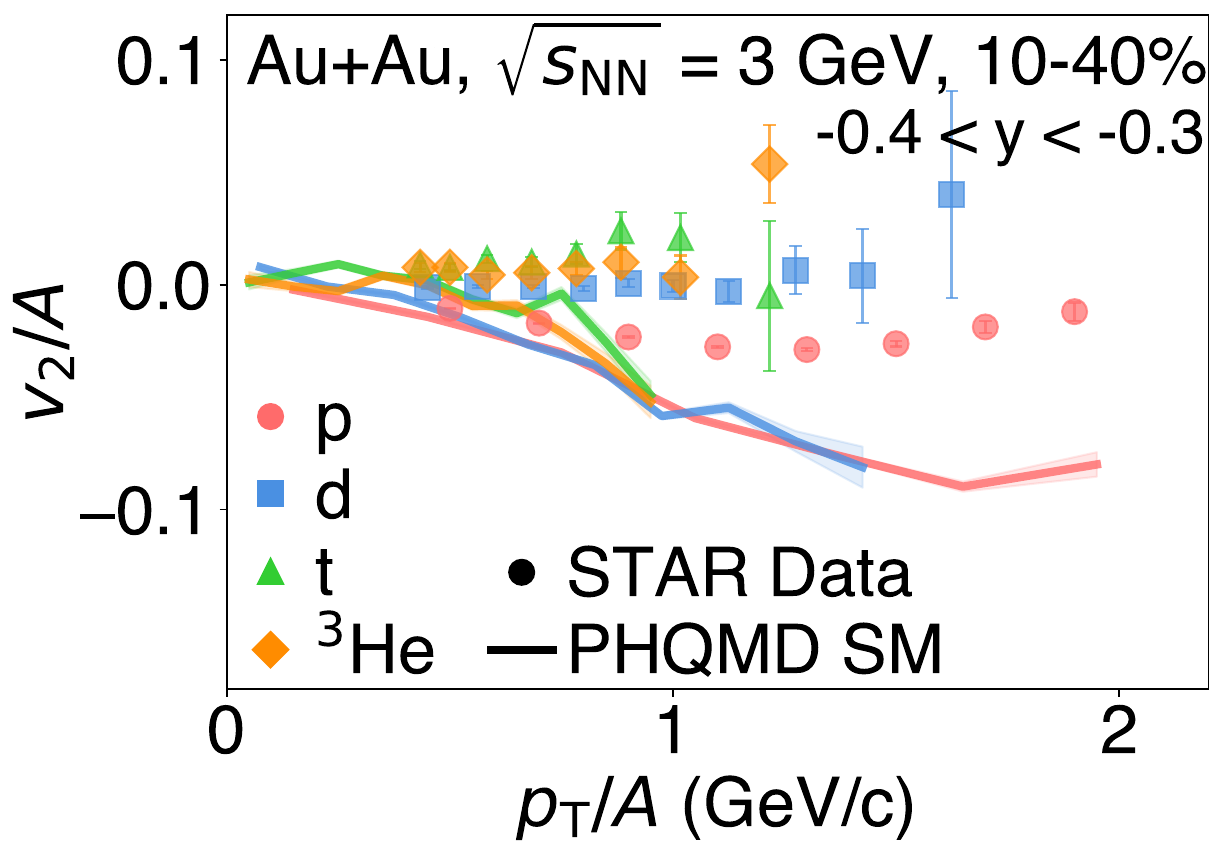}
    \caption{\Yj{The mass number scaled elliptic flow $v_2/A$ of protons, deuterons, tritons, and $^{3}\rm{He}$ as a function of $p_T/A$ in different rapidity intervals in $10-40\%$ mid-central Au$+$Au collisions at $\sqrt{s_{\rm{NN}}}=3$ GeV}. The STAR data~\cite{STAR:2021ozh} are shown as colored markers while PHQMD SM EoS calculations are shown as colored bands.
    \label{fig:v2AptA_pdth3}}
\end{figure}

\Yj{Fig.~\ref{fig:v2AptA_pdth3} shows the compilation of elliptic flow, scaled by the mass number $v_2/A$, for protons, deuterons, tritons, and $^{3}\rm{He}$ in different rapidity intervals for the SM EoS at $\sqrt{s_{\rm{NN}}}=3$ GeV in comparison with the data}. \Yj{Data show although the $v_2$ of the light clusters approximately follows a mass number scaling at midrapidity, in data  as well as in the calculations, the protons do not}.  At $\sqrt{s_{\rm{NN}}}= 2.4$ \cite{HADES:2020lob} a scaling of  $v_2/A(p_2/A)$ for protons and clusters has been observed.  This scaling is predicted if one assumes that $A$ nucleons combine randomly to form a cluster of size $A$. The violation of this scaling indicates that at $\sqrt{s_{\rm{NN}}}$ = 3  GeV clusters are not randomly combined nucleons but that another cluster formation mechanism exist, which are only approximately  captured in the PHQMD approach

However, in data the $v_2(p_T) $ of protons is less than $v_2/A(p_2/A)$ of clusters, for the PHQMD calculations it is opposite.
For the data the fact that  $v_2/A(p_2/A)$ scales for clusters but is different from
$v_2/A(p_2/A)$ for protons continues to larger negative rapidities whereas the PHQMD results of  $v_2/A(p_2/A)$ for protons approaches with increasing negative rapidity that of the clusters.

As already seen in the last figure, the PHQMD calculations reproduce the
experimental $v_2(p_T)$ of protons up to $p_T \approx 1.4$ GeV$/c$  but continue to decrease with increasing $p_T$ whereas in experiment $v_2(p_T)$ of protons bends over. 

\begin{figure}[h!]             
    \centering    

    \includegraphics[width=0.8\linewidth]{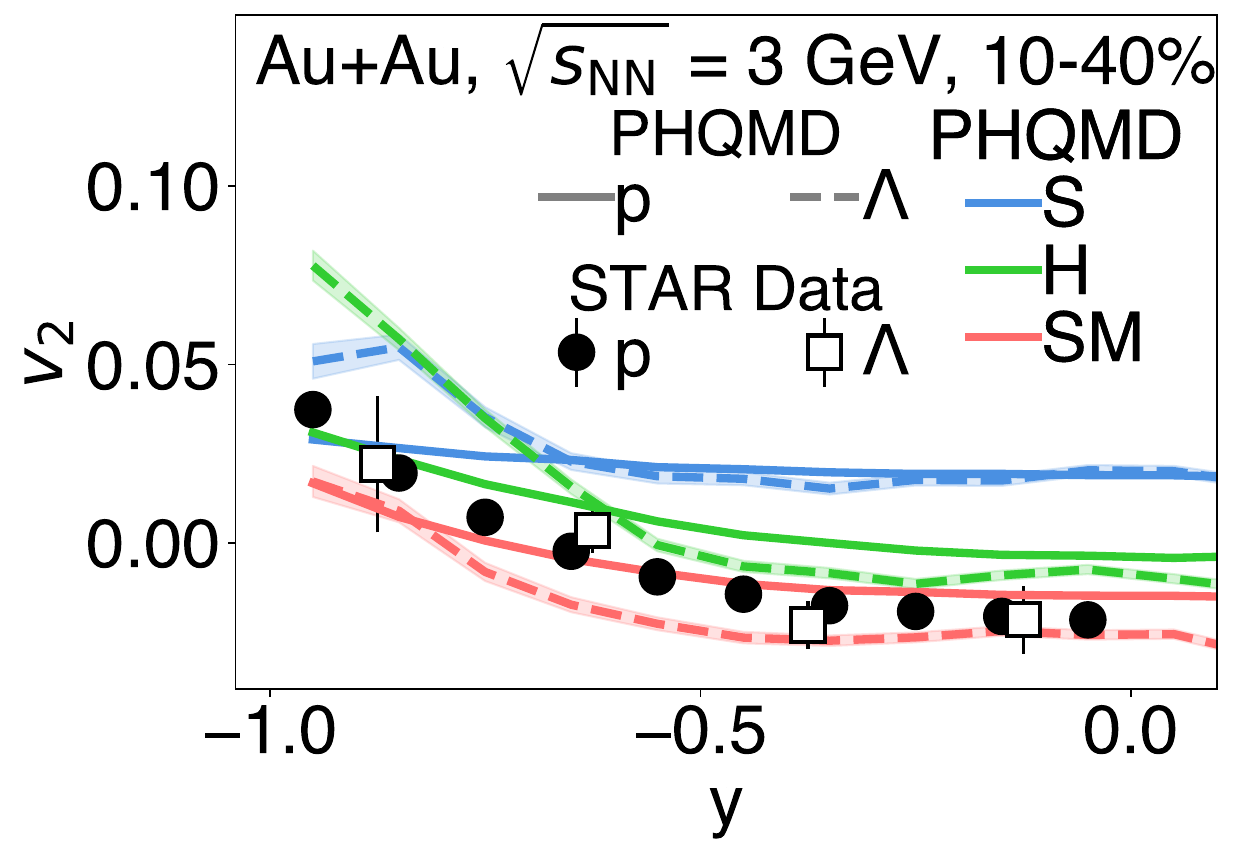}
    \caption{The elliptic flow $v_2$ of proton and $\Lambda$ as a function of rapidity in $10-40\%$ $\sqrt{s_{\rm{NN}}}=3$ GeV Au$+$Au collisions. The STAR data~\cite{STAR:2021yiu} are shown as closed (proton) and open ($\Lambda$) markers while PHQMD calculations are shown as solid (proton) and dashed ($\Lambda$) lines.
    \label{fig:v2ylambdaproton}}
\end{figure}

In Fig.~\ref{fig:v2ylambdaproton} we present the PHQMD results for $v_2(y)$ of protons  and $\Lambda$ in comparison to the STAR data~\cite{STAR:2021ozh} for $\sqrt{s_{\rm{NN}}}=3$ GeV in $10-40\%$ central Au+Au collisions and for the $p_T$ interval $0.4<p_T<2.0$ GeV$/c$ for calculations with a H, S and SM EoS.  We see, first of all, that in experiment $v_2$ for $\Lambda$ and $p$ agree within error bars in the whole rapidity range. This is not the case for the PHQMD calculations. We observe that a soft EoS substantially overestimates the $v_2(y)$ for both, $\Lambda$ and proton, in this $p_T$ interval. The SM EoS gives a good description of the data. For H EoS, the proton $v_2(y)$ is overestimated, particularly at mid-rapidity, while for $\Lambda$ a H EoS can describe the data for $y \ge -0.6$, although this may be partly due to the uncertainty of the data. In PHQMD calculations, the proton $v_2$ is less negative than that of $\Lambda$ at mid-rapidity for all EoS, but the situation appears reversed near target rapidity. We note, however, that in the target
rapidity domain proton emission from spectator remnants may occur in a time scale larger than the time for which we pursue the calculation.

\section{Summary}
\label{sec:summary}

In this study we have compared the results of PHQMD calculations with the most extensive data set (of 3 GeV Au+Au collisions) from the Beam Energy Scan Program phase II of the STAR collaboration, which is presently and in near future available in the energy range of the future FAIR accelerator. The goal was to see how transport calculations for hadrons and, more important, for clusters compare with these high statistics data. For this purpose we studied multiplicities, $p_T$ spectra, rapidity distributions as well as the in-plane ($v_1$) and elliptic ($v_2$) flow for baryons and clusters and applied three different equations of state, a static hard and soft EOS and a momentum dependent soft EoS.

We see, first of all, that at these energies different hadronic EoS gives quite different results for $v_1$ and $v_2$ but influence as well the $p_T$ distributions of elementary baryons and clusters. 

Most of the $\sqrt{s_{\rm{NN}}}=3$ GeV Au+Au data are best described by calculations using a soft momentum dependent EoS (SM), which takes into account the measured momentum dependence of the optical potential. 

The SM calculations are in good agreement with the experimental $p_T$ distributions of protons, $\Lambda$s, clusters and hyper-clusters over 5 orders of magnitude and reproduce the cluster yield at mid rapidity with 30\% accuracy for deuterons and up to a factor of two for $^4 \rm{He}$. Also the rapidity distribution of hypernuclei $^3_\Lambda \rm{H}$ and $^4_\Lambda \rm{H}$ are nicely reproduced. 
\EB{ Although we know from data that the potential between nucleons is momentum dependent we find  that a static hard EoS shows a similar trend as a momentum dependent soft EoS. High quality upcoming  data will hopefully help to disentangle  both for a better understanding of EoS. }

The flow of hadrons and clusters shows some surprising results. The experimental in-plane and directed flow of $p$ and $\Lambda$ agrees within error bars whereas their radial flow differs. In PHQMD the difference in the radial flow is due to the fact that the production of strange hadrons happens more probable close to the center of the reaction. In PHQMD calculations this provokes also a difference in the in-plane flow. 

The PHQMD calculations show that also at this energy the in-plane flow $v_1$ depends on the nuclear equation of state and it is assuring that different extrapolations of the optical potential from the measured beam energy region to the energies required for studying $\sqrt{s_{\rm{NN}}} = 3$ GeV reactions do not influence the result for $v_1$. 
PHQMD calculations show the same scaling and the same functional form of $v_1/A(p_T/A)$ as seen in the data, but the numerical value of $v_1/A(p_T/A)$ is slightly smaller. 

For the elliptic flow the data at $\sqrt{s_{\rm{NN}}} = 3$ GeV do not show anymore the scaling in $v_2/A(p_T/A)$ observed at $\sqrt{s_{\rm{NN}}} = 2.4$ GeV. This is a strong hint that at this energy the clusters are not anymore formed by a random selection of nucleons but that their production is a selective process. The scaling is still present for clusters but the protons deviate. The same is observed for PHQMD calculations.
For $y \leq 0.2$ the data for $v_2/A(p_T/A)$ approaches zero for clusters whereas PHQMD calculations show still a finite negative $v_2$. The understanding of this observation  will be the subject for a future investigation. One has also to take into account that different extrapolations of the optical potential to the energies of interest here influence $v_2$ in a non-negligible way.

The general qualitative and in many kinematical regions quantitative agreement between data and PHQMD calculations, even for the very sensitive flow observables, confirms the maturity of the transport approaches and that the underlying physical processes are to a large extent understood. The above mentioned discrepancies show as well where more intensive studies are necessary to improve these approaches as far as cluster and strangeness production is concerned. 

\clearpage

\appendix{APPENDIX A: }
\begin{figure*}[h!]             
    \centering    \includegraphics[width=0.95\linewidth]{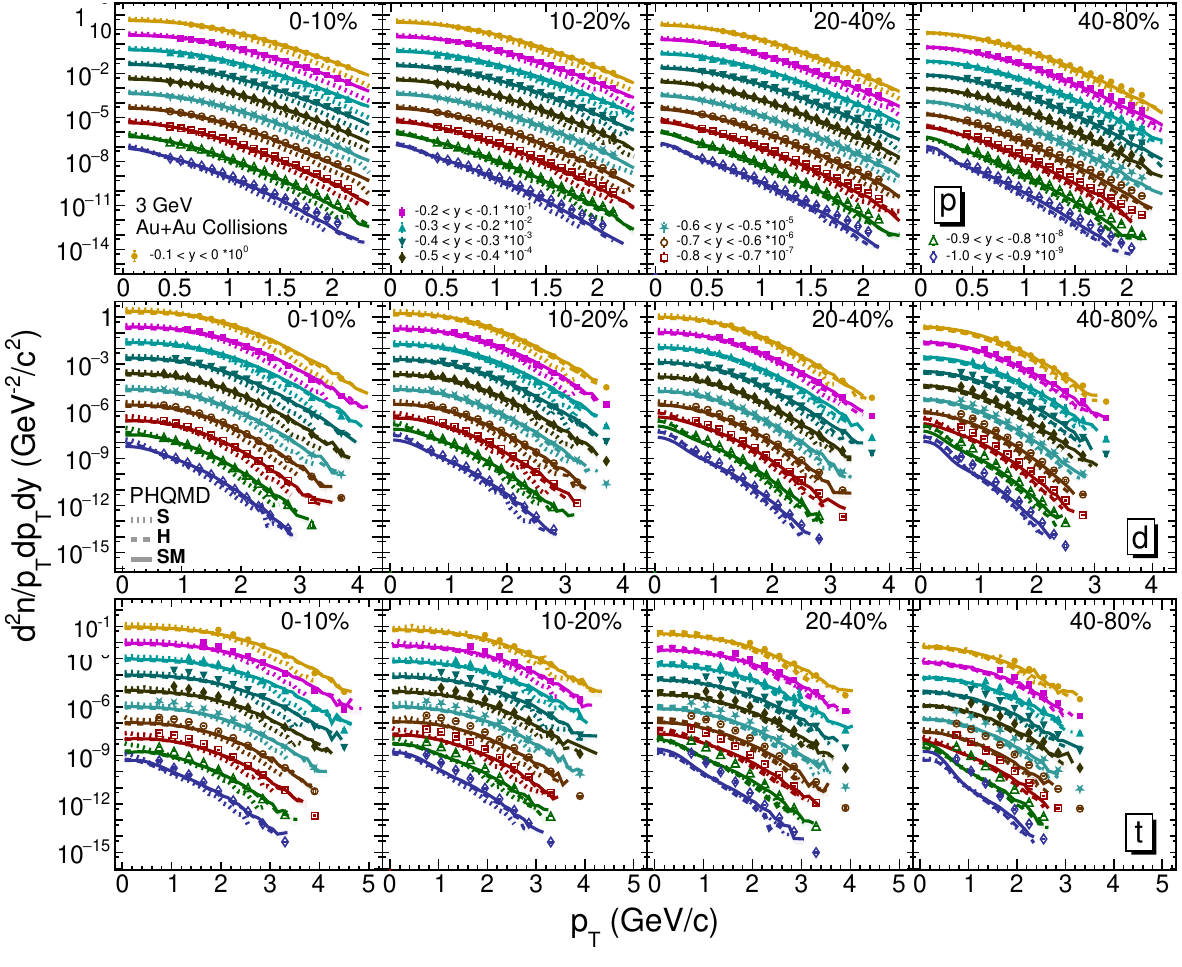}
    \centering    \includegraphics[width=0.95\linewidth]{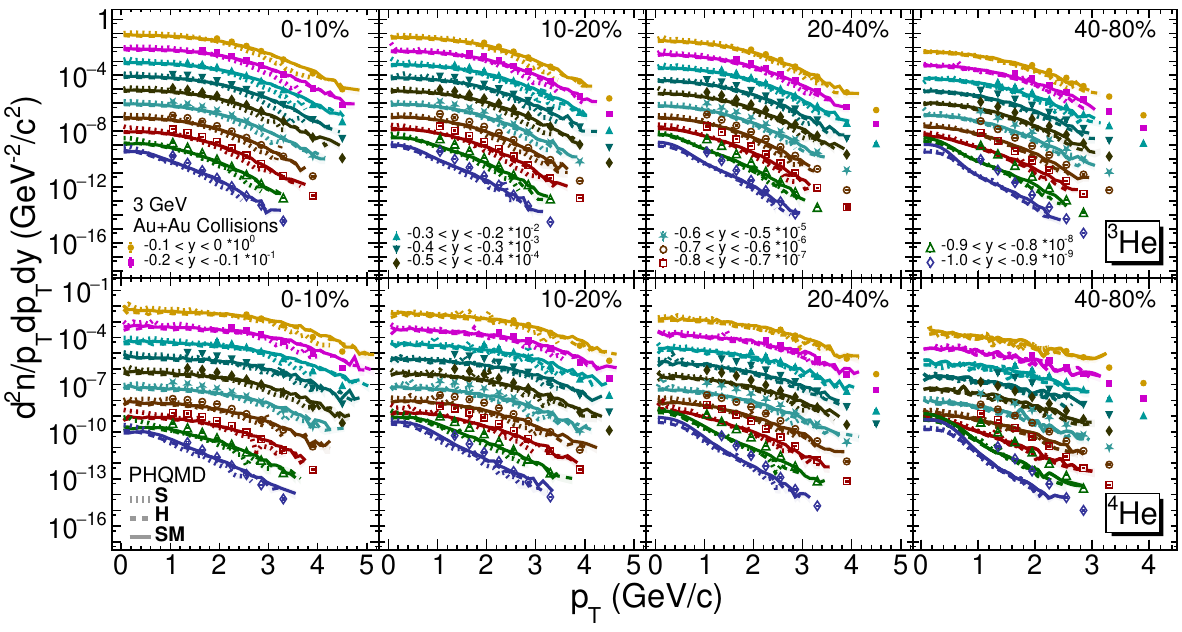}
    \caption{The transverse-momentum spectra of proton, deuteron, triton, ${}^{3}\rm{He}$ and ${}^{4}\rm{He}$ for different rapidities and centralities in Au+Au collisions at $\sqrt{s_{\rm{NN}}}=3$ GeV. The proton spectra have been corrected for the feed-down from $\Lambda$ decays. The measured data points from STAR~\cite{STAR:2023uxk} are shown as colored markers and the calculations from PHQMD using soft EOS, hard EOS, and soft EOS with momentum dependence are shown as dotted, dashed, and solid lines respectively. The data points and model calculations are scaled by factors of 1/10 from mid- to forward rapidities as indicated in the legend. 
    \label{fig:pTspectrapdt}}
\end{figure*}

\begin{figure*}[h!]             
    \centering    \includegraphics[width=0.94\linewidth]{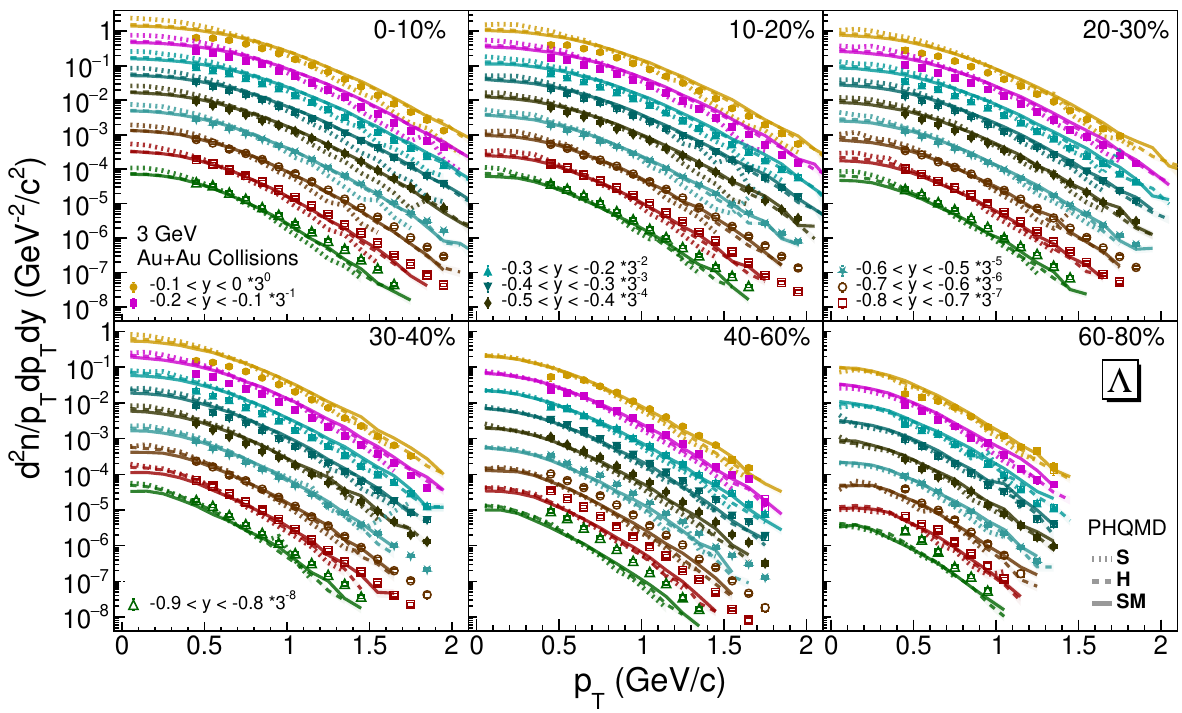}
    \caption{The transverse-momentum spectra of $\Lambda$ for different rapidities and centralities in Au+Au
collisions at $\sqrt{s_{\rm{NN}}}=3$ GeV. The spectra have been corrected for the feed-down from $\Xi^{-}$ and $\Xi^{0}$ decays. The measured data points from STAR~\cite{STAR:2024znc} are shown as colored markers and the calculations from PHQMD using soft EOS, hard EOS, and soft EOS with momentum dependence are shown as dotted, dashed, and solid lines respectively. The data points and model calculations are scaled by factors of 1/3 from mid- to forward rapidities as indicated in the legend. 
    \label{fig:pTspectralambda}}
\end{figure*}

\bibliography{phqmdref} 
\end{document}